\documentclass[
  aps,
  prd,
  reprint,
  superscriptaddress,
  nofootinbib,
  floatfix
]{revtex4-2}

\usepackage[T1]{fontenc}
\usepackage{newtxtext,newtxmath}

\usepackage{amsmath,amssymb,amsthm,amstext}
\usepackage{bm}
\usepackage{braket}

\usepackage[dvipsnames]{xcolor}
\usepackage{graphicx}
\usepackage{array}
\usepackage{multirow}
\usepackage{booktabs}
\usepackage{siunitx}
\usepackage{microtype}
\usepackage{textcomp}
\usepackage{gensymb}
\usepackage{enumerate}
\usepackage{enumitem}
\usepackage[normalem]{ulem}

\usepackage{algpseudocode}

\usepackage[
  breaklinks,
  colorlinks,
  citecolor=blue,
  linkcolor=blue,
  urlcolor=blue
]{hyperref}
\usepackage{orcidlink}

\newcommand{\be}{\begin{equation}}
\newcommand{\ee}{\end{equation}}

\newcommand{\dd}{\mathrm{d}}
\newcommand{\ii}{\mathrm{i}}
\newcommand{\HH}{\mathcal H}
\newcommand{\PPsi}{\Psi}
\newcommand{\EP}{\mathrm{EP}}

\definecolor{revisiongreen}{RGB}{0,128,0}
\definecolor{revisionblue}{RGB}{0,80,190}

\newenvironment{revision}{}{}

\newcommand{\green}[1]{#1}

\newcommand{\blue}[1]{#1}
\newenvironment{bluerevision}{}{}

\newcommand{\Kern}{K}

\newcounter{prdalgorithm}
\newenvironment{prdalgorithm}[1]{%
  \par\medskip\noindent
  \begin{minipage}{\columnwidth}
  \refstepcounter{prdalgorithm}
  \noindent\hrulefill\par\smallskip
  \noindent\textbf{Algorithm \theprdalgorithm. #1}\par\smallskip
}{%
  \smallskip\noindent\hrulefill
  \end{minipage}\par\medskip
}

\graphicspath{{Figure/}}

\newcommand{\draftgraphic}[2]{%
  \IfFileExists{Figure/#1}{%
    \includegraphics[width=\linewidth]{Figure/#1}%
  }{%
    \fbox{%
      \parbox[c][#2][c]{0.94\linewidth}{%
        \centering
        Figure placeholder\\[2pt]
        \texttt{\detokenize{#1}}%
      }%
    }%
  }%
}
  
\begin{document}
\title{Gravitational Waves from Green's Function Decomposition for a Kerr Black Hole: Equatorial Innermost Stable Circular Orbit Plunge}
\author{Junquan Su \orcidlink{0009-0008-1901-533X}}
\affiliation{Department of Astronomy, Tsinghua University, Beijing 100084, China}
\author{Neev Khera \orcidlink{0000-0003-3515-2859}}
\affiliation{Department of Astronomy, Tsinghua University, Beijing 100084, China}
\author{Abhishek Chowdhuri \orcidlink{0000-0003-4474-790X}}
\affiliation{Department of Astronomy, Tsinghua University, Beijing 100084, China}

\author{Marc Casals}
\affiliation{Institut f\"ur Theoretische Physik, Universit\"at Leipzig,\\ Br\"uderstra{\ss}e 16, 04103 Leipzig, Germany}
%\affiliation{School of Mathematics and Statistics, University College Dublin, Belfield, Dublin 4, D04 V1W8, Ireland}
\affiliation{Scoil na Matamaitice agus na Staitistic\'i, An Col\'aiste Ollscoile, Baile \'Atha Cliath,  D04 N2E5, Ireland}
\affiliation{Centro Brasileiro de Pesquisas F\'isicas (CBPF), Rio de Janeiro, CEP 22290-180, Brazil}

\author{Huan Yang \orcidlink{0000-0002-9965-3030}}
\email{hyangdoa@tsinghua.edu.cn}
\affiliation{Department of Astronomy, Tsinghua University, Beijing 100084, China}

\begin{abstract}

We present a decomposition of the Kerr Green's function in the time domain, motivated by the frequency-domain split previously studied in the Schwarzschild limit. We formally show that the identification of a quasinormal-mode contribution, a direct part, and a late-time tail is still available, where the split times are narrowed down to an $\mathcal{O}(a)$ window around the Schwarzschild value with exact location undetermined. We have checked this Green's function with time-domain Teukolsky numerical simulations and find close agreement away from the split time. We also apply this decomposed Green's function in the time domain to a model problem with a test particle plunging into a Kerr black hole. The approximated dynamically excited direct part and quasinormal mode sum are obtained by convolving the Green's function with the particle's source term, which may be viewed as the first order in mass ratio of a spinning  black hole  ringdown.

\end{abstract}

\maketitle

\section{Introduction}

Gravitational-wave (GW) astronomy is entering an era of precision
measurement, as signaled by the recent detection of GW250114 with
a signal-to-noise ratio (SNR) close to 80. In the merger-ringdown
stage of this event, there is significant statistical evidence
not only for the fundamental and first-overtone modes, i.e.,
$(220)$ and $(221)$
\cite{LIGOScientific:2025rid,LIGOScientific:2025wao},
but also for quadratic modes
\cite{Mitman:2022qdl,Cheung:2022rbm,Khera:2023oyf,
Cheung:2023vki,Ma:2024qcv,Bourg:2024jme,Khera:2024bjs,
Wang:2026rev,Sberna:2021eui,Bucciotti:2024zyp,
Lagos:2022otp,Wang:2026rev}
and  the ``direct wave''
\cite{Oshita:2025qmn,Lu:2025vol}.
As the SNR of merger-ringdown signals continues to improve,
the detection of various nonlinear and non-modal signal
components is becoming increasingly feasible.
Therefore, the traditional ``black-hole spectroscopy'' program
should be extended to a more complete ringdown test, in which
different components of the ringdown signal are theoretically
identified, understood, and targeted for detection.

The Green's function plays a central role in classifying the
linear ringdown signal and, in turn, provides a basis for
identifying nonlinear contributions.
Leaver~\cite{Leaver1986} proposed a decomposition of the
Schwarzschild Green's function by applying contour integration
in the complex frequency domain, separating the response into
the direct part, the quasinormal-mode (QNM) contribution, and
the late-time tail.
However, the mathematical implementation faces difficulties
in treating the large-arc contour integral, and as a result
the corresponding direct part was not explicitly obtained
within that formalism.
Recent theoretical developments have reformulated the Green's
function in a way that enables a proper decomposition in the
time domain, as demonstrated by direct calculations of the
different components of the Schwarzschild Green's function
\cite{DeAmicis:2025xuh,arnaudo2025quasinormalmodescompletemode,
arnaudo2025priceslawquasinormalmodes,Su:2026fvj}
(also see
\cite{kuntz2025greenfunctionposchltellerpotential,Zhang:2026xgv}).

In this work, we extend this program to the Kerr black-hole
spacetime and use the Kerr Green's-function decomposition to
compute the corresponding components of the linear GWs generated by
a particle plunging into a Kerr black hole.
This model problem provides a basis for understanding the
generic ringdown of binary black-hole (BH) mergers, which,
from the perturbation-theory point of view, can be expanded
in powers of the symmetric mass ratio.
There are a few major issues associated with the calculation
and decomposition of the Kerr Green's function, in contrast
to the Schwarzschild case.

First, the perturbation
is formulated here using
the Teukolsky equation~\cite{Teukolsky:1973} rather than the
Regge--Wheeler--Zerilli equation.
Even in the Schwarzschild limit, we find that the QNM sum
in the Teukolsky formulation converges slowly
near its onset.
The same difficulty is encountered for the
Kerr scenario considered here.
Since computing the excitation factors of high-overtone QNMs
for generic Kerr black holes is computationally challenging,
a practical summation over only a limited number of overtones
can leave a poorly converged interval near the transition between the direct part and the QNM response.

Second, the large-frequency exponential in the angular factor
of an individual Kerr Green's-function mode affects both the
convergence of the direct-part branch-cut integral and whether
the large-arc contribution can be neglected.
For comparison, in Schwarzschild, for an observer outside the
source at $r_*>0$, the direct part of the time-domain radial
Green's function starts at $t-t'=r_*-r_*'$ for $r_*'>0$ and is absent for $r_*'<0$~\cite{mark2017recipe,
aruquipa2026greenfunctionsreggewheelerteukolsky}. The QNM contribution starts at $t-t'=|r_*|+|r_*'|$~\cite{PhysRevD.86.024021,Su:2026fvj}.
For Kerr, our WKB analysis in Appendix~\ref{app:kerr_split_time}
shows that the angular exponent depends on the source and observer
angles in the asymptotic directions considered. The large-frequency asymptotics of the separation constant also depend on $\arg\omega$. Conversion from curvature to metric changes the polynomial frequency
weights but leaves the exponential factors unchanged.
These asymptotics suggest a possible gap of width $O(a)$ around $t-t'=|r_*|+|r_*'|$ between the time ranges in which the direct-part and QNM contour prescriptions can be applied.

Outside the window, the radial phases suggest how to close
the contour, though the angular factors still affect convergence. Since the Kerr split times are hard to obtain simply, we use the Schwarzschild values as an approximation in this work. But it must be determined in the future.

Third, angular branch cuts arising from exceptional points obstruct the usual contour argument for an isolated Kerr $(\ell,m)$ mode.
In Schwarzschild, the retarded radial Green's function vanishes before
the radial arrival time $t-t'=|r_*-r_*'|$ of a (radial) null geodesic connecting the  points with $(t,r_*)$ and  $(t',r_*')$ and same angles~\cite{Su:2026fvj}.
In Kerr, the upper-half-plane angular cuts prevent the same argument
from establishing that an individual mode vanishes before this time.
These discontinuities cancel in the complete sum over $\ell$ at fixed
$m$, as shown in Appendix~A of Ref.~\cite{PhysRevD.94.124053}, removing
the angular obstruction to the causality argument for the fixed-$m$
retarded Green's function. To recover the angle-dependent arrival
structure of the full $3+1$ Green's function from its mode expansion,
one must also sum over $m$.

We use the Kerr Green's function to compute waveforms from
spatially extended sources with prescribed spherical-harmonic profiles for a first simple test. We compare the same spherical-harmonic components with those from time-domain Teukolsky simulations of the same sources. Away from the causal fronts, the two calculations agree closely in amplitude, phase, and decay.

In the contour calculation, we keep the direct part and the
QNM residues. We leave out the radial branch-cut term that
produces the tail because we expect it to be small over the
times shown. We also leave out the angular-cut integrals:
obtaining reliable values on both sides of the cuts at large complex frequencies takes too long with our current methods. For the real-frequency inversion, we use the unsplit response, so no separate cut integrals are needed.

We compute the $(\ell,m)=(2,2)$ metric waveform for a particle
that starts near the innermost stable circular orbit (ISCO)
and plunges into a Kerr black hole in the equatorial plane.
We first invert the radiative metric kernel along the real
frequency axis. For each source point, we split the resulting
time-domain kernel at the approximate transition time
$t-t'=|r_*'|+|r_*|$ and then integrate each part along the
particle's trajectory. This split shows how the direct response
gives way to QNM-dominated ringing in the plunge waveform.
How much each physical component contributes to the total
waveform remains an open question. We also compare our
approximate direct part with the phenomenological direct-wave
model of Ref.~\cite{Oshita:2025qmn} (see also
Refs.~\cite{Lu:2025vol,DeAmicis:2026tus,Kankani:2026kst,
Han:2026cgd,Kankani:2026byb,Dyer:2026yex,Chung:2026eph})
to show how the two differ.

This paper is organized as follows.
In Sec.~\ref{sec:sch}, we introduce the Teukolsky equation
and fix the notation used throughout the paper.
In Sec.~\ref{sec:green-function-validation}, we analyze
the analytic structure of the Kerr Teukolsky Green's function,
generalize the contour strategy of our previous
work~\cite{Su:2026fvj}, and compare the contour results with
time-domain simulations.
In Secs.~\ref{sec:kernel-decomposition}
and~\ref{sec:numerical-implementation}, we introduce the radiative kernel, discuss its formal direct-part--QNM--tail decomposition, and compare the direct-part contour integrals and QNM pole sum with the full real-frequency inverse transform.
In Sec.~\ref{sec:orbit-waveform}, we integrate these kernels along finite-start ISCO plunge trajectories and obtain the direct-part and post-front contributions to the plunge waveform. We conclude in Sec.~\ref{sec:con}.
Throughout this paper, we use units $G=c=M=1$.

\section{Teukolsky Equation}\label{sec:sch}

A spin-$s$ field $\psi$ on a Kerr background of mass $M$ and angular momentum $a$ in Boyer--Lindquist coordinates $(t,r,\theta,\phi)$, expressed in the Kinnersley tetrad, satisfies the Teukolsky equation~\cite{Teukolsky:1973}
\begin{align}
	&\left[\frac{\left(r^2+a^2\right)^2}{\Delta} -a^2 \sin^2\theta\right]\frac{\partial^2 \psi}{\partial t^2}
	+
	\frac{4Mar}{\Delta}\frac{\partial^2 \psi}{\partial t \partial \phi}
    \nonumber \\
	&+
	\left[\frac{a^2}{\Delta}-\frac{1}{\sin^2 \theta}\right]\frac{\partial^2\psi}{\partial\phi^2}
     - \Delta^{-s} \frac{\partial}{\partial r}
	\left(\Delta^{s+1}\frac{\partial \psi}{\partial r}\right)
    + \left(s^2 \cot^2 \theta -s\right) \psi
    	\nonumber \\
	&
	-\frac{1}{\sin \theta} \frac{\partial}{\partial \theta} \left(\sin \theta \frac{\partial \psi}{\partial \theta}\right)
	-2s\left[\frac{a\left(r-M\right)}{\Delta} + \frac{i \cos \theta}{\sin^2 \theta}\right]\frac{\partial\psi}{\partial\phi}
    \nonumber \\
    &
	-2s\left[\frac{M\left(r^2-a^2\right)}{\Delta}-r-ia\cos\theta\right]\frac{\partial \psi}{\partial t} 
     = 4\pi \Sigma T,
	\label{AllTeu}
\end{align}
where $\Delta=r^2-2Mr+a^2$, $\Sigma=r^2+a^2\cos^2\theta$ and $T=T(t,r,\theta,\phi)$ is a source term.

We separate variables via 
\begin{equation}
\psi(t,r,\theta,\phi)=\frac{1}{\sqrt{2\pi}}\int_{\mathbb{R}} d\omega\sum_{\ell=|s|}^{\infty}\sum_{m=-\ell}^{\ell}e^{-i\omega t+im\phi}R_{\ell m\omega}(r)\,{}_sS_{\ell m\omega}(\theta).
\label{eq:Teu_sep}
\end{equation}
Of particular relevance in this paper is the spin $s=-2$, in which case $\psi=(r-ia\cos\theta)^4\,\psi_4$, where $\psi_4$ is a Weyl scalar.

The Teukolsky equation \eqref{AllTeu} decomposes into the radial Teukolsky equation satisfied by $R_{\ell m\omega}(r)$ and the angular Teukolsky equation satisfied by 
%the spin-weighted spheroidal harmonics 
${}_sS_{\ell m\omega}(\theta)$. 

In the following subsections, we introduce the analytical structure of the radial and angular solutions to guide the design of the integration contours.

\subsection{ Angular equation} \label{subsection::Angular_equation}

The function ${}_sS_{\ell m\omega}$ in Eq.~\eqref{eq:Teu_sep}
is a spin-weighted spheroidal harmonic.  Introducing the dimensionless
parameter $c=a\omega$, the angular Teukolsky equation reads
\begin{equation}
    \frac{1}{\sin\theta}\frac{d}{d\theta}
    \left(
    \sin\theta\frac{d{}_sS_{\ell m\omega}}{d\theta}
    \right)
    +\mathcal V_\theta(\theta;c)\,{}_sS_{\ell m\omega}=0,
    \label{eq:angular_teukolsky_full}
\end{equation}
where
\begin{align}
\mathcal V_\theta(\theta;c)={}&
    c^2\cos^2\theta
    -\frac{m^2}{\sin^2\theta}
    +{}_sA_{\ell m}(c)
\nonumber\\
&
    -2cs\cos\theta
    -\frac{2ms\cos\theta}{\sin^2\theta}
    -s^2\cot^2\theta+s .
    \label{eq:angular_potential_full}
\end{align}
Here, ${}_sA_{\ell m}(c)$ is the angular eigenvalue.
The dependence of the angular eigenvalue and
eigenfunction on $a$ and $\omega$ enters through $c=a\omega$.
It is also convenient to define
\begin{equation}
\lambda_{\ell m\omega}
:=
    {}_sA_{\ell m}+a^2\omega^2-2am\omega.
    \label{eq:lambda_def_large_omega}
    \end{equation}
    
For fixed $s$, $m$, and $\omega$, the regular spin-weighted spheroidal
harmonics satisfy a bilinear orthogonality relation.  To obtain it, we
multiply the angular equation for ${}_sS_{\ell m\omega}$ by
${}_sS_{\ell' m\omega}$, subtract the corresponding expression with
$\ell$ and $\ell'$ exchanged, and integrate over $\theta$.  The boundary
terms vanish because both eigenfunctions are regular at $\theta=0,\pi$.
For distinct angular eigenvalues, this gives
\begin{equation}
 \int_0^\pi
 {}_sS_{\ell m\omega}(\theta)\,
 {}_sS_{\ell' m\omega}(\theta)\,
 \sin\theta\,\dd\theta
 =
 \alpha_{\ell m}(\omega)\,\delta_{\ell\ell'},
 \label{eq:angular-bilinear-orthogonality}
\end{equation}
where
\begin{equation}
 \alpha_{\ell m}(\omega)=\int_0^\pi
 \bigl[{}_sS_{\ell m\omega}(\theta)\bigr]^2
 \sin\theta\,\dd\theta.
 \label{eq:angular-bilinear-normalization}
\end{equation}

No complex conjugation is taken in these expressions.  This bilinear
pairing preserves the analytic dependence on $\omega$ on the chosen
modal sheet and is therefore the appropriate orthogonality relation
for the complex-frequency continuation used in the later discussion on Green's function.
For real $a\omega$, the eigenfunctions can be chosen real, and the
bilinear relation reduces to the usual orthogonality relation.
At complex frequencies, $\alpha_{\ell m}(\omega)$ can be complex and
need not equal unity.  

The angular Teukolsky equation can be solved and analytically continued
to complex frequencies using a pseudospectral expansion on a basis of
spin-weighted spherical harmonics, as described in
Appendix.~\ref{app:angular-spectral}.
In this study, we implemented our own angular solver because, in our tests, the default complex-frequency outputs of the Black Hole Perturbation Toolkit~\cite{BHPToolkit} and \texttt{SpinWeightedSpheroidalHarmonics.jl} did not consistently satisfy the Cauchy--Riemann conditions.

\begin{figure}[t]
 \centering
 \includegraphics[width=\columnwidth]
 {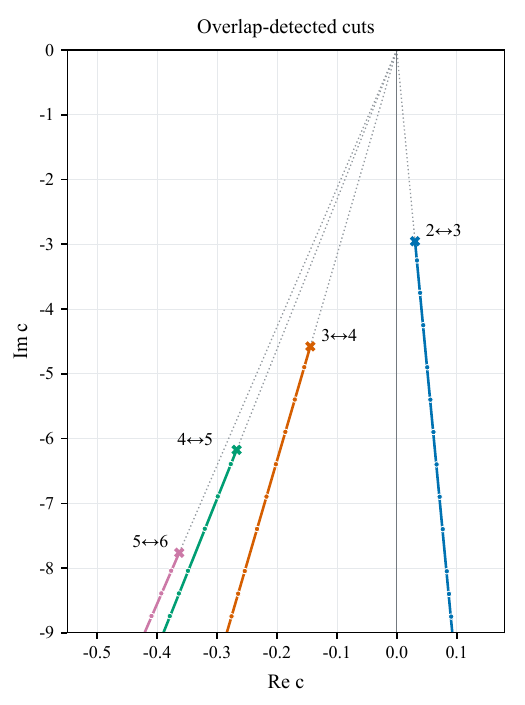}
 \caption{ 
 Angular cuts obtained after analytic overlap continuation
 for $s=-2$ and $m=2$.  Crosses mark exceptional points; solid curves show
 the detected cuts, and the faint radial guides are the rays induced by
 Eq.~\eqref{eq:radial-continuation}.
 A label $p\leftrightarrow q$ identifies the pair of spheroidal eigenvalue sheets, seeded by $\ell=p$ and $\ell=q$ at $c=0$, that exchange at the corresponding exceptional point.}
 \label{fig:angular-cuts}
\end{figure}

Although the coefficients of the differential operator of the angular Teukolsky equation are entire in
$c$, an individually labeled eigenpair is not necessarily globally single-valued.
There exist \cite{Kato1995}  exceptional points  at $c=c_\EP$, where $\lambda=\lambda_\EP$, near which the eigenvalue is given by
\begin{equation}
 \lambda_\pm(c)=\lambda_\EP
 \pm\alpha\sqrt{c-c_\EP}+O(c-c_\EP),
 \qquad \alpha\neq0.
 \label{eq:puiseux}
\end{equation}
One loop about $c_\EP$ exchanges the two sheets.  The branch point at the exceptional point $c=c_\EP$ is
intrinsic and our reference continuation uses radial paths
\begin{equation}
 \gamma_c(t)=tc,\qquad 0\leq t\leq1.
 \label{eq:radial-continuation}
\end{equation}

This path passes through an exceptional point exactly when
\begin{equation}
 c=\rho c_\EP,\qquad \rho\geq1.
 \label{eq:radial-angular-cut}
\end{equation}
The cuts generated by this convention are therefore outward rays beginning at each
exceptional point. In this paper, the notation
$p\leftrightarrow q$ means that analytic continuation once around the exceptional point exchanges the two spheroidal sheets seeded at $c=0$ by $\ell=p$ and $\ell=q$.

For $s=-2$ and $m=2$, the first lower-half-plane exceptional points
identified by overlap continuation in the complex $c$ plane are listed
in Table~\ref{tab:angular-ep-locations}.

\begin{table}[h]
\centering
\caption{Angular exceptional points in the complex $c=a\omega$ plane
for $s=-2$ and $m=2$.  The first column identifies the two modes whose
angular eigenvalues coalesce, using their spherical-limit labels.}
\label{tab:angular-ep-locations}
\begin{ruledtabular}
\begin{tabular}{ccc}
Coalescing modes
& $\operatorname{Re}c_{\EP}$
& $\operatorname{Im}c_{\EP}$ \\
\hline
$2\leftrightarrow3$ & $ 0.0303725113$ & $-2.9520981348$ \\
$3\leftrightarrow4$ & $-0.1443170101$ & $-4.5776424264$ \\
$4\leftrightarrow5$ & $-0.2672407660$ & $-6.1756773558$ \\
$5\leftrightarrow6$ & $-0.3630298423$ & $-7.7634394438$
\end{tabular}
\end{ruledtabular}
\end{table}

The angular exceptional points occur in complex-conjugate pairs,
as a consequence of the complex-conjugation symmetry of the angular Teukolsky equation. 
For real $\theta$, $s$, and $m$, complex conjugation maps a regular
solution with parameters $(c,{}_sA_{\ell m})$ to a regular solution
with parameters $(c^*,{}_sA_{\ell m}^*)$, preserving the boundary
conditions at $\theta=0,\pi$.  Starting from the same spherical-limit
mode labels at $c=0$ and continuing along conjugate paths therefore gives
\begin{equation}
 {}_sA_{\ell m}(c^*)
 =
 \left[ {}_sA_{\ell m}(c) \right]^*
 \label{eq:angular-conjugation-symmetry}
 \end{equation}
on the corresponding modal sheets.  Consequently, an exceptional
point at $c_{\EP}$ has a conjugate partner at $c_{\EP}^*$, involving
the same pair of mode labels.
The cuts shown in Fig.~\ref{fig:angular-cuts} extend radially from these paired exceptional points and are therefore also symmetric under reflection about the real axis.

Note that Eq.~\eqref{eq:angular-bilinear-orthogonality} should be used away from exceptional points, where angular eigenvalues coalesce and the individual-mode normalization can become singular.

\subsection{Radial equation}

The radial Teukolsky equation can be written as
\begin{equation}
    \Delta^{-s}\frac{d}{dr}
    \left(
        \Delta^{s+1}\frac{dR_{\ell m\omega}}{dr}
    \right)
    +V_{\ell m\omega}^{(s)}(r)R_{\ell m\omega}=0,
    \label{eq:radial_teukolsky_equation}
\end{equation}
where
\begin{align}
    K&=(r^2+a^2)\omega-am,
    \label{eq:K_def_large_omega}\\
    V_{\ell m\omega}^{(s)}(r)
    &=
    \frac{K^2-2is(r-M)K}{\Delta}
    +4is\omega r
    -\lambda_{\ell m\omega}.
    \label{eq:radial_potential_full}
\end{align}

The homogeneous radial Teukolsky equation
\eqref{eq:radial_teukolsky_equation} possesses two regular
singular points at the inner and outer horizons
$r_\pm=M\pm\sqrt{M^2-a^2}$ and one irregular singular point
at radial infinity. There are four standard
solutions distinguished by their asymptotic boundary conditions, all in the same two-dimensional solution space
of this second-order equation.
For radiative modes ($\omega\neq0$), the two solutions of
primary physical interest are the ingoing (IN)
$R^{\rm in}_{\ell m\omega}(r)$ and the upgoing (UP)
$R^{\rm up}_{\ell m\omega}(r)$.

The  (DOWN) solution $R^{\rm down}_{\ell m\omega}(r)$ is purely ingoing at radial infinity, while the outgoing (OUT) solution $R^{\rm out}_{\ell m\omega}(r)$ is purely outgoing at the outer horizon. 
To tell whether the wave moves inward or outward, we combine the radial solution with the time and azimuthal factor $e^{-i\omega t+im\phi}$.

Here $k=\omega-ma/(2Mr_+)$ is the horizon frequency,
$r_*$ is the tortoise coordinate defined by
$dr_*/dr=(r^2+a^2)/\Delta$, and
$A^{\rm in/up/down/out}_{\rm trans/ref/inc}$
are complex scattering coefficients. We omit the frequency
label in the asymptotic forms below and suppress the subscript
$\ell m$ elsewhere when no ambiguity arises.
The four solutions have the asymptotic forms
\begin{align}
    R^{\rm in}_{\ell m}(r) &\sim
    \begin{cases}
        A^{\rm in}_{\rm trans}\,\Delta^{-s}e^{-ikr_*}
        & \text{as } r\to r_+,\\[4pt]
        A^{\rm in}_{\rm ref}\,r^{-2s-1}e^{i\omega r_*}
        +
        A^{\rm in}_{\rm inc}\,r^{-1}e^{-i\omega r_*}
        & \text{as } r\to\infty,
    \end{cases}
    \label{eq:IN_asymptotic}\\
    R^{\rm up}_{\ell m}(r) &\sim
    \begin{cases}
        A^{\rm up}_{\rm inc}\,e^{ikr_*}
        +
        A^{\rm up}_{\rm ref}\,\Delta^{-s}e^{-ikr_*}
        & \text{as } r\to r_+,\\[4pt]
        A^{\rm up}_{\rm trans}\,r^{-2s-1}e^{i\omega r_*}
        & \text{as } r\to\infty,
    \end{cases}
    \label{eq:UP_asymptotic}\\
    R^{\rm down}_{\ell m}(r) &\sim
    \begin{cases}
        A^{\rm down}_{\rm inc}\,\Delta^{-s}e^{-ikr_*}
        +
        A^{\rm down}_{\rm ref}\,e^{ikr_*}
        & \text{as } r\to r_+,\\[4pt]
        A^{\rm down}_{\rm trans}\,r^{-1}e^{-i\omega r_*}
        & \text{as } r\to\infty,
    \end{cases}
    \label{eq:DOWN_asymptotic}\\
    R^{\rm out}_{\ell m}(r) &\sim
    \begin{cases}
        A^{\rm out}_{\rm trans}\,e^{ikr_*}
        & \text{as } r\to r_+,\\[4pt]
        A^{\rm out}_{\rm inc}\,r^{-2s-1}e^{i\omega r_*}
        +
        A^{\rm out}_{\rm ref}\,r^{-1}e^{-i\omega r_*}
        & \text{as } r\to\infty.
    \end{cases}
    \label{eq:OUT_asymptotic}
\end{align}

In this paper, the tortoise coordinate $r_*$ is chosen as
\begin{equation}
    r_* =
    r
    +\frac{2Mr_+}{r_+-r_-}
    \ln\left(\frac{r-r_+}{2M}\right)
    -\frac{2Mr_-}{r_+-r_-}
    \ln\left(\frac{r-r_-}{2M}\right)
\end{equation}
to avoid ambiguity in the integration constant.

Motohashi and Suichi~\cite{Motohashi_2026} analyzed the poles of the Kerr radial solutions, their asymptotic amplitudes, and the radial Green's function in the complex-$\omega$ plane. We complement that analysis on the analytical structure by describing the branch cuts of these quantities here.

The UP solution has a branch point at $\omega=0$ associated
with the long-range behavior of the radial equation at
infinity. Its Mano--Suzuki--Takasugi (MST)
representation (see Ref.~\cite{SasakiTagoshi2003} for a review; see App.~\ref{app:mst} for details)
contains Tricomi confluent hypergeometric functions
$\mathrm U(\alpha,\beta,z)$ with
\begin{equation}
    z=-2i\omega(r-r_-).
    \label{eq:up-tricomi-argument}
\end{equation}

For generic parameters, $\mathrm U$ has a branch point at $z=0$
and a standard cut along the negative real $z$ axis. At a real
exterior radius, $\omega=-i\sigma$ with $\sigma>0$ gives
$z=-2\sigma(r-r_-)<0$. The cut therefore runs from $\omega=0$
along the negative imaginary axis (NIA); see Sec.~V of
Ref.~\cite{PhysRevD.94.124053}.

On consistently chosen angular sheets, the UP and DOWN solutions are related by~\cite{Motohashi_2026}
\begin{equation}
    \frac{{}_sR^{\rm down}_{\ell m\omega}(r)}{{}_sA^{\rm down}_{\rm trans}}
    =
    \Delta^{-s}(r)\,
    \left[
    \frac{{}_{-s}R^{\rm up}_{\ell m\omega^*}(r)}
         {{}_{-s}A^{\rm up}_{\rm trans}(\omega^*)}
    \right]^*.
    \label{eq:up-down-spin-conjugation}
\end{equation}

Frequency conjugation maps the NIA cut of this
UP solution to the positive imaginary axis (PIA). Since
$\Delta^{-s}$ is independent of frequency, DOWN inherits
the PIA cut.

The branch cuts of the asymptotic amplitudes follow from
their Wronskian representations.
Define the conserved radial Wronskian by
\begin{equation}
    W[f,g]
    = \Delta^{s+1} \left(f\,\partial_r g-g\,\partial_r f\right).
    \label{eq:radial-weighted-wronskian}
\end{equation}
The evaluation at  infinity above give
\begin{equation}
    W[R^{\rm down}, R^{\rm up}]
    = 2i\omega A^{\rm down}_{\rm trans} A^{\rm up}_{\rm trans}.
    \label{eq:up-down-wronskian}
\end{equation}
Taking Wronskians of the scattering identity
\begin{equation}
    R^{\rm in}
    =
    \frac{A^{\rm in}_{\rm inc}}{A^{\rm down}_{\rm trans}}
    R^{\rm down}
    +
    \frac{A^{\rm in}_{\rm ref}}{A^{\rm up}_{\rm trans}}
    R^{\rm up}.
    \label{eq:radial-scattering-identity}
\end{equation}
gives, for $\omega\neq0$,
\begin{align}
    W[R^{\rm in},R^{\rm up}]
    &=2i\omega A^{\rm up}_{\rm trans}A^{\rm in}_{\rm inc},
    \\
    W[R^{\rm down},R^{\rm in}]
    &=2i\omega {A^{\rm down}_{\rm trans}}A^{\rm in}_{\rm ref}.
\end{align}
With a horizon normalization analytic in frequency, the IN
solution is meromorphic on the chosen angular sheet. Away
from isolated poles and angular cuts, the radial jumps are
\begin{align}
    \operatorname{Disc}_{\rm NIA}A^{\rm in}_{\rm inc}
    &=
    \frac{
    W\!\left[R^{\rm in},\operatorname{Disc}_{\rm NIA}
    \left(\frac{R^{\rm up}}{A^{\rm up}_{\rm trans}}\right)
    \right]}{2i\omega},\\
    \operatorname{Disc}_{\rm PIA}A^{\rm in}_{\rm ref}
    &=
    \frac{
    W\!\left[\operatorname{Disc}_{\rm PIA}
    \left(\frac{R^{\rm down}}{A^{\rm down}_{\rm trans}}\right),
    R^{\rm in}\right]}{2i\omega}.
    \label{eq:amplitude-radial-cut-jumps}
\end{align}
Here $\operatorname{Disc}$ is the value on the right minus the value on the left of the indicated cut. Dividing by the
respective transmission amplitudes removes the arbitrary
normalizations of the UP and DOWN solutions. The incident
amplitude inherits the NIA cut from the normalized UP solution, and the reflected amplitude inherits the PIA cut from the normalized DOWN solution. The radial Green's function inherits the NIA cut through the ratio
$R^{\rm up}/W[R^{\rm in},R^{\rm up}]$.

The radial solutions, their asymptotic amplitudes, and the
separated Green-function modes can also inherit angular cuts through their dependence on $\lambda_{\ell m\omega}$. We use the same angular-cut convention for
all these quantities~\cite{PhysRevD.94.124053}.

\section{Green's function decomposition and validation}
\label{sec:green-function-validation}

In this section, we introduce the Green's function of the Teukolsky equation. We analyze the analytic structure of individual $(\ell,m)$ modes for both the radial and angular parts of the Green's function, and that of the full $(3+1)$-dimensional Green's function. Finally, we generalize the contour construction strategy from our previous work~\cite{Su:2026fvj} to the Teukolsky equation.

For the full $(3+1)$-dimensional Teukolsky equation, the Green's function is 
\begin{align}
 G(x,x')={}&\frac{1}{(2\pi)^2}
 \sum_{\ell=|s|}^{\infty}\sum_{m=-\ell}^{\ell}
 \int_{-\infty+\ii\delta}^{\infty+\ii\delta}\dd\omega\,
 e^{-\ii\omega(t-t')+\ii m(\phi-\phi')}
 \nonumber\\
 &\times
 \frac{{}_sS_{\ell m\omega}(\theta)
 {}_sS_{\ell m\omega}(\theta')}
 {\alpha_{\ell m}(\omega)}
 \tilde{G}_{\ell m}(\omega;r,r'),
 \qquad \delta>0.
 \label{TotalKerrGF}
\end{align}
where \(\tilde{G}_{\ell m}(\omega;r,r')\) is the radial Green's function defined in the following subsection.
To avoid clutter, the $\ell m$ indices are suppressed in $\tilde{G}_{\ell m}$ and in some other quantities in what follows.

Eq.~\eqref{TotalKerrGF} admits a natural analytic continuation from real frequencies to complex frequencies.
Note that at complex frequency the two angular factors in Eq.~\eqref{TotalKerrGF} must be paired bilinearly, without complex-conjugating either spheroidal harmonic.
Replacing the source-side factor by ${}_sS^*_{\ell m\omega}$ in the way some papers do would make the integrand depend on both $\omega$ and $\omega^*$, so it would no longer be an analytic function of the contour variable $\omega$.  Such an expression cannot be used for analytic continuation or contour deformation in the complex frequency plane.

The radial Green's function admits the same
Wronskian construction as in Schwarzschild.
Splitting it into two pieces and applying contour deformations
analogous to those used in Schwarzschild allows us to identify
the direct part, the QNM contribution, and the radial branch-cut tail.
For an isolated $(\ell,m)$ mode, however, a complete contour
representation must additionally include the angular-cut integrals
arising from the frequency dependence of the spheroidal harmonics.
Any nonvanishing large-arc contributions must also be retained,
with a consistent treatment of the zero-frequency terms.

\subsection{Radial Green's function and its analytical structure}
\label{subsec:schwarzschild-green-function}

The frequency-domain radial Green's function can be written as 
\begin{align}
 \widetilde G(\omega;r,r')&=
 \frac{R^{\rm in}(r_<) R^{\rm up}(r_>)}
 {W},
 \nonumber\\
 r_<&\equiv\min(r,r'),\,
 r_>\equiv\max(r,r').
 \label{eq:GF-decomp}
\end{align}
 The conserved Teukolsky Wronskian of the radial equation is
\begin{equation}
 W=\Delta^{s+1}
 \left(R^{\rm in}\partial_r R^{\rm up}
 - R^{\rm up}\partial_rR^{\rm in}\right).
 \label{eq:wronskian-definition}
\end{equation}
Evaluation at infinity gives
$W=2\ii\omega A^{\rm in}_{\rm inc}$.  
In this paper we use the normalization that $A^{\rm in}_{\rm trans} = A^{\rm down}_{\rm trans} = A^{\rm up}_{\rm trans}=1$.
A different normalization of either homogeneous radial solution must be
accompanied by a corresponding change in the Wronskian and scattering
amplitudes to ensure consistency.

For the case $r>r'$ of interest in this paper, we use the scattering identity \eqref{eq:radial-scattering-identity}.
We can decompose the frequency domain Green's function
Eq.~\eqref{eq:GF-decomp} as follows:
\begin{equation}
 \widetilde G=\widetilde G^++\widetilde G^-,
 \label{eq:G-decomposition}
\end{equation}
where
\begin{align}
 \widetilde G^+(\omega;r,r')&=\frac{A^{\rm in}_{\rm ref}}{W}
  R^{\rm up}(r) R^{\rm up}(r'),
 \label{schwarzschildGplus}\\
 \widetilde G^-(\omega;r,r')&=\frac{A^{\rm in}_{\rm inc}}{W}
  R^{\rm up}(r) R^{\rm down}(r').
 \label{schwarzschildGminus}
\end{align}

This decomposition was originally motivated by the observation that the cosmological
horizon of Schwarzschild--de Sitter spacetime produces discrete Matsubara
modes on the imaginary-frequency axis
\cite{arnaudo2025quasinormalmodescompletemode,arnaudo2025priceslawquasinormalmodes}.
As the cosmological constant tends to zero, these modes coalesce into branch
cuts on the PIA and NIA of $\widetilde G^+$ and $\widetilde G^-$.  
The QNM poles, which are the zeros of $A^{\rm in}_{\rm inc}(\omega)$ (or,
equivalently, of $W$), occur only in $\widetilde G^+(\omega; r, r')$, while
$\widetilde G^-(\omega; r, r')$ is free of QNM poles because its common
$A^{\rm in}_{\rm inc}(\omega)$ factor cancels.  Although $\widetilde G^+(\omega; r, r')$ and
$\widetilde G^-(\omega; r, r')$ separately have cuts
on both the PIA and NIA, their sum $\widetilde G(\omega; r, r')$ has only the NIA cut
(see Fig.~1 of Ref.~\cite{Su:2026fvj}).

\subsection{Angular Projector in the separated
\texorpdfstring{$(\ell,m)$}{(ell,m)} mode}
\label{subsec:angular-cut-brief}

In the separated representation of the full $3+1$ dimensional Green's
function in Eq.~\eqref{TotalKerrGF}, the angular dependence of a fixed
$(\ell,m)$ mode is carried by
\begin{equation}
 \Pi_{\ell m}(\theta,\theta';\omega)
 =
 \frac{{}_sS_{\ell m\omega}(\theta)
 {}_sS_{\ell m\omega}(\theta')}
 {\alpha_{\ell m}(\omega)} .
 \label{angularfactor}
\end{equation}
This factor is the integral kernel of the projector onto the
selected spheroidal mode.  Neither this projector nor its angular
eigenvalue is single-valued and analytic throughout the complex
$\omega$ plane~\cite{PhysRevD.94.124053}.

Consequently, for an individual $(\ell,m)$ mode, angular branch cuts
extend above every finite horizontal Bromwich contour.  The usual
mode-by-mode proof of strict pre-arrival vanishing cannot therefore be
applied, because it assumes that the frequency-domain Green's function
is analytic everywhere above the contour.

This is a limitation of the separated modal representation, not a
violation of causality by the complete retarded Green's function.  For
each fixed $m$, the angular-cut discontinuities cancel when all $\ell$
modes are summed, as shown in Appendix~A of
Ref.~\cite{PhysRevD.94.124053}.  The complete Green's function thus
recovers the upper-half-plane analyticity required by the standard
causality argument.  An isolated $(\ell,m)$ contribution, however, need
not vanish strictly before the arrival time.

The Schwarzschild case is simpler.  In the limit $a\to0$, one has
$c=a\omega\to0$ at every fixed finite frequency, so the spheroidal
harmonics reduce to spin-weighted spherical harmonics and different
$\ell$ modes decouple.  No finite-frequency angular exceptional points
or angular cuts arise, and the usual mode-by-mode causality argument
remains applicable~\cite{Su:2026fvj}.  This agrees with our previous
Schwarzschild results.

\begin{figure*}[t]
 \centering
 \draftgraphic{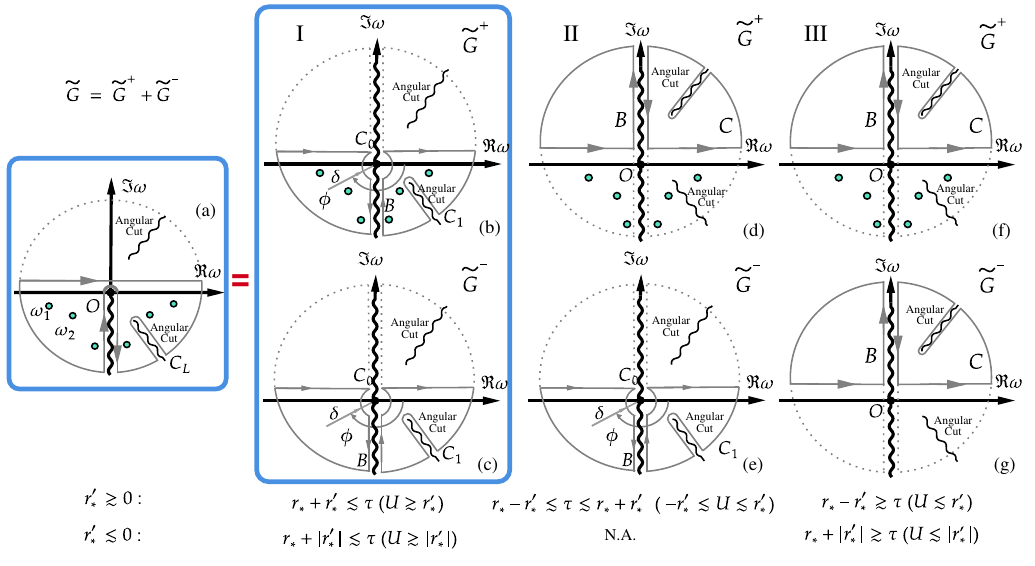}{5.3cm}
 \caption{
         Contours for calculating the Kerr Green's function or radiative kernels in different time intervals for a fixed $(\ell,m)$ mode.
        The cyan dots represent QNMs, and the zig-zag lines denote branch cuts.
        $C_L$ denotes the large arc in the Leaver-like contour, $C$ and $C_1$ the upper- and lower-half-plane large arcs, and $C_0$ the small zero-frequency detour. $B$ represents the radial branch cuts; the angular cuts are explicitly indicated.
         The idea here is similar to Fig.~1 of our previous work.
         These angular cuts  follow conjugate-symmetric rays from exceptional points to infinity.
         After the complete sum over $\ell$ at fixed $m$, the angular-cut discontinuities cancel,
         yielding a causal complete response. The radial contour organization remains unchanged, with its contributions summed over $\ell$.
         Regions I--III indicate approximate radial-support boundaries for an isolated $(\ell,m)$ contribution.
         In region I, we close both contours of
         $\widetilde G^+$ and $\widetilde G^-$
         in the lower half-plane [panels (b) and (c)], equivalent to the Leaver-like contour in panel (a).
         In region II, we close $\widetilde G^+$ in the upper half-plane and $\widetilde G^-$ in the lower half-plane;
         this interval is absent for negative $r_*'$ in the leading radial-support approximation. In region III, both contours close in the upper half-plane; the radial PIA contributions cancel, while angular-cut contributions can remain for an isolated $(\ell, m)$ mode.
 }
 \label{fix_ell_m_contour}
\end{figure*}

\subsection{Contour strategy}
\label{subsection:Contour_Strategy}

In our previous work~\cite{Su:2026fvj}, we developed a contour
decomposition of the Schwarzschild Green's function that avoids the
difficulties associated with Leaver's original large semicircle in the
lower half-plane~\cite{Leaver1986}.  We now extend this strategy to Kerr.

Let $\tau\equiv t-t'$, $U\equiv u-t'$, $u\equiv t-r_*$. 

Thus \(U=\tau-r_*\) at future null infinity.  The leading
large-frequency radial phases suggest three approximate time regions,
shown in Fig.~\ref{fix_ell_m_contour}.  These regions provide a useful
contour organization, but their boundaries are not exact support
boundaries for an isolated Kerr mode: the full large-frequency behavior
also depends on the angular functions.

\begin{itemize}

\item \textbf{Region I: late times}
(\(\tau\gtrsim |r_*|+|r_*'|\), or
\(U\gtrsim |r_*'|\) at future null infinity).
We use a Leaver-like contour obtained by deforming the Bromwich
contour into the lower half-plane.  The leading large-frequency
phases of both \(\widetilde G^+\) and \(\widetilde G^-\), after
multiplication by \(e^{-\ii\omega\tau}\), decay on the lower-half-plane
large arc in this region, motivating this contour closure.  As in
Leaver's construction, the deformed contour encloses the QNM poles
and the radial NIA cut, producing the damped, oscillatory QNM response
and the branch-cut tail.  For an isolated \((\ell,m)\) mode in Kerr,
it also encloses the lower-half-plane angular branch cuts of
\(\widetilde G\), whose contributions must be included in the
\((\ell,m)\)-mode response.

 \item \textbf{Region II: the direct-part interval}
 (\(r_*-r_*'\lesssim\tau\lesssim r_*+r_*'\), or
 \(-r_*'\lesssim U\lesssim r_*'\) at future null infinity, for
  \(r_*' \gtrsim 0\)).
 Separate contours are used for the two radial pieces:
 \(\widetilde G^+\) is closed on the upper half-plane, whereas
 \(\widetilde G^-\) is closed on the lower half-plane, with a small
 detour around \(\omega=0\).  This choice is motivated by the leading
 radial phases
 \(\widetilde G^+\sim e^{\ii\omega(r_*+r_*')}\) and
 \(\widetilde G^-\sim e^{\ii\omega(r_*-r_*')}\), together with the
 Fourier factor \(e^{-\ii\omega\tau}\). 
 %%%
In this approximate radial organization, the radial direct part consists
of the PIA jump of \(\widetilde G^+\), the NIA jump of
\(\widetilde G^-\), and the zero-frequency detour.  For an isolated
\((\ell,m)\) mode in Kerr, the two contour closures also enclose the
corresponding angular branch cuts: the upper-half-plane cuts for
\(\widetilde G^+\) and the lower-half-plane cuts for
\(\widetilde G^-\).  Their discontinuity integrals must therefore be
included in the complete modal contribution.  Although the separate
\(\widetilde G^+\) and \(\widetilde G^-\) contour contributions depend
on the radius of the small detour around \(\omega=0\), their sum in the
time-domain Green's function does not.  For negative \(r_*'\), the
interval \(-r_*'\lesssim U\lesssim r_*'\) is empty, so the radial
direct part is absent within this approximation, although the angular-cut
contributions of an isolated mode must still be considered separately.

\item \textbf{Region III: before the radial arrival front}
(\(\tau\lesssim r_*-r_*'\), or
\(U\lesssim-r_*'\) at future null infinity).
The leading large-frequency radial phases, including
\(e^{-\ii\omega\tau}\), decay when both radial pieces are closed in
the upper half-plane.  In Schwarzschild, their PIA discontinuities
cancel when summing up \(\tilde{G}^+,\,\tilde{G}^-\) and the large-arc integrals vanish.  Each \((\ell,m)\) contribution is therefore strictly zero for
\(\tau<r_*-r_*'\), and we call
\(\tau_{\rm arr}=r_*-r_*'\), or \(U_{\rm arr}=-r_*'\), the
\emph{arrival time}.
For an isolated Kerr \((\ell,m)\) mode, the contour also encloses
upper-half-plane angular cuts, whose contributions need not vanish
before this radial arrival time.  The complete retarded Green's
function remains causal because these angular-cut discontinuities
cancel in the sum over \(\ell\) at fixed \(m\).

\end{itemize}

We call the boundaries separating the three contour regions
\emph{radial causal fronts}.  They are determined by the leading
large-frequency radial phases and organize the transition between the
direct response and the QNM--branch-cut response.  For an isolated Kerr
mode, however, angular-cut contributions can extend across these
fronts, so they are not strict boundaries of the time-domain support.

The role of high frequencies becomes increasingly important near these
fronts.  Away from a front, the combined radial, angular, and Fourier
phases generally provide exponential suppression at large
\(\lvert\omega\rvert\).  As a front is approached, this suppression
weakens, and the relevant frequency scale moves to progressively larger
\(\lvert\omega\rvert\).  Consequently, the high-frequency portions of
the PIA and NIA integrals, the upper- and lower-half-plane angular-cut
integrals, and the high-overtone part of the QNM sum can all contribute
to the near-front behavior.  Approximations or truncations that are
accurate away from a front therefore need not remain uniform as the
front is approached.

The same loss of high-frequency suppression affects the large-arc
integrals.  As discussed in Appendix~\ref{app:kerr_split_time}, there
may be an \(O(a)\) transition window around
\(t-t'=|r_*|+|r_*'|\), corresponding to \(U=|r_*'|\) in the radiative
limit, in which neither of the relevant large-arc integrals is
guaranteed to be negligible.  A contour representation obtained by
dropping both arcs is therefore not justified uniformly inside this
window.  The missing contribution is expected to be localized near the
front, but its precise time-domain representation remains to be
established.

A controlled decomposition near a causal front must therefore treat
the PIA and NIA integrals, the angular-cut integrals in both
half-planes, the large-overtone behavior of the QNM sum, and the
relevant large-arc contributions within the same
large-complex-frequency analysis.  The leading radial phases determine
the basic contour strategy, but the detailed near-front behavior
depends on the combined radial, angular, and Fourier asymptotics.

\subsection{ Causal indicators}

For later convenience, we introduce the following \emph{causal indicators}
for the radiative response at future null infinity:
\begin{align}
 I_D(U,r_*')&=
 \Theta(r_*')\Theta(U+r_*')\Theta(r_*'-U),
 \label{eq:direct-indicator}\\
 I_L(U,r_*')&=\Theta(U-|r_*'|).
 \label{eq:late-indicator}
\end{align}

Here \(I_D\) selects the approximate direct-response interval, while \(I_L\) selects the late-time region containing the QNM and branch-cut
response.

In a complete contour decomposition, the causal indicators reflect the convergence regions of the contributions associated with poles, branch cuts, and the remaining contour terms. They encode the analytic support properties of the specified
Green-function decomposition, rather than serving as arbitrary windows chosen for numerical convenience.
For an isolated Kerr \((\ell,m)\) mode, angular-cut contributions may
extend across these radial fronts, so a complete modal decomposition
may require additional indicators.

A subtlety arises when this decomposition is applied to the radiative responses of the Green's function and to
the radiative kernels introduced in Subsec.~\ref{subsec:kernel-definition}: the curvature-level and metric-level waveform decompositions are generally different.

Let \(\widetilde G_{\PPsi}(\omega;r')\) denote the radiative curvature responses extracted from \(\widetilde G^\pm\) at future
null infinity, and let
\(\widetilde G_{\HH}(\omega;r')\) denote the corresponding
metric responses.  Mode labels and angular arguments are suppressed
here for clarity.
The frequency-domain responses satisfy
\begin{equation}
 \widetilde G_{\HH}(\omega;r')
 =
 -\frac{2}{\omega^2}
 \widetilde G_{\PPsi}(\omega;r'),
 \qquad \omega\neq0.
 \label{eq:metric-scattering-conversion}
\end{equation}
The conversion factor changes the algebraic frequency dependence, but not the radial exponential factors that determine the contour-closure thresholds.

In this paper, all waveform decompositions are
defined at the metric level: the frequency-domain responses are first converted to metric quantities, and the contour contributions are then
organised according to their time regions.
This should be distinguished from first selecting a curvature
contribution within a given time interval and then
integrating that contribution twice with respect to the retarded time \(U\).  Time windowing and retarded-time integration do not generally commute.

To see this distinction, let \(G_{\PPsi}(U;r')\) and \(G_{\HH}(U;r')\) denote the complete time-domain radiative responses, which satisfy \(\partial_U^2G_{\HH}=2G_{\PPsi}\). For a contour component \(X\), let \(G_{\PPsi}^{X}\) and
\(G_{\HH}^{X}\) denote the corresponding curvature and metric contributions, with
\(\partial_U^2G_{\HH}^{X}=2G_{\PPsi}^{X}\).
If the metric contribution is restricted by an indicator \(I_X\), then
\begin{align}
 \frac12\partial_U^2\left[I_XG_{\HH}^{X}\right]
 ={}&
 I_XG_{\PPsi}^{X}
 +(\partial_UI_X)\partial_UG_{\HH}^{X}
 \nonumber\\
 &+
 \frac12(\partial_U^2I_X)G_{\HH}^{X}.
 \label{eq:metric-split-front-terms}
\end{align}
Note that \(I_X\) need not be either \(I_D\) or \(I_L\); the same identity applies to other combinations of step functions that may follow from a more complete analysis of the causal-front region. The derivatives of the step functions generate delta-function and delta-function-derivative contributions at the causal fronts.

Therefore, the curvature associated with a time-windowed metric contribution is not simply \(I_XG_{\PPsi}^{X}\); it also contains terms supported at the causal fronts.

\label{subsec:kerr-green-function}

\subsection{Spherical-mode transfer Green's function}
\label{subsec:spherical-transfer-green-function}

We wish to compare the Green's-function construction with a time-domain simulation.  A direct comparison of the full \(3+1\)-dimensional Kerr Green's function is computationally
impractical, because resolving an angular delta-function source requires an infinite sum over \((\ell,m)\) modes.  In any numerical calculation, this angular-mode sum must be truncated.

We therefore introduce the \emph{spherical-mode transfer Green's function}, obtained by projecting both the source and the response onto specified spin-weighted spherical modes.  This provides a well-defined comparison with the corresponding time-domain calculation without reconstructing the angular delta function or summing over a prohibitively large number of \(\ell\) modes.

To distinguish the spin-weighted spherical-harmonic basis from the spheroidal-harmonic basis, we use $\ell$ for the spheroidal (and radial) channel and $L$ for the spin-weighted spherical-harmonic index and introduce spherical-spheroidal mixing coefficients ${}_s\mathcal A_{L\ell m}(a\omega)$ through the expansion  
\begin{equation}
 \frac{{}_sS_{\ell m\omega}(\theta)e^{\ii m\phi}}{\sqrt{2\pi}}
 =\sum_{L=L_{\min}(m)}^\infty
 {}_s\mathcal A_{L\ell m}(a\omega){}_sY_{Lm}(\theta,\phi),
\qquad
 \label{eq:spheroidal-spherical-expansion}
\end{equation}
where $L_{\min}(m)=\max(|s|,|m|)$.
 By orthonormality of the spin-weighted spherical harmonics these coefficients are given by 
\begin{equation}
    {}_s\mathcal A_{L\ell m}(a\omega)=\frac{1}{\sqrt{2\pi}}\int d\Omega\,{}_sY_{Lm}^{*}(\theta,\phi)\,{}_sS_{\ell m\omega}(\theta)e^{im\phi}.
\end{equation}
Their numerical computation using a spectral expansion of the angular equation is described in Appendix~\ref{app:angular} (see also \cite{Cook_2014} for further details).
Substituting this expansion into
Eq.~\eqref{TotalKerrGF} gives
%\begin{widetext}
\begin{align}
 G(x,x')={}&
 \sum_{\ell=|s|}^{\infty}\sum_{m=-\ell}^{\ell}
 \sum_{L=L_{\min}(m)}^\infty
 \sum_{L'=L_{\min}(m)}^\infty
 \int_{-\infty+\ii\delta}^{\infty+\ii\delta}
 \frac{\dd\omega}{2\pi}
 e^{-\ii\omega(t-t')}
 \nonumber\\
 &\times\frac{G_{\ell m\omega}(r,r')}
 {\alpha_{\ell m}(\omega)}
 {}_s\mathcal A_{L\ell m}(a\omega)
 {}_s\mathcal A_{L'\ell m}(a\omega)
 \nonumber\\
 &\times{}_sY_{Lm}(\theta,\phi){}_sY^*_{L'm}(\theta',\phi').
 \label{eq:kerr-green-spherical-expansion}
\end{align}
%\end{widetext}
Consider the impulsive source
\begin{equation}
 T(x')=\delta(t')\delta(r'-r_1)
 {}_sY_{L_1m_1}(\theta',\phi').
 \label{SphericalSourceTerm}
\end{equation}

Projection of the response at \(r=r_2\) onto the spherical mode
\((L_2,m_2)\) gives, as a function of the elapsed time
\(\tau \equiv t-t'\), the following
expression.  We choose the time origin at the impulsive source event, \(t'=0\), so that \(\tau=t\):
\begin{align}
 \psi_{L_2m_2}(\tau,r_2)={}&\delta_{m_1m_2}
 \sum_{\ell=L_{\min}(m_2)}^\infty
 \int_{-\infty+\ii\delta}^{\infty+\ii\delta}
 \frac{\dd\omega}{2\pi}\,e^{-\ii\omega \tau}
 \nonumber\\
 &\times
 \frac{G_{\ell m_2\omega}(r_2,r_1)}
 {\alpha_{\ell m_2}(\omega)}
 {}_s\mathcal A_{L_2\ell m_2}(a\omega)
 \nonumber\\
 &\times
 {}_s\mathcal A_{L_1\ell m_1}(a\omega).
 \label{eq:psil2m2}
\end{align}

The expression under the sum over $\ell$ actually describes the mixing of spherical modes during radial propagation through the spheroidal/radial channel $\ell$, from $L_1$ to $L_2$. We therefore define it as the \emph{spherical-mode transfer Green's function}
%%%
%%%
%%%
\begin{align}
 \mathcal G_{L_2\ell L_1m}(r,r';\tau)={}&
 \int_{-\infty+\ii\delta}^{\infty+\ii\delta}
 \frac{\dd\omega}{2\pi}\,e^{-\ii\omega\tau}
 \frac{G_{\ell m\omega}(r,r')}{\alpha_{\ell m}(\omega)}
 \nonumber\\
 &\times{}_s\mathcal A_{L_2\ell m}(a\omega)
 {}_s\mathcal A_{L_1\ell m}(a\omega).
 \label{eq:l-GFfinter}
\end{align}
The complete transfer from $L_1$ to $L_2$ is obtained by summing over the intermediate spheroidal/radial index $\ell$. At future null infinity, the radiative limit is taken as $r\to\infty$ at fixed retarded time $u=t-r_*$.
In the frequency domain, this extraction is implemented by removing the outgoing radial phase $e^{\ii\omega r_*}$.
We therefore define
\begin{equation}
 G^{\mathscr I^+}_{\ell m\omega}(r')
 \equiv
 \lim_{r\to\infty}
 r^{2s+1}e^{-\ii\omega r_*}
 G_{\ell m\omega}(r,r').
 \label{eq:radial-green-scri-limit}
\end{equation}
This definition includes the transmitted amplitude multiplying the
asymptotic UP solution.  The corresponding transfer Green's function is
\begin{align}
 \mathcal G^{\mathscr I^+}_{L_2\ell L_1m}(r';U)={}&
 \int_{-\infty+\ii\delta}^{\infty+\ii\delta}
 \frac{\dd\omega}{2\pi}\,e^{-\ii\omega U}
 \frac{G^{\mathscr I^+}_{\ell m\omega}(r')}
 {\alpha_{\ell m}(\omega)}
 \nonumber\\
 &\times{}_s\mathcal A_{L_2\ell m}(a\omega)
 {}_s\mathcal A_{L_1\ell m}(a\omega).
 \label{eq:l-GFscri}
\end{align}
Here \blue{$U=u-t'$}.

The contour-deformation strategy introduced in subsection \ref{subsection:Contour_Strategy} separates each transfer function into its direct part,
QNM contribution, and tail.  
\blue{
The QNM poles and the radial branch cuts on the imaginary axis actually come from the radial Green's function and hence depend on the intermediate
radial index $\ell$, not on $L_1$ or $L_2$. By contrast, the angular separation constant and the factor ${}_s\mathcal{A}_{L_2\ell m}(a\omega)
 {}_s\mathcal A_{L_1\ell m}(a\omega)/\alpha_{\ell m}(\omega)$ may additionally introduce angular branch cuts into the spherical-mode transfer Green’s function.
}

\subsection{Fixed-source validation against time-domain simulations}
\label{subsec:fixed-source-validation}

In this subsection, we compare the contour-reconstructed
spherical-mode transfer Green's function with a time-domain simulation.
Both calculations use the same fixed radial source and the same input
and output spherical modes.

For the contour calculation, the PIA and NIA integrals are evaluated
over a finite range of complex frequency.  We omit the angular-cut
integrals in this comparison because they extend into the same
large-complex-frequency regime and are difficult to evaluate reliably.
Our own solvers, the \texttt{Teukolsky} package in the Black Hole
Perturbation Toolkit, and
\texttt{GeneralizedSasakiNakamura.jl} do not yet provide sufficiently
robust and efficient results in this regime.

The present comparison therefore tests the radial contour
reconstruction with a finite high-frequency cutoff, rather than the
complete isolated-mode Kerr response.  The omitted high-frequency and
angular-cut contributions are expected to be most important near the
causal fronts.  A uniform asymptotic treatment could help control these
terms, but lies beyond the scope of the present comparison.  We now
turn to the time-domain simulation.

Having constructed the spherical-mode transfer Green's function, we now
validate it against independent time-domain evolutions. For the same localized impulsive source, we compare the time-domain waveform with the frequency-domain reconstruction obtained from the direct and QNM contributions. The radial solutions and asymptotic amplitudes required for these contour integrals were computed using the UP and IN solutions: in particular, the UP solution was evaluated via the MST series~\cite{Mano_1996a,Mano_1996b}, while the IN solution was obtained from the Jaff\'e series~\cite{Leaver:1986a,CO13}. For further details, see Appendix~\ref{app:mst} for MST and Appendix~\ref{app:jaffe} for Jaff\'e.

We first test a single $s=-2$, $\ell=2$ mode of the Schwarzschild Teukolsky
equation against a $(1+1)$-dimensional time-domain evolution.  The delta
source is approximated by a Gaussian in $t$ and $r_*$, centered at
\blue{$t=0$ and $r_0=10M$}, with width $0.3M$ in both directions.  The
domain is $r_*\in[-500M,2500M]$ with $30000$ grid points.

\begin{figure}[t]
 \centering
 \draftgraphic{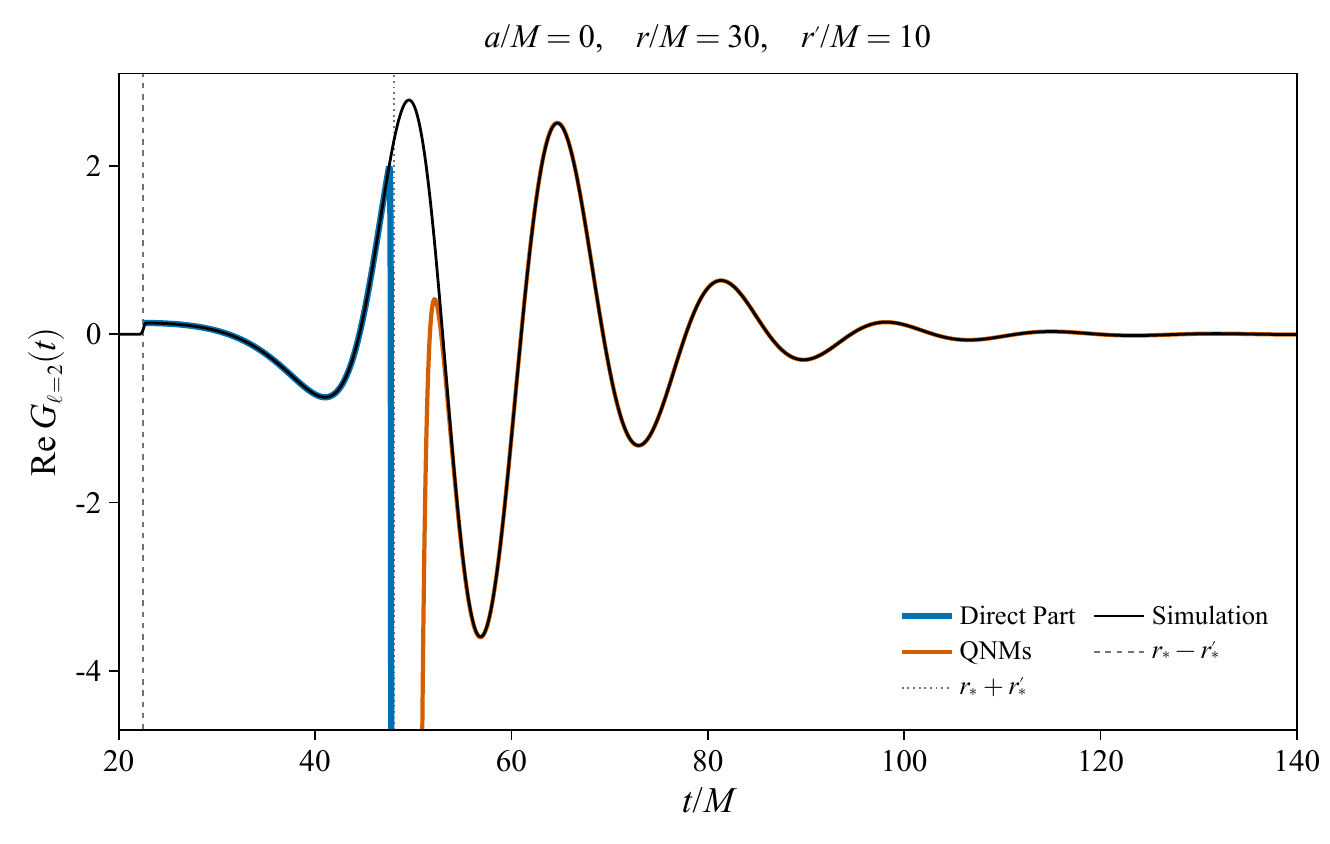}{4.5cm}
 \caption{
 Schwarzschild Teukolsky Green's function waveform for an observer at
 $r=30M$, a source at $r'=10M$, and $(s,\ell)=(-2,2)$.  The blue curve is
 the direct part, the orange curve is the sum of QNM overtones $0-7$,
 and the black curve is the $(1+1)$-dimensional numerical evolution.
 The tail is negligible over the displayed interval.}
 \label{fig:Schwarzschild_Teukolsky}
\end{figure}

As shown in Fig.~\ref{fig:Schwarzschild_Teukolsky}, the time-domain waveform nearly overlaps the sum of the direct part and the
QNM contribution.  A narrow transition window remains between the end of
the direct-part interval and the time at which the truncated QNM sum has
converged.  This gap was absent in the analogous Regge-Wheeler comparison in ~\cite{Su:2026fvj}.  Adding overtones systematically narrows the
gap, and the complete overtone series is expected to reconstruct the onset
in the appropriate distributional sense but computing sufficiently many
Teukolsky overtones is nevertheless impractical.

\begin{figure*}[t]
 \centering
 \includegraphics[width=\textwidth]
 {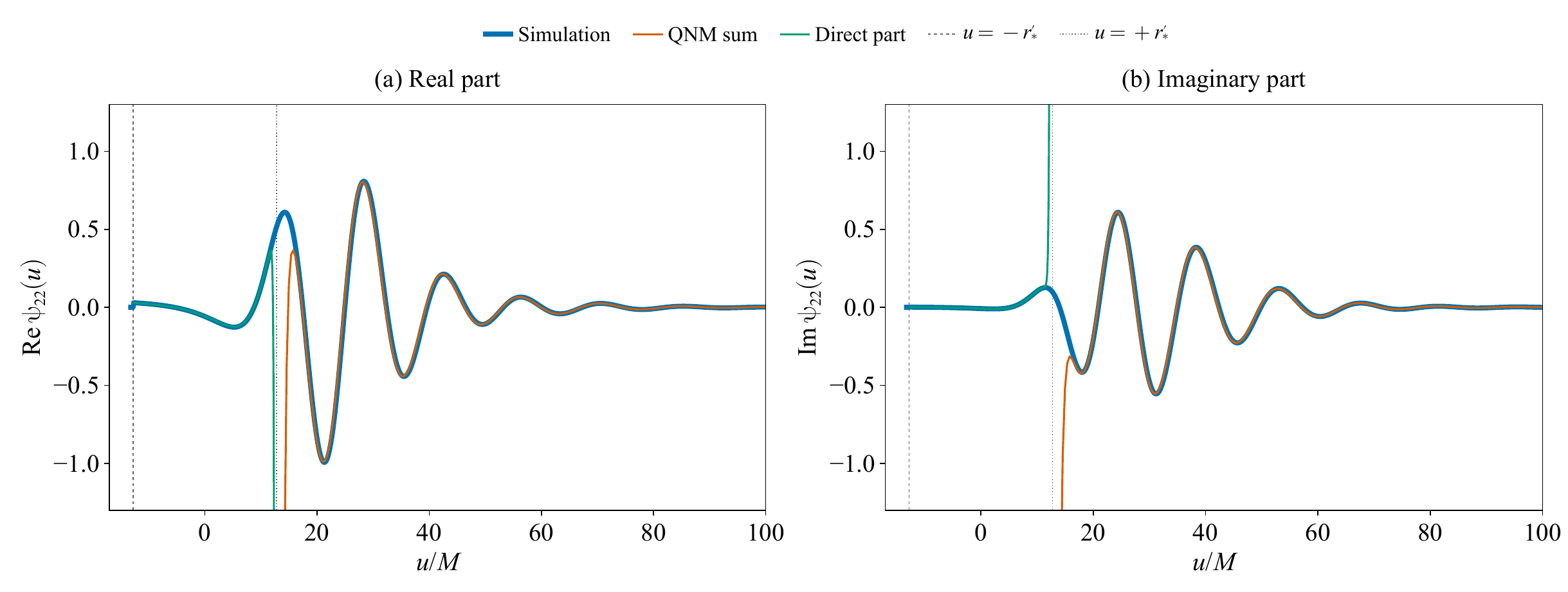}
 \caption{
Real (a) and imaginary (b) parts
of the radiative response $\psi$
at future null infinity for $s=-2$ and $a=0.4M$.
The source \eqref{SphericalSourceTerm} is injected at $r_1=10M$ in the
$(L_1,m_1)=(2,2)$ spherical mode.
The blue solid curve shows the full time-domain 2+1-D evolution.
The green and orange solid curves show,
respectively, the direct part and the sum of the first 32
QNMs (overtone $0-31$) associated
with the $(L_1,m_1)=(2,2)$ to $(L_2,m_2)=(2,2)$ transfer channel.
Their close agreement with the full time-domain response
over the respective displayed intervals except near $u=r_*'$ indicates that this $\mathcal G^{\mathscr I^+}_{2,2,2,2}$ channel dominates the waveform.
The grey dashed and dotted lines mark
${u}=-r_*'$ and ${u}=r_*'$,
respectively.}
\label{fig:ellspaceGreensFunctionLinearScale}
\end{figure*}

For Kerr, azimuthal modes decouple but the evolution remains
$(2+1)$-dimensional.  We use a compactified time-domain Teukolsky solver and
inject an $s=-2$, $L_1=m_1=2$ pulse at $r_1=10M$ that approximates
Eq.~\eqref{SphericalSourceTerm}.  We extract the spherical
$(L_2,m_2)=(2,2)$ mode at future null infinity for $a=0.4M$ red and plot the comparison in Fig.~\ref{fig:ellspaceGreensFunctionLinearScale}.

The comparison uses the direct part and the first 32 QNM overtones of
$\mathcal G^{\mathscr I^+}_{2,2,2,2}$; their frequencies are taken from the
Kerr QNM data set from
\cite{Cook_2014,PhysRevD.107.044043,Cook_2025_KerrModes}.  In
principle, the spherical response
also contains the channels with intermediate $\ell=3,4,\ldots$, but the
$\ell=2$ channel alone agrees closely with the evolution because the mixing
matrix is strongly diagonal.  As in Fig.~\ref{fig:ellspaceGreensFunctionLinearScale} (a) and Fig.~\ref{fig:ellspaceGreensFunctionLinearScale} (b),
the remaining gap near the front
reflects slow convergence of the truncated
QNM overtone series.

These fixed-source tests validate the analytic decomposition before a
worldline integral is introduced.  We now construct the radiative metric
kernel needed for the ISCO plunge and use the same direct-part/QNM+tail
organization at each source radius.

\section{Radiative Kernels and Their Decomposition}
\label{sec:kernel-decomposition}

The preceding part of this paper tested the contour representation of a
fixed-source Green's function against an independent time-domain evolution and
showed the slow, nonuniform convergence at a causal front.  We now transfer
the same analytic decomposition to the radiative metric kernel required for
an extended source, beginning with the response to one fixed source point.

We work with the leading complex radiative metric mode at future null
infinity,
\begin{equation}
 \HH(u,\theta,\phi)=\lim_{r\to\infty}r(h_+-\ii h_\times),
 \qquad
 \PPsi=\frac12\partial_u^2\HH,
 \label{eq:radiative-observable}
\end{equation}
where $h_+$ and $h_\times$ are, respectively, the `plus' and `cross' metric polarizations and $\PPsi=\lim_{r\to\infty}r\psi_4$.  With inverse-transform convention
$e^{-\ii\omega u}$, we focus on the contribution of one separated
$(s,\ell,m)=(-2,2,2)$ mode (on spin-weighted spheroidal harmonic basis) at a fixed observer direction.  In the following subsections, we first define
the \emph{radiative kernel} precisely and then discuss its contour decomposition.

\begin{revision}
This separation is needed because the contouring strategy in subsection \ref{subsection:Contour_Strategy} is naturally a
fixed-source statement: at each source radius it identifies the direct part, QNM, and tail sectors and their distinct time supports.  Only after this fixed-source structure has been established do we integrate it along the plunging worldline, where the fronts become moving boundaries. Treating the radiative kernel first also allows us to compare the real-axis inversion directly with the contour representation before performing any source integration.
\end{revision}

\subsection{Finite-start equatorial ISCO orbit}
\label{subsec:isco-orbit}
The source points used in Figs.~\ref{fig:kernel-direct} and
\ref{fig:kernel-no-direct} are sampled from the finite-start ISCO plunge used
throughout this section.  We therefore summarize the orbit before presenting
those fixed-source slices.  For the radius $r_{\rm ISCO}$ of the prograde equatorial ISCO, with
$\chi=a/M$, one has \cite{BardeenPressTeukolsky1972}
\begin{align}
 Z_1&=1+(1-\chi^2)^{1/3}
 \left[(1+\chi)^{1/3}+(1-\chi)^{1/3}\right],\\
 Z_2&=(3\chi^2+Z_1^2)^{1/2},\\
 \frac{r_{\rm ISCO}}{M}
 &=3+Z_2-\sqrt{(3-Z_1)(3+Z_1+2Z_2)}.
 \label{eq:isco-radius}
\end{align}
At $a/M=0.4$, $r_{\rm ISCO}/M=4.6143$, with intrinsic energy
$E=0.9249$ and intrinsic azimuthal angular momentum $L_z=3.0341M$.  Because the exactly marginal
ISCO is a multiple root of the radial potential, we start the inward branch
at the radius
\begin{equation}
 r_{\rm start}=r_{\rm ISCO}-0.05M,
 \qquad t_p(r_{\rm start})=\phi_p(r_{\rm start})=0,
 \label{eq:finite-start}
\end{equation}
while retaining the ISCO values of $E$ and $L_z$.  Define
\begin{align}
 \mathcal B&=L_z-aE,&
 \mathcal P&=E(r^2+a^2)-aL_z,\\
 \mathcal C&=r^2+\mathcal B^2,&
 \mathcal R&=\mathcal P^2-\Delta\mathcal C.
\end{align}
The infalling branch then obeys
\begin{equation}
 \dot r=-\frac{\sqrt{\mathcal R}}{r^2},\qquad
 J_t(r)\equiv-\frac{\dd t_p}{\dd r}
 =\frac{a\mathcal B+(r^2+a^2)\mathcal P/\Delta}
 {\sqrt{\mathcal R}}.
 \label{eq:plunge-jacobian}
\end{equation}
The numerical worldline ends at $r=r_++10^{-6}M$, with
$r_+/M=1.9165$, without treating the Boyer--Lindquist horizon as a
finite coordinate-time endpoint.

\subsection{Fixed-source kernel and its construction}
\label{subsec:kernel-definition}

Consider a worldline event at source radius $r'$, with coordinates
$(t_p(r'),r',\pi/2,\phi_p(r'))$, and define $U=u-t_p(r')$. 
We write \(\phi_p(t)\) when parametrizing the worldline by time and \(\phi_p(r')\equiv\phi_p(t_p(r'))\) when using the source radius; both denote the same orbital phase.

We define the radiative kernel $\Kern_{\PPsi,\ell m}$ (for curvature variable) and $\Kern_{\HH,\ell m}$ (for metric variable) as the response extracted at future null infinity for this fixed point-particle source event, before integration over the worldline.
That is, the curvature contribution from the complete plunge is
\begin{equation}
 \PPsi_{\ell m}(u)
 =
 \int_{r_+}^{r_{\rm start}}\dd r'\,
 J_t(r')\,
 \Kern_{\PPsi,\ell m}\bigl(u-t_p(r');r'\bigr),
 \label{eq:kernel-operational-definition}
\end{equation}
with an analogous expression obtained by replacing
$\PPsi\to\HH$ and
$\Kern_{\PPsi,\ell m}\to\Kern_{\HH,\ell m}$.
Here $J_t(r')=-\dd t_p/\dd r'>0$ is the Jacobian introduced in
Eq.~\eqref{eq:plunge-jacobian}.
It appears because the point-particle source is initially integrated
over source coordinate time. Along the infalling trajectory,
$\dd t_p=-J_t(r')\,\dd r'$; reversing the radial integration limits
then gives Eq.~\eqref{eq:kernel-operational-definition}.
The finite starting radius $r_{\rm start}$ is specified in
Sec.~\ref{subsec:isco-orbit}.
This definition fixes all factors that might otherwise be assigned to the Green's function, the source amplitude, or the final worldline integral.
Below, we construct the kernel explicitly from the frequency-domain Green's function and the point-particle source.

We use the previously defined $s=-2$ IN solution
$R^{\rm in}_{\ell m\omega}$ and separation constant
$\lambda_{\ell m\omega}$.
Mode labels are suppressed below whenever no ambiguity arises.
The orbit-dependent quantities $\mathcal B$, $\mathcal P$,
$\mathcal C$, and $\mathcal R$ are defined in
Sec.~\ref{subsec:isco-orbit}.

Before proceeding, we define the following quantities:
\begin{align}
 V&=\sqrt{\mathcal R},
 &
 \mathcal T&=
 a\mathcal B+\frac{(r^2+a^2)\mathcal P}{\Delta},
 \nonumber\\
 b&=a\omega-m,
 &
 S_0&={}_{-2}S_{\ell m\omega}(\pi/2),
 \nonumber\\
 S_1&=
 \left.
 \partial_\theta{}_{-2}S_{\ell m\omega}
 \right|_{\theta=\pi/2},
 &
 \mathcal L&=S_1+bS_0.
 \label{eq:source-angular-abbreviations}
\end{align}
Following Ref.~\cite{SasakiTagoshi2003}, the source coefficients
for the equatorial plunge are
\begin{align}
 A_{nn0}(r)
 ={}&
 -\frac{\mathcal C^2}
 {2\sqrt{2\pi}\,\mathcal T(\mathcal P+V)^2}
 \left[
 \left(2b-\frac{2\ii a}{r}\right)\mathcal L
 -\lambda S_0
 \right],
 \label{eq:ann0}\\
 A_{\bar mn0}(r)
 ={}&
 \frac{\ii\mathcal B\mathcal C}
 {\sqrt{2\pi}\,\mathcal T(\mathcal P+V)}
 \mathcal L
 \left(\frac{\ii K}{\Delta}+\frac{2}{r}\right),
 \label{eq:amn0-source}\\
 A_{\bar m\bar m0}(r)
 ={}&
 \frac{\mathcal B^2S_0}
 {2\sqrt{2\pi}\,\mathcal T}
 \left[
 -\ii\left(\frac{K}{\Delta}\right)_{,r}
 -\frac{K^2}{\Delta^2}
 +\frac{2\ii K}{r\Delta}
 \right],
 \label{eq:amm0-source}\\
 A_{\bar mn1}(r)
 ={}&
 \frac{\ii\mathcal B\mathcal C}
 {\sqrt{2\pi}\,\mathcal T(\mathcal P+V)}
 \mathcal L,
 \label{eq:amn1-source}\\
 A_{\bar m\bar m1}(r)
 ={}&
 \frac{\mathcal B^2S_0}
 {\sqrt{2\pi}\,\mathcal T}
 \left(\frac{\ii K}{\Delta}+\frac{1}{r}\right),
 \label{eq:amm1-source}\\
 A_{\bar m\bar m2}(r)
 ={}&
 \frac{\mathcal B^2S_0}
 {2\sqrt{2\pi}\,\mathcal T}.
 \label{eq:amm2-source}
\end{align}
Here
\begin{equation}
 \left(\frac{K}{\Delta}\right)_{,r}
 =
 \frac{2r\omega\Delta-2(r-M)K}{\Delta^2}.
 \label{eq:K-over-Delta-derivative}
\end{equation}

For convenience, we combine these coefficients as
\begin{align}
 A_0&=A_{nn0}+A_{\bar mn0}+A_{\bar m\bar m0},
 \nonumber\\
 A_1&=A_{\bar mn1}+A_{\bar m\bar m1},
 \nonumber\\
 A_2&=A_{\bar m\bar m2}.
 \label{eq:source-A-combinations}
\end{align}

With these coefficients, the separated point-particle source for
the radial Teukolsky equation can be written, following
Ref.~\cite{SasakiTagoshi2003}, as
\begin{align}
 T_{\ell m\omega}&(r)
 =
 \frac{\mu}{{\alpha_{\ell m}(\omega)}}\int_{-\infty}^{\infty}\dd t\,
 e^{\ii\omega t-\ii m\phi_p(t)}\Delta^2
 \Bigl[
 A_0\delta\bigl(r-r_p(t)\bigr)
 \nonumber\\
 &+
 \bigl\{A_1\delta\bigl(r-r_p(t)\bigr)\bigr\}_{,r}
 +
 \bigl\{A_2\delta\bigl(r-r_p(t)\bigr)\bigr\}_{,rr}
 \Bigr],
 \label{eq:radial-particle-source}
\end{align}
where $\mu$ is the mass of the plunging particle.
The source is understood to vanish before the finite starting time
specified in Sec.~\ref{subsec:isco-orbit}.

Let \(R(r)\) denote either of the homogeneous radial solutions used to
construct the radial Green function.  The Green-function solution contains
an integral of \(R(r)T_{\ell m\omega}(r)/\Delta^2\) over the source radius.
Consequently, the factor \(\Delta^2\) in
Eq.~\eqref{eq:radial-particle-source} cancels before the delta functions
are integrated.  At fixed \(t\), and suppressing factors independent of
\(r\), the relevant integral is
\begin{align}
 &\int_{r_+}^\infty \dd r\,R(r)
 \Bigl[
 A_0\delta\bigl(r-r_p(t)\bigr)
 +\bigl\{A_1\delta\bigl(r-r_p(t)\bigr)\bigr\}_{,r}
 \nonumber\\
 &\hspace{3.2cm}
 +\bigl\{A_2\delta\bigl(r-r_p(t)\bigr)\bigr\}_{,rr}
 \Bigr]
 \nonumber\\
 &\qquad =
 \left[
 A_0R-A_1R_{,r}+A_2R_{,rr}
 \right]_{r=r_p(t)} .
 \label{eq:source-r-integration}
\end{align}
The minus sign in the second term and the plus sign in the third follow from integrating by parts once and twice, respectively.  We use \(\mathcal W[R]\) as shorthand for the resulting combination,
\begin{equation}
 \mathcal W[R]
 =A_0R-A_1R_{,r}+A_2R_{,rr}.
 \label{eq:source-action}
\end{equation}
After the \(r\) integral is carried out, all quantities in
\(\mathcal W[R]\) are evaluated at \(r=r_p(t)\).  The derivatives in
\(\mathcal W[R]\) act on \(R\), not on \(A_0\), \(A_1\), or \(A_2\).

Using the homogeneous radial Teukolsky equation,
Eq.~\eqref{eq:radial_teukolsky_equation}, to eliminate $R_{,rr}$,
the source operator can equivalently be written as
\begin{equation}
 \mathcal W[R]=Q_0R+Q_1R_{,r},
 \label{eq:reduced-source-action}
\end{equation}
where
\begin{align}
 Q_0={}&A_{nn0}+A_{\bar mn0}
 +\frac{\mathcal B^2S_0}
 {2\sqrt{2\pi}\,\mathcal T}\,\Xi_0,
 \label{eq:q0-explicit}\\
 Q_1={}&-A_{\bar mn1}
 +\frac{\mathcal B^2S_0}
 {\sqrt{2\pi}\,\mathcal T}\,\Xi_1,
 \label{eq:q1-explicit}
\end{align}
with
\begin{align}
 \Xi_0(r)
 ={}&
 -\frac{2K^2+2\ii(r-M)K}{\Delta^2}
 +\frac{8\ii\omega r+\lambda+2\ii ab/r}{\Delta},
 \label{eq:xi0-source}\\
 \Xi_1(r)
 ={}&
 \frac{r-M-\ii K}{\Delta}-\frac{1}{r}.
 \label{eq:xi1-source}
\end{align}

With the Green-function normalization adopted in the preceding
section, the frequency-domain radial response to this source is
\begin{equation}
 R_{\ell m\omega}(r)
 =
 \int_{r_+}^{\infty}\dd r'\,
 G_{\ell m\omega}(r,r')
 \frac{T_{\ell m\omega}(r')}{\Delta^2(r')}.
 \label{eq:radial-response-from-gf}
\end{equation}
For $s=-2$, the radiative limit defined in
Eq.~\eqref{eq:radial-green-scri-limit}, corresponding to
$r\to\infty$ at fixed retarded time $u=t-r_*$, gives
\begin{equation}
 G^{\mathscr I^+}_{\ell m\omega}(r')
 =
 +\frac{R^{\rm in}_{\ell m\omega}(r')}
 {2\ii\omega A^{\rm in}_{\rm inc}},
 \label{eq:radiative-gf-explicit}
\end{equation}
with the Green-function normalization adopted above.

The retarded inhomogeneous radial solution therefore has the
outgoing asymptotic form
\begin{align}
 R_{\ell m\omega}(r)
 \sim{}&
 \frac{r^3e^{\ii\omega r_*}}
 {2\ii\omega A^{\rm in}_{\rm inc}}
 \int_{r_+}^{\infty}\dd r'\,
 \frac{T_{\ell m\omega}(r')R^{\rm in}_{\ell m\omega}(r')}
 {\Delta^2(r')}
 \nonumber\\
 =: {}&
 \widetilde Z^\infty_{\ell m\omega}
 r^3e^{\ii\omega r_*},
 \qquad r\to\infty .
 \label{eq:radial-infinity-amplitude}
\end{align}
Substituting Eq.~\eqref{eq:radial-particle-source} and performing
the radial integration in \eqref{eq:radial-infinity-amplitude} gives
\begin{equation}
 \begin{aligned}
 \widetilde Z^\infty_{\ell m\omega}
 =
\mu
 \int_{-\infty}^{\infty}\dd t\,
 \frac{
 e^{\ii\omega t-\ii m\phi_p(t)}
 \left.
 \mathcal W\!\left[R^{\rm in}_{\ell m\omega}\right]
 \right|_{r=r_p(t)} 
 }
 {2\ii\omega A^{\rm in}_{\rm inc}{\alpha_{\ell m}(\omega)}}
 .
 \end{aligned}
 \label{eq:zinfinity-source-action}
\end{equation}

The corresponding Weyl scalar is reconstructed from
\begin{equation}
 \rho^{-4}\psi_4
 =
 \sum_{\ell m}\int_{\mathbb{R}}\dd\omega\,
 e^{-\ii\omega t}
 \frac{{}_{-2}S_{\ell m\omega}(\theta)e^{\ii m\phi}}
 {\sqrt{2\pi}}\,
 R_{\ell m\omega}(r),
 \label{eq:psi4-mode-expansion}
\end{equation}
where $ \rho=-(r-\ii a\cos\theta)^{-1} $
is the usual Kinnersley-tetrad spin coefficient.
At fixed retarded time $u=t-r_*$, the radiative curvature at
future null infinity is therefore
\begin{equation}
 \PPsi(u,\theta,\phi)
 =
 \sum_{\ell m}\int_{\mathbb{R}}\dd\omega\,
 \widetilde Z^\infty_{\ell m\omega}
 \frac{{}_{-2}S_{\ell m\omega}(\theta)e^{\ii m\phi}}
 {\sqrt{2\pi}}\,
 e^{-\ii\omega u}.
 \label{eq:radiative-curvature-infinity}
\end{equation}

The corresponding mode expansion of the metric waveform is
\begin{equation}
 \HH(u,\theta,\phi)
 =
 \sum_{\ell m}\int_{\mathbb{R}}\dd\omega\,
 \widetilde H^\infty_{\ell m\omega}
 \frac{{}_{-2}S_{\ell m\omega}(\theta)e^{\ii m\phi}}
 {\sqrt{2\pi}}\,
 e^{-\ii\omega u},
 \label{eq:radiative-metric-infinity}
\end{equation}
where
\begin{equation}
 \widetilde H^\infty_{\ell m\omega}
 =
 -\frac{2}{\omega^2}
 \widetilde Z^\infty_{\ell m\omega}, \quad \omega \neq 0.
 \label{eq:metric-amplitude-from-curvature}
\end{equation}
The retarded prescription at $\omega=0$ fixes the integration
constants in converting curvature to metric.

We now read off the fixed-source kernels from these expressions.
Because the plunge is monotonic in radius, the source-time
integral in Eq.~\eqref{eq:zinfinity-source-action} can be written as
\begin{align}
 \widetilde Z^\infty_{\ell m\omega}
 ={}&
 \frac{\mu}{2\ii\omega A^{\rm in}_{\rm inc}\alpha_{\ell m}(\omega)}
 \int_{r_+}^{r_{\rm start}}\dd r'\,
 J_t(r')e^{\ii\omega t_p(r')-\ii m\phi_p(r')}
 \nonumber\\
 &\times
 \mathcal W\!\left[
 R^{\rm in}_{\ell m\omega}
 \right]_{r=r'}.
 \label{eq:zinfinity-radial-integral}
\end{align}
Define the observer angular factor
\begin{equation}
 \mathcal O_{\ell m}(\omega)
 :=
 \frac{{}_{-2}S_{\ell m\omega}(\theta_{\rm obs})
 e^{\ii m\phi_{\rm obs}}}{\sqrt{2\pi}}.
 \label{eq:observer-projection}
\end{equation}

Substituting Eq.~\eqref{eq:zinfinity-radial-integral} into
Eq.~\eqref{eq:radiative-curvature-infinity} and Eq.~\eqref{eq:radiative-metric-infinity}, evaluating the angular
dependence at $(\theta_{\rm obs},\phi_{\rm obs})$, and interchanging
the radial and frequency integrations gives
\begin{align}
 \PPsi_{\ell m}(u)
 ={}&
 \int_{r_+}^{r_{\rm start}}\dd r'\,
 \frac{J_t(r')}{2\pi}\int_{\mathbb{R}}\dd\omega\,
 e^{-\ii\omega[u-t_p(r')]}
 \tilde{\Kern}_{\PPsi,\ell m}(\omega,r').
 \label{eq:curvature-kernel-read-off}
 \\
\HH_{\ell m}(u)
 ={}&
 \int_{r_+}^{r_{\rm start}}\dd r'\,
 \frac{J_t(r')}{2\pi}\int_{\mathbb{R}}\dd\omega\,
 e^{-\ii\omega[u-t_p(r')]}
 \tilde{\Kern}_{\HH,\ell m}(\omega,r').
 \label{eq:metric-kernel-read-off}
\end{align}

The relation between the radiative kernel and the Green's function  $G^{\mathscr I^+}_{\ell m\omega} (r')$ is explicitly
\begin{align}
 \tilde{\Kern}_{\PPsi,\ell m}(\omega,r')
 =
 \frac{\mathcal O_{\ell m}(\omega)e^{-\ii m\phi_p(r')}}{{\alpha_{\ell m} (\omega) } \cdot  \mathcal N_F^{-1} \mu^{-1}}
 \mathcal W_{r'}\left[G^{\mathscr I^+}_{\ell m\omega} (r')\right].
 \label{eq:kernel-from-gf}
\end{align}
The constant factor $\mathcal N_F\equiv 2\pi$ accounts for writing the
frequency inversion with the prefactor $1/(2\pi)$, whereas
Eq.~\eqref{eq:radiative-curvature-infinity} has no such prefactor.
The subscript $r'$ specifies that the source operator acts on the source radial argument of the Green's function. Its coefficients are evaluated at $r'$, and its derivatives act only on the Green's function. 

Substituting Eq.~\eqref{eq:radiative-gf-explicit} into Eq.~\eqref{eq:kernel-from-gf} gives the full fixed-source curvature kernel:
%}
\begin{align}
 \tilde{\Kern}_{\PPsi,\ell m}(\omega,r')
 ={}&
 \frac{\mu\,\mathcal O_{\ell m}(\omega)
 e^{-\ii m\phi_p(r')}}
 {   2\ii\omega A^{\rm in}_{\rm inc} {\alpha_{\ell m}(\omega)}\mathcal N_F^{-1}}
 %\nonumber\\
 %&\times
 \mathcal W\!\left[
 R^{\rm in}_{\ell m\omega}
 \right]_{r=r'}.
 \label{eq:full-curvature-kernel}
\end{align}
%%%
%%%
Using Eq.~\eqref{eq:reduced-source-action}, the source operator
can equivalently be evaluated as
\begin{equation}
 \mathcal W\!\left[R^{\rm in}\right]_{r=r'}
 =
 Q_0(r')R^{\rm in}(r')
 +
 Q_1(r')R^{\rm in}_{,r}(r').
 \label{eq:kernel-reduced-source}
\end{equation}

The corresponding metric kernel follows from
Eq.~\eqref{eq:metric-amplitude-from-curvature}:
\begin{equation}
 \tilde{\Kern}_{\HH,\ell m}(\omega,r')
 =
 -\frac{2}{\omega^2}
 \tilde{\Kern}_{\PPsi,\ell m}(\omega,r'),\qquad \omega \neq 0.
 \label{eq:full-metric-kernel}
\end{equation}
For either radiative quantity $X\in\{\PPsi,\HH\}$, the
fixed-source time-domain kernel is
\begin{equation}
 \Kern_{X,\ell m}(U;r')
 =
 \frac{1}{2\pi}
 \int_{\mathbb{R}}\dd\omega\,
 e^{-\ii\omega U}
 \tilde{\Kern}_{X,\ell m}(\omega,r'),
 \label{eq:fixed-source-time-kernel}
\end{equation}
and its worldline integral gives the complete waveform as in
Eq.~\eqref{eq:kernel-operational-definition}.

The radiative kernel is thus the Green-function response with
the point-particle source applied and the observer dependence
included, rather than a separate propagator.
Since the source operator and radiative extraction are linear,
the contour decomposition of the kernel follows by applying
these operations to the Green-function decomposition discussed
in Sec.~\ref{subsec:schwarzschild-green-function}

Following the decomposition strategy used for the Green's function in
Sec.~\ref{subsec:schwarzschild-green-function}, applying the scattering
identity \eqref{eq:radial-scattering-identity} to the IN solution in the
radiative kernel gives
\begin{equation}
 \tilde{\Kern}_{\PPsi}
 =
 \tilde{\Kern}_{\PPsi}^{-}
 +
 \tilde{\Kern}_{\PPsi}^{+},
 \label{eq:kernel-plus-minus}
\end{equation}
where the two pieces have the schematic frequency
dependence
\begin{align}
 \tilde{\Kern}_{\PPsi}^{-}
 &\propto
 \frac{\mathcal O_{\ell m}(\omega)e^{-\ii m\phi_p}}
 {2\ii\omega\alpha_{\ell m}(\omega)}
 \,\mathcal W[R^{\rm down}],
 \nonumber\\
 \tilde{\Kern}_{\PPsi}^{+}
 &\propto
 \frac{\mathcal O_{\ell m}(\omega)e^{-\ii m\phi_p}}
 {2\ii\omega\alpha_{\ell m}(\omega)}
 \frac{A^{\rm in}_{\rm ref}}{A^{\rm in}_{\rm inc}}
 \,\mathcal W[R^{\rm up}].
 \label{eq:kernel-scattering-pieces}
\end{align}
We suppress the \((\ell,m)\) labels when no ambiguity can arise.  We also
write \(\tilde{\Kern}_X^\pm\) when the discussion applies to a general
radiative variable \(X\), such as \(X=\PPsi\) or \(X=\HH\).
The analytic structure of
\(\tilde{\Kern}_X=\tilde{\Kern}_X^-+\tilde{\Kern}_X^+\) parallels that of
\(\tilde G=\tilde G^-+\tilde G^+\).  We therefore define the direct, QNM,
and tail contributions using the same contour deformations and time interval as for the
Green function.  In all calculations below, \(X=\HH\): unless stated
otherwise, the numerical results and waveforms are constructed from the
metric kernel \(\tilde{\Kern}_{\HH}\), or its time-domain inverse
\(\Kern_{\HH}\).

\section{ Numerical Implementation}
\label{sec:numerical-implementation}

In this section, we first implement the contour integration of the radiative kernel, decompose it into contributions from the different nonanalytic structures enclosed by the contour, and compare the result with the radiative kernel obtained by a real-frequency inverse Fourier transform. We find that the two calculations agree closely except near the causal fronts, similar to the behavior shown in Fig.~\ref{fig:ellspaceGreensFunctionLinearScale}.

A natural question then arises: how do the QNM and branch-cut direct contributions combine to produce the waveform from a plunging particle?
Despite the difficulties near the causal fronts, we can approximately divide the radiative kernel obtained from the real-frequency contour into two parts. We associate the part before the arrival time with the branch-cut direct contribution and approximate the later part by the QNM sum, neglecting the small late-time tail. This provides a rough estimate of the contribution from each part. A complete treatment of the region near the causal front is left for future work.

Subsections~\ref{subsec:contour-numerics} and~\ref{subsec:real-axis-numerics} describe the numerical details of the contour and real-frequency calculations, respectively. Readers who are not interested in the technical details may skip these two subsections.
Subsection~\ref{subsec:numerical-kernel-results} compares the radiative kernels obtained from contour integration and real-frequency inversion. Subsection~\ref{subsec:numerical-kernel-results} then presents the ISCO-plunge waveforms and their decomposition based on these radiative kernels.

\subsection{ Numerical Implementation of the Contour Calculation}
\label{subsec:contour-numerics}

We use the contours shown in Fig.~\ref{fix_ell_m_contour} to separate the
frequency-domain response into an early-time direct contribution and a
sum over quasinormal modes (QNMs). The QNM sum includes overtone labels
\(n=0,\ldots,41\). We use Cook's numerical frequencies \cite{Cook_2025_KerrModes} as starting
values and refine them with our MST code. The same code is used to
calculate the amplitude of each mode. The frequency derivative entering
the amplitude is obtained by Richardson extrapolation as in Ref.~\cite{Lo:2025njp}, while the ingoing radial solution
is evaluated with a Jaff\'e series. Further details are given in
Appendices~\ref{sec:qnm-subtleties}, \ref{app:mst}, and~\ref{app:jaffe}.

The direct part contribution comes from integrating along the two sides of
the cuts that extend upward and downward from \(\omega=0\) along the
imaginary-frequency axis. To avoid evaluating directly on a cut, we
place the two sides at
\(\operatorname{Re} (M\omega)=\pm10^{-8}\) and take the difference between them. The
upper integral is terminated at \(\operatorname{Im}(M\omega)=40\), and the lower one
at \(\operatorname{Im}(M\omega)=-2\). Near the origin, the two sides are joined by
semicircles of radius \(|M\omega|=0.05\). We use 6001 points on the
upper cut, 301 points on the lower cut, and 300 points on each
semicircle.

The MST accuracy is adjusted separately at each frequency. The
calculation begins with 96-bit arithmetic and 80 recurrence terms. If
the result has not converged, both the precision and the number of terms
are increased, up to 768 bits and 2240 terms. We stop once two
successive refinements agree within the chosen tolerance. The resulting
frequency-domain values are interpolated with piecewise quintic
polynomials, and the contour integrals are then evaluated numerically.

We omit the separate late-time tail integral because it is small over
the time interval studied here. We also omit additional cut
contributions associated with the angular eigenfunctions. These terms
are difficult to follow consistently through the complex-frequency
plane, as discussed in Sec.~\ref{subsec:fixed-source-validation}. The angular-cut contribution has little effect away from the causal fronts, but becomes more important near them, which will be discussed in Sec.~\ref{subsec:numerical-kernel-results}.

\subsection{Numerical evaluation of the real-axis inverse transform}
\label{subsec:real-axis-numerics}

The positive-frequency angular data and the radial quantities entering
Eq.~\eqref{eq:full-curvature-kernel} are calculated directly. In the range
$0<M\omega\leq0.4$, we use the MST formalism
\cite{Mano_1996a}  (see App.~\ref{app:mst} for details) and Jaff\'e series \cite{Leaver:1986a} (see Appendix~\ref{app:jaffe} for details) at
160-bit precision, with angular-basis cutoff 
%\MC{Is this the number of $\ell$-modes %used in the equivalent of %\eqref{eq:psil2m2}?}
%\JS{No, they are irrelevant. %\eqref{eq:psil2m2} is for the $\ell$ %space Green's function while here is for %the radiative kernel \MC{That's why I %said *the equivalent of* %\eqref{eq:psil2m2}}. And this \%%(L_{\max}^{\rm basis}\) here is the %truncation of the pseudo spectral %calculation for the angular equation.}
$L_{\max}^{\rm basis}=60$ (the truncation in Eq.~\eqref{eq:Spheroidal_Spherical_Expansion}), 400 MST-series terms, and 8000 Jaff\'e-series terms. 
At relatively large real frequencies, the MST solver is usually inefficient. Therefore, when $M\omega>0.4$, we switch to using
a Riccati solver from the modified open-source \texttt{Julia} package \texttt{
GeneralizedSasakiNakamura.jl}
\cite{Lo:2023fvv}. The angular eigenvalues
and spin-weighted spheroidal harmonics supplied to both radial solvers are
computed using the method described in Appendix~\ref{app:angular}
\green{\cite{FackerellCrossman1977}}.

%\MC{Why is this  the
%overlap region when the Riccati solver %for SN is used only for $M\omega>0.4$?} %\JS{See the last paragraph. For large %real frequencies, the MST calculation is %not efficient, whereas the %\texttt{GeneralizedSasakiNakamura.jl} %Riccati solver is.}
%\MC{Precisely: so for $M\omega>0.4$ %Riccati is used, not MST, and so %$M\omega>0.4$ is not an overlap region %(there's no overlap between %$0<M\omega\leq0.4$ and $M\omega>0.4$)}
In the spectra for generating the final waveform, the
Riccati/Sasaki--Nakamura (see Ref.~\cite{Lo:2023fvv} for details) radial solution is used only for $M\omega>0.4$.
As a handoff check, however, we also run the Riccati solver in the validation
band $0.1\leq M\omega\leq0.4$, where the MST solution is available.  In this
band we compare $R^{\rm in}/A^{\rm in}_{\rm inc}$,
$R^{\rm in}_{,r}/A^{\rm in}_{\rm inc}$, and the assembled full kernel between
the two radial methods.

The zero-frequency node requires a separate check because
the conversion from the curvature kernel to the metric kernel contains a factor $1/\omega^2$, which cannot be evaluated directly at $\omega=0$, while the real-axis interpolation requires a value at that node.

Extrapolation from positive and negative frequencies gives
$\tilde{\Kern}_{\HH}=O(\omega^2)$ and
$\tilde{\Kern}_{\PPsi}=O(\omega^4)$ as $\omega\to0$.
We therefore set the stored $\omega=0$
values of both kernels to zero before constructing the
real-axis inverse transform.

The spectra used to compute the final waveforms are sampled at 26,465 frequencies on three nested grids. The base grid covers $|M\omega|\leq40$
with spacing $M\Delta\omega=0.005$. A finer low-frequency grid covers $|M\omega|\leq0.4$ with spacing
$M\Delta\omega=7.8125\times10^{-5}$, and the finest central grid covers $|M\omega|\leq0.005$ with spacing $M\Delta\omega=1.953125\times10^{-5}$.

Duplicate nodes from different grids are removed after confirming that the corresponding values agree within numerical accuracy. The finite frequency interval is closed smoothly with the following smooth Planck taper:
\begin{equation}
 W_{40,5}(\omega)=
 \begin{cases}
 1,&|M\omega|\leq35,\\
 w(z),&35<|M\omega|<40,\\
 0,&|M\omega|\geq40,
 \end{cases}
 \qquad z=\frac{|M\omega|-35}{5},
 \label{eq:planck-regulator}
\end{equation}
where
\begin{equation}
 w(z):=\left[1+\exp\!\left(\frac{1}{1-z}-\frac{1}{z}\right)\right]^{-1}.
 \label{eq:planck-transition}
\end{equation}

Because the frequency grid is nonuniform, we evaluate the inverse transform
interval by interval using a degree-11 Hermite interpolant
\cite{StoerBulirsch2002}, rather than a periodic FFT.

Since $\omega=0$ is a branch point, the positive- and negative-frequency
sides are interpolated separately.

The derivatives needed for the Hermite interpolation are computed by finite
differences, with local polynomial fits used on the refined low-frequency
grid.  The inverse transform is obtained by analytically integrating the
interpolant over each frequency interval.  We use \texttt{Double64}
arithmetic and, at early times, check an accelerated chirp-$z$ evaluation
against the direct sum.  Their discrepancy is required to remain below
$10^{-7}$ in relative $L^2$ norm and below $10^{-8}$ relative to the
waveform peak.

%%%
%%

\subsection{Numerical Comparison of the Radiative Kernels}
\label{subsec:numerical-kernel-results}

In this subsection, we compare the radiative kernel from real frequency inverse Fourier transform with the result obtained by contour integration. Both  one with a direct-part interval and one without such an interval.

\begin{figure*}[t]
 \centering
 \includegraphics[width=\textwidth]
{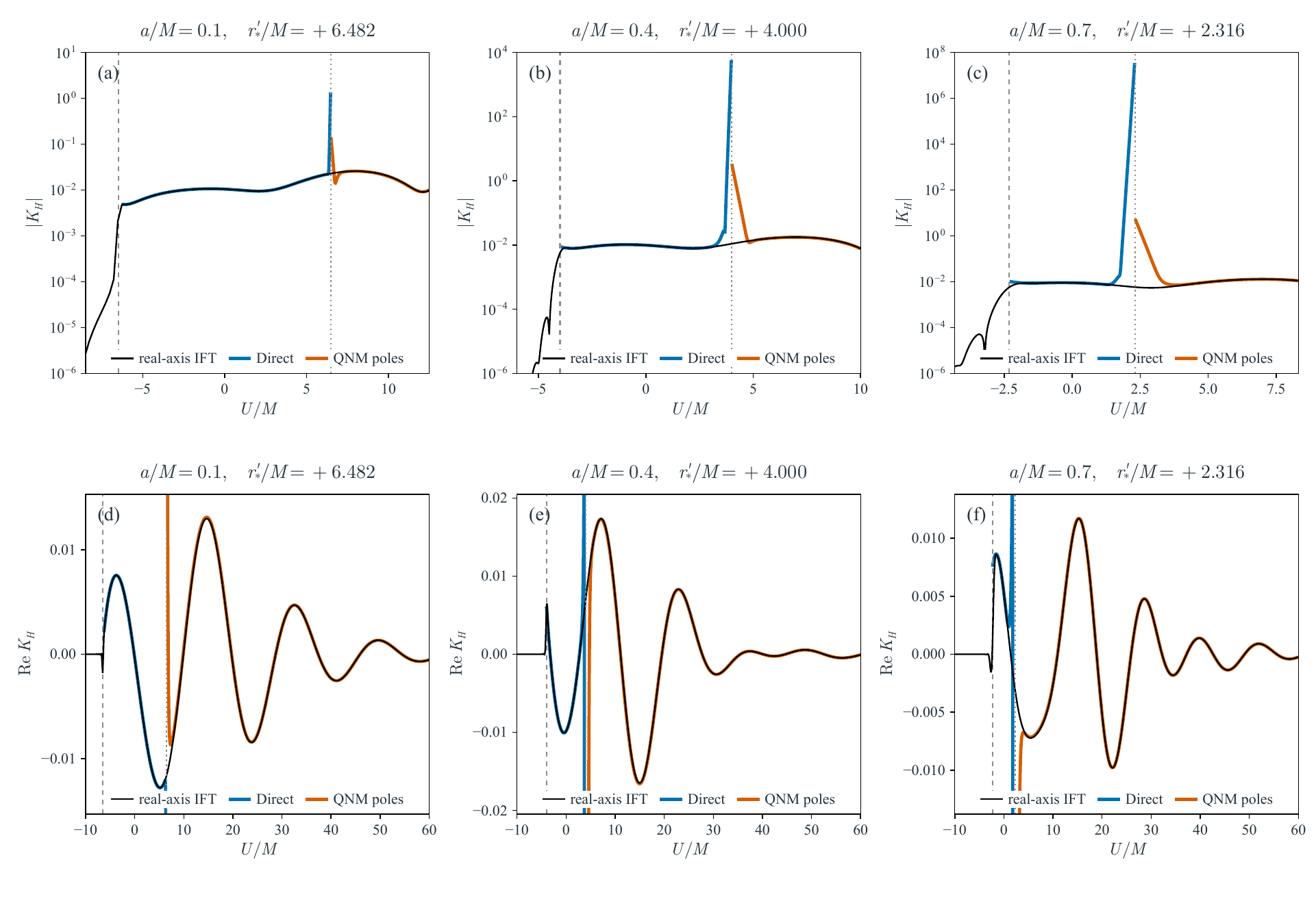}

\caption{ 
 Comparison of the $(\ell,m)=(2,2)$ radiative-metric kernel $\Kern_{\HH}$ for Kerr spins $a/M=0.1,\ 0.4,\ 0.7$ (left to right), with positive source positions $r_*'/M=6.482,\ 4.000,\ 2.316$, respectively, and observer angle $\theta_{\rm obs}=1.1$.
 Black, blue, and orange curves show, respectively, the real-axis inverse Fourier transform, the direct part computed from the truncated contour integrals, and the QNM pole sum. The PIA and NIA integrals are truncated at
 $\sigma_{\max}^{\rm PIA}M=40$ and $\sigma_{\max}^{\rm NIA}M=2$, respectively. The integration paths on the two sides of each branch cut lie a distance $|\operatorname{Re}\omega|M=10^{-8}$ from the imaginary axis.
 Panels (a)--(c) show $|\Kern_{\HH}|$ with a logarithmic vertical axis and a linear horizontal axis. Panels (d)--(f) show $\operatorname{Re}\Kern_{\HH}$ with both axes linear. The vertical limits in the lower panels are chosen to show the black curves clearly, so the larger blue values near $U=r_*'$ lie outside the plotted range.
 Dashed and dotted vertical lines mark $U=-r_*'$ and $U=r_*'$, respectively.
 The QNM sums include both prograde and retrograde poles with overtone labels $n=0,\ldots,41$. For $a/M=0.7$, the retrograde $n=14$ pole is omitted because its continuation lies on a different Riemann sheet; see Appendix~\ref{sec:qnm-subtleties}.
}

 \label{fig:kernel-direct}
\end{figure*}

For the case with a direct part, the waveforms are shown in  Fig.~\ref{fig:kernel-direct}, with the parameters specified in the caption. In all panels, the direct part shown by the blue curve—constructed from the PIA contribution of \(\tilde{K}^{+}_{\HH,\ell m}\), the small detour around the origin, and the NIA contribution of \(\tilde{K}^{-}_{\HH,\ell m}\)—agrees well with the real-frequency inversion except near the dashed and dotted gray lines. The finite-overtone QNM sums also agree well with the real-frequency inversion, except near the dotted gray line marking the causal font. The mismatch region is narrow for small \(a\) and becomes wider as \(a\) increases.

The mismatch around the dashed line is mainly due to the neglected angular cut, whereas the mismatch near the dotted line has a more complicated origin. There are several contributing factors.

First, the angular cuts are neglected near these boundaries. On the direct-part side, the angular cuts lie in the upper half-plane of \(\tilde{K}^{+}_{\HH,\ell m}\) and the lower half-plane of \(\tilde{K}^{-}_{\HH,\ell m}\). On the QNM side, the relevant angular cut lies in the lower half-plane of the Leaver-like contour, as shown in panel (a) of Fig.~\ref{fix_ell_m_contour}. Near the gray lines, the Fourier factor no longer strongly suppresses the angular-cut contributions, so these contributions can produce visible deviations.

Second, the PIA integral and the QNM sum converge slowly near the causal font. In calculating the direct part, the PIA integral involving \(\tilde{K}_{\HH, \ell m}^+\) is truncated at \(|M\omega|=40\), and the QNM sum is also truncated at a finite overtone number. Away from the dotted gray line, the Fourier factor suppresses the omitted high-frequency contributions. Near the dotted line, however, this suppression becomes much weaker and the truncated parts can contribute appreciably to the deviation. Figure~\ref{fig:PositiveSourceQNMConvergence} in Appendix~\ref{sec:qnm-subtleties} shows the contributions of different overtones at several times near the beginning of the QNM signal. Very close to \(U=|r_*'|\), the plotted contribution does not even decrease as the overtone number increases. After a very small shift in time, it begins to decrease, but only slowly. After a larger shift, it decays rapidly. This clearly illustrates the slow convergence of the QNM sum near its starting time.

Third, there may be an \(O(a)\) window of the type discussed in Appendix~\ref{app:kerr_split_time}. The exponential behavior of the angular functions modifies the large-frequency exponential behavior of the Green function in the complex-frequency plane. Consequently, the convergence boundaries of both the PIA integral and the QNM sum may differ from the value expected from \(U\sim |r_*'|\). The angular factor also modifies the large-arc contribution, which may therefore remain non-negligible within an \(O(a)\) interval around \(U\sim |r_*'|\). This provides a possible explanation for the widening of the mismatch region in Fig.~\ref{fig:kernel-direct} as \(a\) increases.

There are several other noteworthy features. In particular, the waveform does not vanish before the dashed line. One might suspect that this is the Gibbs effect from the inverse Fourier transform. However, the cutoff used in the real-frequency inversion is sufficiently large that the observed structure cannot be explained by the frequency truncation alone. We interpret it instead as a consequence of the angular cut in the upper half-plane of \(\tilde{K}_{\HH,\ell m}\), which leads to a violation of causality for an individual fixed \((\ell,m)\) mode. The structure before the dashed gray line becomes more pronounced as \(a\) increases. We believe this occurs because the magnitude of the angular exceptional-point frequency decreases as \(a\) increases. The corresponding contribution is then less strongly suppressed by the Fourier factor and leaves a more visible imprint on the waveform. This behavior supports the angular-cut interpretation.

\begin{figure*}[t]
 \centering
 \includegraphics[width=\textwidth]
 {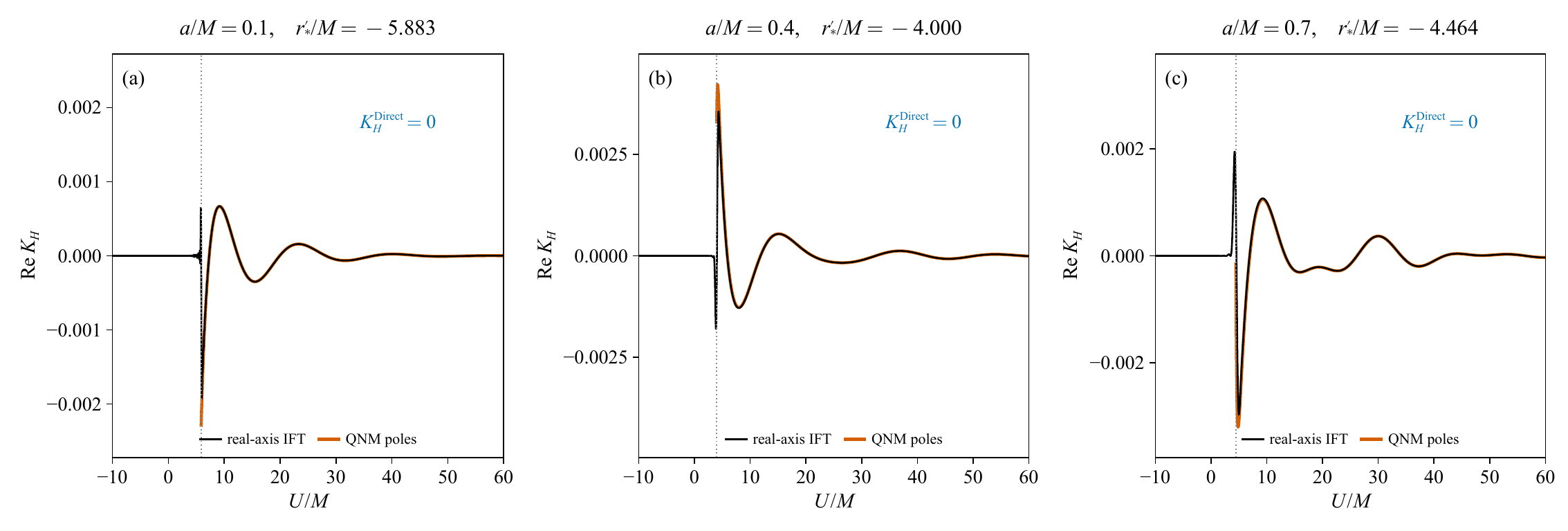}
 \caption{
 Real part of the $(\ell,m)=(2,2)$ radiative-metric kernel $\Kern_{\HH}$  for Kerr spins $a/M=0.1,\ 0.4,\ 0.7$ in panels (a)--(c), with negative source positions $r_*'/M=-5.883,\ -4.000,\ -4.464$, respectively, and observer angle $\theta_{\rm obs}=1.1$.  Black curves show the real-axis inverse Fourier transform,
 while orange curves show the combined prograde and retrograde QNM pole sum with overtone labels $n=0,\ldots,41$.
 The retrograde $n=14$ pole is omitted for $a/M=0.7$ because its continuation lies on a different Riemann sheet; see Appendix~\ref{sec:qnm-subtleties}. Both axes are linear.
 Dotted vertical lines mark $U=|r_*'|$, the onset of the
 plotted QNM contribution.
 Within the approximate causal support prescription
 discussed in Sec.~\ref{subsection:Contour_Strategy}, the direct-part contribution vanishes for these negative-source cases.
 }
 \label{fig:kernel-no-direct}
\end{figure*}

Figure~\ref{fig:kernel-no-direct} shows the case without a direct-part interval. The direct-part interval is absent because \(r_*'\lesssim0\), so the direct part interval $-r_*'\lesssim U \lesssim |r_*'|$
disappears. In this case, the finite QNM sum agrees well with the real-frequency inversion except within a small region close the dotted gray line. Nevertheless, the peak value of the QNM pole sum differs from that of the real-frequency inversion. The real-frequency result also contains structures before the dotted gray line, which may again be caused by the angular cut in the upper half-plane of \(\tilde{K}_{\HH,\ell m}\).

The deviations on the right-hand side may receive contributions from both the lower-half-plane angular cut of \(\tilde{K}_{\HH,\ell m}\) and the slow convergence of the QNM sum. Figure~\ref{fig:NegativeSourceQNMConvergence} in Appendix~\ref{sec:qnm-subtleties} shows that, although the low-overtone contributions decrease rapidly with overtone number, the higher-overtone contributions decay much more slowly near the starting time. This again reveals a convergence problem for the QNM sum close to its onset.

Even if the Green function could be computed numerically at arbitrarily large
frequencies, the individual contour contributions would still have restricted
domains of convergence in observer time.  The QNM sum and the integrals along
the PIA, NIA, and angular cuts converge only within bounds determined by their
large-frequency behavior.  These bounds reflect the analytic structure of the
Green function rather than limitations of the numerical calculation.
For Kerr, determining the exact convergence domain of the full Green function,
including its angular dependence, is particularly important for studying the
dynamical excitation of QNMs.  To our knowledge, a consistent treatment of
dynamical excitation, including the appropriate convergence bound, was first
developed for Schwarzschild spacetime in Ref.~\cite{Ma:2026qbq}.  Recent work
has made substantial progress in understanding the convergence of the QNM
representation~\cite{oshita2024reconstructionringdownexcitationfactors,Oshita:2026vxh}
and the tail contribution~\cite{gupta2026earlytimetailskerrblack} of the radial
Green function.  However, the corresponding convergence bounds for the full
radiative kernel have not yet been established.  Such bounds are needed for a
reliable calculation of both modal contributions, such as QNMs, and nonmodal
contributions, such as the direct part and the tail.  We leave their
determination to our future work.

These difficulties may be addressed systematically.  For the angular cut, the
contributions from different $\ell$ modes are expected to cancel when a
sufficiently complete angular sum is included.  The deviations near the dotted
lines should then decrease and the causal behavior of the full field should be
recovered, as discussed in Sec.~\ref{subsec:angular-cut-brief}.  For an
individual separated $(\ell,m)$ mode, however, the corresponding angular-cut
contribution must instead be included explicitly.

The remaining high-frequency contributions may be treated by deriving
asymptotic expansions for the discontinuities along the PIA, NIA, and angular
cuts, together with expansions for the large-arc contribution, the QNM
frequencies, and the QNM residues.  These results could be used to construct
semi-analytic expressions for the unresolved high-frequency parts of each
component and to determine their convergence domains.  We leave this analysis
to our future work.

\section{Waveform from the plunge worldline}
\label{sec:orbit-waveform}

Despite the subtleties near the causal fronts, we have decomposed the radiative kernel into a direct-part contribution and a QNM-sum contribution in the last section.
A natural question is how these components contribute to the gravitational
waveform generated by a plunging point particle. A direct evaluation based
on contour integration would be computationally expensive, and the behavior
near $U\sim |r_*'|$ is difficult to treat accurately. We therefore introduce
an approximate decomposition motivated by the close agreement between the
direct-part and QNM results and the real-frequency inverse Fourier transform in most of the time regions.

For each fixed source radius, we approximately divide the metric kernel
obtained from the real-frequency inverse Fourier transform at split time $U=|r_*'|$. The portion before this
time is identified with the direct part, while the portion after it is
identified with the QNM-plus-tail contribution. Strictly speaking, the direct part and QNM+tail kernel near the causal fonts also receives contributions from the angular cut and the large arc. In the present approximation, these contributions are not evaluated separately but are  absorbed into the two-part decomposition for simplicity.

Let the metric kernel obtained by the real-frequency inversion in
Eq.~\eqref{eq:fixed-source-time-kernel} be denoted by
$K^{\rm IFT}_{\ell m}(U,r')$. The time-separated fixed-source kernels are
then defined by
\begin{align}
 \Kern^{\rm direct}_{\ell m}(U;r')
 &\equiv
 I_D(U;r')K^{\rm IFT}_{\ell m}(U,r'),
 \\
 \Kern^{\rm QNM+tail}_{\ell m}(U;r')
 &\equiv
 I_L(U;r')K^{\rm IFT}_{\ell m}(U,r').
\end{align}

We next integrate the fixed-source radiative kernel
$\Kern_{\HH,\ell m}$ along the worldline of an equatorial plunge to obtain
the metric waveform $\HH_{\ell m}$. Following
Eq.~\eqref{eq:metric-kernel-read-off}, we have
\begin{align}
 \HH_{\ell m}(u)
 &=
 \int_{r_+}^{r_{\rm start}}\dd r'\,
 J_t(r')K^{\rm IFT}_{\ell m}
 \bigl(u-t_p(r');r'\bigr)
 \nonumber\\
 &=
 \int_{r_+}^{r_{\rm start}}\dd r'\,
 J_t(r')\Kern_{\HH,\ell m}
 \bigl(u-t_p(r');r'\bigr).
 \label{eq:orbit-integral}
\end{align}

The corresponding decomposition of the waveform is then defined by
\begin{align}
 \HH_{\ell m}^{\rm direct}(u)
 &=
 \int_{r_+}^{r_{\rm start}}\dd r'\,
 J_t(r')\Kern_{\ell m}^{\rm direct}
 \bigl(u-t_p(r');r'\bigr),
 \\
 \HH_{\ell m}^{\rm QNM+tail}(u)
 &=
 \int_{r_+}^{r_{\rm start}}\dd r'\,
 J_t(r')\Kern_{\ell m}^{\rm QNM+tail}
 \bigl(u-t_p(r');r'\bigr).
\end{align}
These quantities provide an approximate measure of how the direct and QNM-plus-tail components contribute to the total waveform.

The source time and source radius are related by the plunge trajectory. Consequently, the fixed-source causal fronts become moving boundaries in
the worldline integral. Along the inward branch,
$J_t\,\dd r'=-\dd t_p$, so the source integral can equivalently be evaluated using the monotonic plunge time $t_p$ as the integration variable.

\begin{figure*}[t]
 \centering
 \includegraphics[width=\textwidth]
 {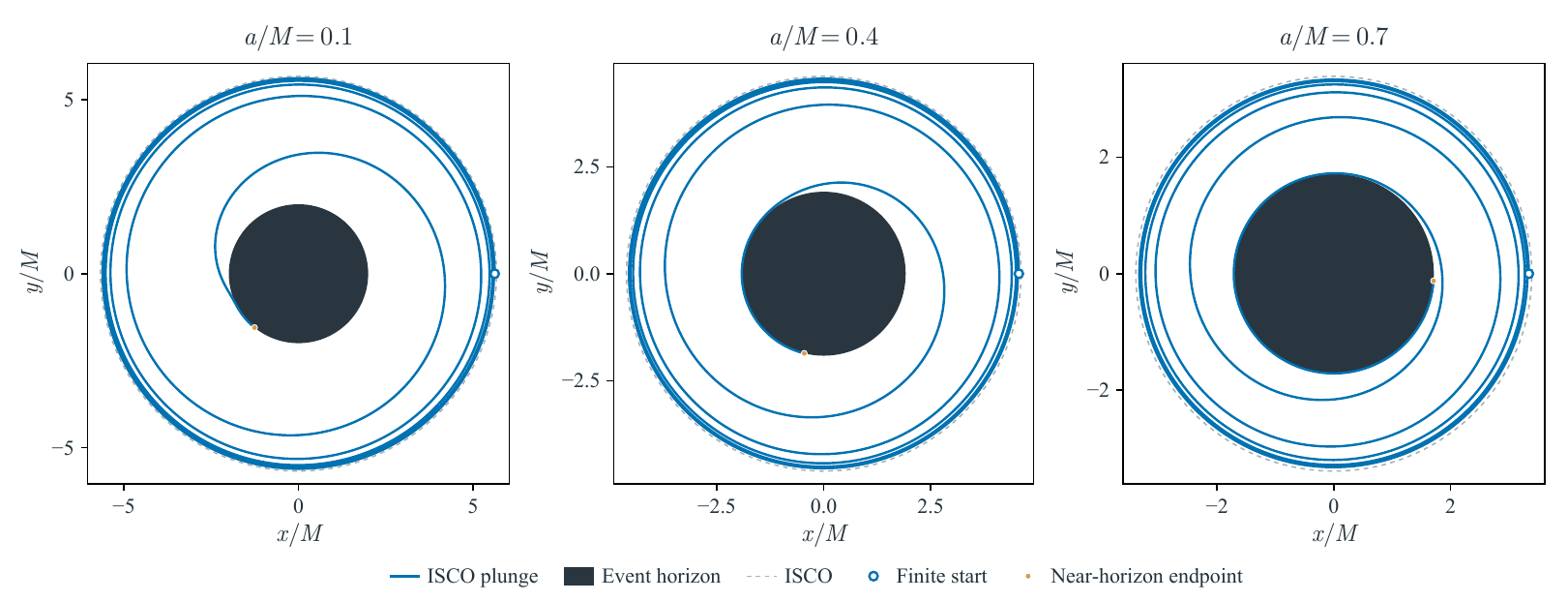}
 \caption{ 
 Equatorial ISCO-plunge trajectories in Kerr spacetime for
 $a/M=0.1$, $0.4$, and $0.7$, from left to right.
 The particles follow prograde geodesics with the ISCO energy
 and angular momentum, starting at
 $r_{\mathrm{start}}=r_{\mathrm{ISCO}}-0.05M$.
 The trajectories are displayed using the planar projection.
 Dark disks represent the event horizons, while dashed
 circles indicate the ISCO radii.
 Open circles mark the initial positions, and gold markers
 denote the endpoints at $r=r_++10^{-6}M$.
 Each panel preserves equal horizontal and vertical scales,
 but the coordinate ranges differ between panels.
 }
 \label{fig:trajectory}
\end{figure*}

The plunging particle has the energy and angular momentum of the Kerr innermost stable circular orbit (ISCO), as described in Sec.~\ref{subsec:isco-orbit}. Figure~\ref{fig:trajectory} shows the resulting
orbits for $a/M=0.1$, $0.4$, and $0.7$.  In each case, the particle follows
several nearly circular revolutions before the radial plunge accelerates.
The plotted plane,
$(x,y)=(r\cos\phi_p,r\sin\phi_p)$, is a projection in Boyer--Lindquist
coordinates rather than a representation of proper spatial distance.
The source radii examined in Figs.~\ref{fig:kernel-direct} and
\ref{fig:kernel-no-direct} are taken from these trajectories.

The next two paragraphs describe how we evaluate the source integral
numerically.  Readers interested only in the physical results may skip these
implementation details.

We sample the plunge at 1793, 1331, and 896 source times for
$a/M=0.1$, $0.4$, and $0.7$, respectively.  Consecutive samples are normally
separated by $\Delta t_p=0.25M$.  For each waveform component
$b\in\{\mathrm{direct\,part},\mathrm{QNM+Tail}\}$, it is convenient to define
\begin{equation}
 \widehat{\Kern}^b_{\ell m}(u,t_p)
 =\Kern^b_{\ell m}\bigl(u-t_p;r(t_p)\bigr).
\end{equation}
We then evaluate the source integral with the trapezoidal rule.  At observer
time $u_i$, this gives
\begin{equation}
 \HH^b_{\ell m}(u_i)\simeq
 \sum_{q\in\mathcal G_i}
 \frac{t_{q+1}-t_q}{2}
 \left[
 \widehat{\Kern}^b_{\ell m}(u_i,t_q)
 +\widehat{\Kern}^b_{\ell m}(u_i,t_{q+1})
 \right],
 \label{eq:discrete-orbit-quadrature}
\end{equation}
where $\mathcal G_i$ denotes the adjacent source-time intervals used at
that observer time.  Before forming these intervals, we add to the original
$t_p$ grid every source time at which a causal front reaches $u_i$, and then
sort the enlarged grid.  For a given source position, if the required value
of the kernel time $\tau=u_i-t_p$ lies between two stored values, we obtain
the kernel by linear interpolation in $\tau$.

\begin{figure*}[t]
 \centering
 \includegraphics[width=0.98\textwidth]
 {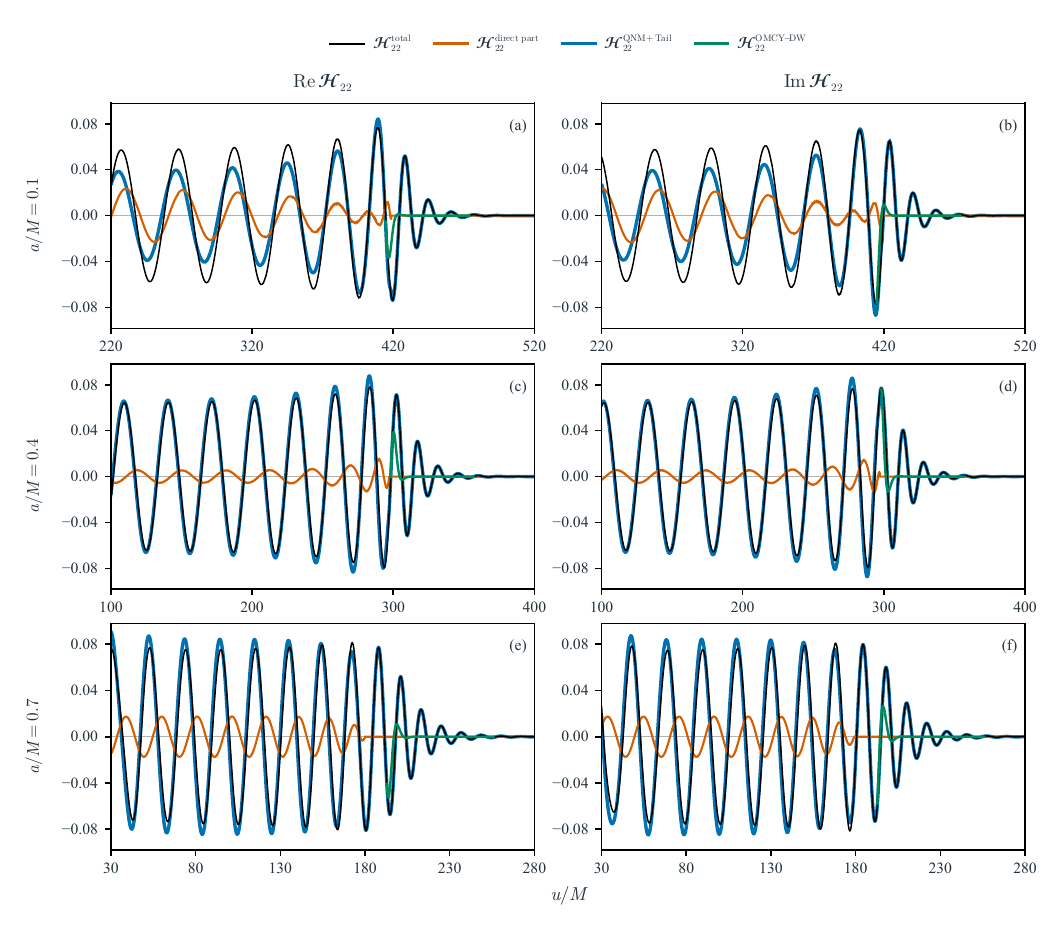}
 \caption{
 Trajectory-integrated $(s,\ell,m)=(-2,2,2)$ radiative
 metric mode $\HH_{22}$ for $a/M=0.1$, $0.4$, and $0.7$
 (top to bottom), for the trajectories in Fig.~\ref{fig:trajectory}.
 The left and right columns show the real and imaginary
 parts, respectively.
 Because the plunge durations differ, the rows use
 $220\leq u/M\leq520$, $100\leq u/M\leq400$, and
 $30\leq u/M\leq280$.
 All panels have linear axes and common vertical limits.
 The orange curve is the approximately time-classified
 direct part, the blue curve is the approximately time-classified QNM+Tail, and the black curve is their sum.
 Here QNM+Tail denotes the post-front sector
 selected from the full real-axis kernel before worldline
 integration; its tail contribution is included implicitly.
 The green curve, labeled $\HH_{22}^{\rm OMCY-DW}$,
 represents the direct-wave prescription of
 Ref.~\cite{Oshita:2025qmn};
 it is distinct from the direct-part contribution
 defined in this work.
 }
 \label{fig:orbit-waveform}
\end{figure*}

The observer times are separated by $\Delta u=0.05M$.  The two moving fronts,
 $u=t_p-r_*'$ and
 $u=t_p+|r_*'|$
divide the source integral into smooth pieces.  By adding their crossing
times to the $t_p$ grid, we ensure that each trapezoidal interval remains on
one side of a front; no interval crosses a discontinuity.  As a convergence test, refining the
source-time grid produces no visible change over the time ranges shown in
Fig.~\ref{fig:orbit-waveform}.

\begin{figure*}[t]
 \centering
 \includegraphics[width=\textwidth]
 {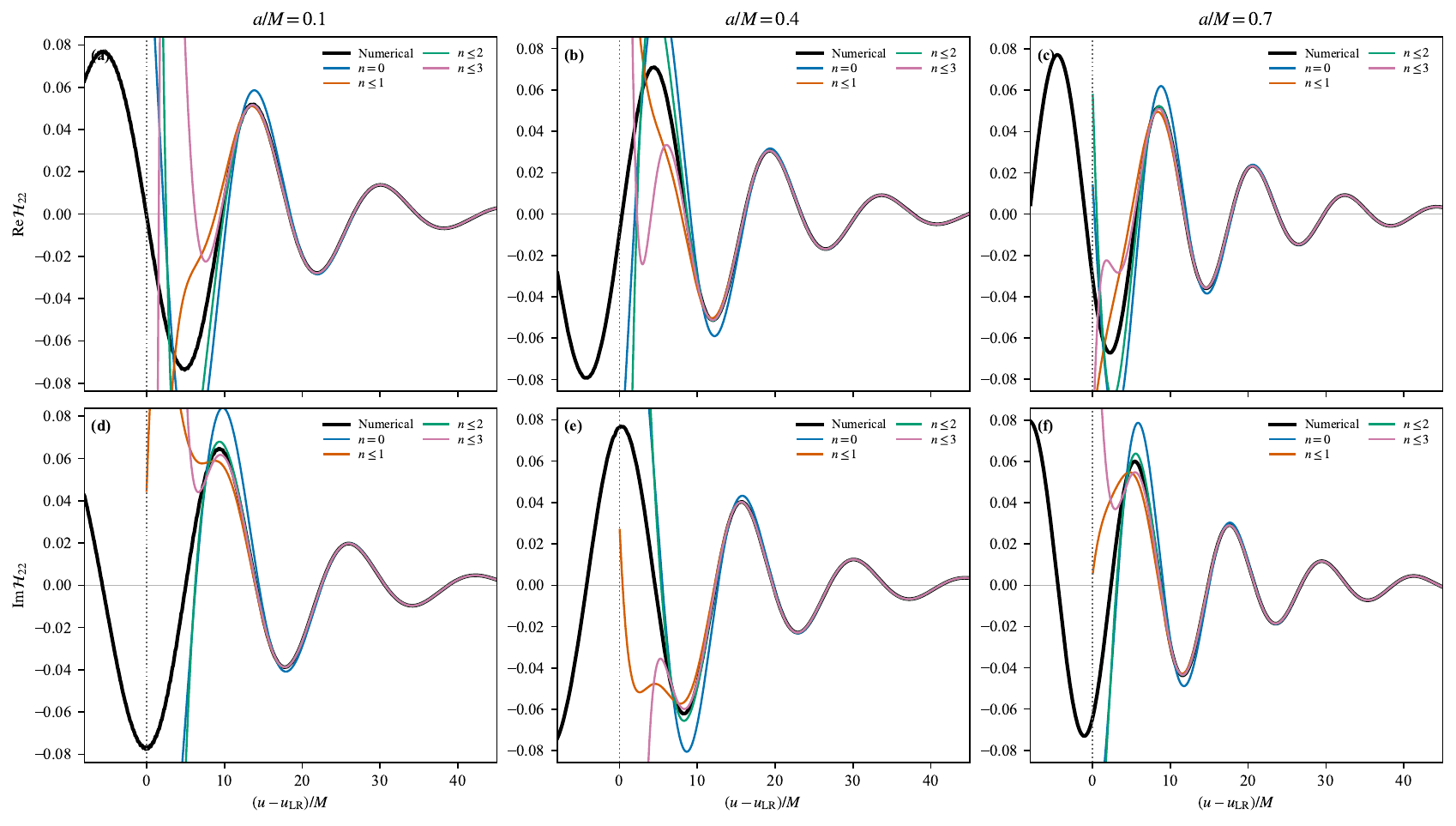}
 \caption{
 Late-time comparison between the trajectory-integrated
 metric waveform and cumulative QNM reconstructions for
 $a/M=0.1$, $0.4$, and $0.7$ (left to right), for the trajectories in Fig.~\ref{fig:trajectory}.
 The upper and lower rows show the real and imaginary
 parts.
 The reconstructions include the fundamental mode
 and successively the first three overtones,
 $n=0,\ldots,3$, with excitation coefficients computed
 following Ref.~\cite{DellaRocca:2025zbe}.
 The horizontal coordinate is
 $(u-u_{\rm LR})/M$, where $u_{\rm LR}$ is the retarded
 time of the particle's prograde light-ring crossing.
 The late-time agreement checks the real-frequency
 waveform against a separately computed QNM reconstruction.
 }
 \label{fig:lateqnm}
\end{figure*}

As a separate validation, Fig.~\ref{fig:lateqnm} compares the waveform with a QNM reconstruction whose excitation coefficients are calculated from the particle source following Ref.~\cite{DellaRocca:2025zbe}.
The cumulative sums through overtone number $n=3$ closely reproduce
the late-time amplitude, phase, and decay in all three cases. This supports the reliability of the trajectory-integrated waveform and its QNM-dominated ringdown.

Figure~\ref{fig:orbit-waveform} illustrates how the
source-driven waveform evolves from nearly periodic motion to ringdown.
During the near-ISCO cycles, both components exhibit driven oscillations,
as the approximately periodic motion of the particle continually modulates
the radiative response. In particular, a dynamically excited QNM
contribution can track the orbital motion while its excitation amplitudes
continue to evolve; consequently, its early-time waveform need not resemble
a freely decaying QNM sum with constant amplitudes
\cite{DeAmicis:2025xuh}. As the plunge accelerates, the waveform's envelope
and phase depart from this nearly periodic behavior. The orange direct-part
contribution then terminates, after which the total waveform coincides with
the blue post-front response. Over the displayed ringdown intervals, we
interpret the remaining signal as being dominated by QNM radiation, with a
small tail contribution neglected.

This termination occurs because, under the adopted time splitting, a source
point contributes to the direct part only when
\begin{equation}
 r_*'>0,\qquad
 t_p(r')-r_*'<u<t_p(r')+r_*'.
 \label{eq:orbit-direct-condition}
\end{equation}
For the plunging trajectories considered here, the upper endpoint
$t_p(r')+r_*'$ increases as the particle moves inward, until the particle
reaches $r_*'=0$, where the allowed interval shrinks to zero. The
termination time of the direct-part contribution in this approximated calculation is therefore estimated as
\begin{equation}
 u_{\rm D,end}\simeq t_p(r_0),
 \qquad r_*(r_0)=0.
 \label{eq:direct-end-time}
\end{equation}
The corresponding termination times are approximately $418.8M$, $297.5M$,
and $180.2M$ for $a/M=0.1$, $0.4$, and $0.7$, respectively. Under the
adopted time splitting approximation, the orange contribution is set to zero after this
crossing. These values therefore estimate the termination of the direct
part within the approximated calculation. Note that they should not be see as exact Kerr termination times for direct part contribution.

We emphasize, however, that this termination is not merely an artifact of
approximating the direct part through the time-split radiative kernel
obtained from the real-frequency calculation. As discussed in
Sec.~\ref{subsec:numerical-kernel-results}, for a fixed-source kernel there
are time boundaries beyond which the PIA integral defining the direct part
ceases to converge and the QNM sum begins to converge. These boundaries
mark the transition between the direct part and the QNM contribution.
Correspondingly, in the exact treatment, one also expects a definite time
at which the direct part disappears as the source passes through the region
$r_*'\sim0$. After integration along the plunging trajectory, this behavior
should produce a termination of the direct-part contribution similar to
that found in the present approximate treatment.

Finally, we note that the concept termed the ``direct wave'' in
Ref.~\cite{Oshita:2025qmn} has recently been widely discussed.  In
discussions of our previous work~\cite{Su:2026fvj} and the progress of the
present study, we have found that this concept is often confused with our
\emph{branch-cut direct part} by other people.  We therefore clarify the distinction here.

The green curve in Fig.~\ref{fig:orbit-waveform} shows the direct-wave
contribution defined in Ref.~\cite{Oshita:2025qmn}.  In that work, the direct
wave is associated with ``prompt'' radiation emitted as the source plunges
through the strong-field region.  From the viewpoint of Green-function
and radiative-kernel decomposition, this signal is mainly associated with
the initial portion of the QNM sum shown in
Fig.~\ref{fig:kernel-no-direct} for source positions near the horizon,
(therefore satisfying $r_*'\lesssim0$).  The initial portion of the \emph{branch-cut direct part}
shown in Fig.~\ref{fig:kernel-direct} should therefore not be included in
the ``direct wave.''  This contribution is driven by the particle's motion
near the horizon.  Analytically, the ``direct wave'' in
Ref.~\cite{Oshita:2025qmn} is obtained from a saddle point of the inverse Fourier integral, with each observer time mapped to an emission time along
the plunging trajectory.

By contrast, our \emph{branch-cut direct part} is defined by the branch-cut
contribution along the imaginary-frequency axis that appears when the decomposed
Green-function contour is deformed in the complex-frequency plane.  In our
decomposition, it appears only for $r_*'\gtrsim0$.  The similar names
therefore refer to different mathematical constructions and physical
pictures and should not be used interchangeably.

\section{Conclusion}\label{sec:con}

In this work, we extended the direct-part--QNM--tail decomposition developed in our previous study of the Schwarzschild Green's function~\cite{Su:2026fvj} to Kerr spacetime.
We used this decomposition to guide an approximate
metric-level separation of the plunge waveform into direct-part and QNM$+$tail contributions.

We formally analyzed the analytic structure of the Green's function and of the fixed-source radiative kernel, identifying the contributions associated with the direct part, the QNM residues, and the branch-cut tail. 
The Kerr Teukolsky problem is more subtle than its Schwarzschild Regge--Wheeler counterpart. We also numerically compared the radiative kernel at several
source radii and for several black-hole spins, as obtained from the computed contour contributions and from the real-frequency integral. The two calculations agreed closely away from the causal fronts. We then analyzed the origin of the remaining mismatch near the causal fronts.

For simplicity, the final plunge waveform
was calculated using the full real-axis inverse transform
of the kernel rather than a contour integral. The resulting
fixed-source response was then approximately partitioned
in time using causal indicators.

We computed the trajectory-integrated waveform and its
decomposition into the direct part and the post-front sector for several black-hole spins. The early part of the decomposed waveform exhibits slowly evolving oscillations associated with the near-ISCO cycles, followed by a more rapidly varying plunge signal and a damped ringdown. The direct part contributes to the total waveform over a finite time interval.

We also clarified that the ``direct wave''
proposed in~\cite{Oshita:2025qmn} 
is distinct from the direct-part contribution
defined in the present work. We believe
that this distinction will help avoid possible confusion.

To our knowledge, the present calculation provides the
first numerical evaluation of the branch-cut direct part
in Kerr spacetime. It also represents a first step toward a systematic
time-domain decomposition of Kerr plunge waveforms
based on their frequency-domain analytic structure. A
more complete treatment of the Kerr convergence bounds and front-localized terms,
isolated-$(\ell,m)$ angular cuts, explicit radial branch-cut tails,
additional modes, and other plunge trajectories is left for future work.

In particular, the differences arising from changes in
the ordering of source differentiation, time partitioning, and curvature-to-metric conversion have not yet been quantified separately.
Waveform decompositions at finite observer radius should
also be investigated in future studies of nonlinear
coupling among these frequency-domain structures and their role in a complete nonlinear ringdown model.

Decomposing the gravitational-wave radiation from a particle plunging into a Kerr black hole provides a first-principles framework for analyzing ringdown in the small-mass-ratio limit.
Compared with the ringdown of a comparable-mass-ratio binary black hole merger, this point-particle plunge solution may be viewed as a description at leading order in the (symmetric) mass ratio in the small-mass-ratio limit.
Nonlinear interactions between various kinds of modal and non-modal signals need to be theoretically understood before constructing a full nonlinear ringdown model.

\begin{acknowledgments}

We thank Yanbei Chen and Sizheng Ma for interesting discussions.
This work makes use of the Black Hole Perturbation Toolkit.
HY is supported by the Natural Science Foundation of China (Grant 12573048). NK is supported by the Shuimu fellowship of Tsinghua University. 
  AC is funded by Beijing Natural Science Foundation (BNSF) International Scientists Project (Grant No. IS25033) and also from a postdoctoral fellowship (Grant
No. 202504) through the Department of Astronomy, Tsinghua University.
MC is grateful to Tsinghua University for visit support and hospitality.
 ChatGPT (OpenAI) and Kimi Code (Moonshot AI) have been used to assist with code development, debugging, numerical verification, figure preparation, and manuscript editing. All AI-assisted content was carefully reviewed and verified by the authors.
The Black Hole Perturbation Toolkit (BHPT) was used as a benchmark during the development of our own code.

%\MC{Did we use the BHPT to get results? If so, we should write sth like "This work makes use of the Black Hole Perturbation Toolkit"}

\end{acknowledgments}

\appendix

\section{MST solution of the radial Teukolsky equation}
\label{app:mst}

\subsection{Theoretical framework}
\label{app:mst-framework}

To solve the radial Teukolsky equation for a field of spin weight $s$ in
Kerr spacetime, we employ the Mano--Suzuki--Takasugi (MST) formalism
\cite{Mano_1996a,Mano_1996b}.  Here we collect the
variables and formulas needed for the radial Green's function and its
asymptotic amplitudes.

\paragraph{Auxiliary variables.}
The MST series depend on the following dimensionless combinations:
\begin{align}
 q&=\frac{a}{M},&
 \epsilon&=2M\omega,
 \\
 \kappa&=\sqrt{1-q^2},&
 \tau&=\frac{\epsilon-mq}{\kappa},
 \label{eq:mst-auxiliary-parameters}\\
 x&=\frac{r_+-r}{r_+-r_-}
   =\frac{r_+-r}{2M\kappa},&
 \hat z&=\epsilon\kappa(1-x),
 \label{eq:mst-radial-variables}\\
 \epsilon_+&=\frac{\epsilon+\tau}{2},&
 \epsilon_-&=\frac{\epsilon-\tau}{2}.
 \label{eq:mst-epsilon-pm}
\end{align}
Here $r_\pm=M(1\pm\kappa)$.  On the exterior domain
$r\in[r_+,\infty)$, the variable $x$ runs from $0$ to $-\infty$;
$\hat z$ is the frequency-scaled distance from the inner horizon.
The combinations $\epsilon_\pm$ satisfy
$\epsilon_++\epsilon_-=\epsilon$ and
$\epsilon_+-\epsilon_-=\tau$.

\paragraph{Upgoing solution.}
The UP solution is represented by a bilateral series of irregular confluent
hypergeometric functions $U(a,b,z)$:
\begin{align}
R^{\rm up}_{\ell m\omega}
={}&2^\nu e^{-\pi\epsilon-\ii\pi(\nu+1+s)}e^{\ii\hat z}
\hat z^{\nu+\ii\epsilon_+}
(\hat z-\epsilon\kappa)^{-s-\ii\epsilon_+}
\nonumber\\
&\times\sum_{n=-\infty}^{\infty}\ii^n
{a_n^\nu}
\frac{(\nu+1+s-\ii\epsilon)_n}
     {(\nu+1-s+\ii\epsilon)_n}(2\hat z)^n
\nonumber\\
&\times \mathrm{U}(n+\nu+1+s-\ii\epsilon,2n+2\nu+2,-2\ii\hat z),
\label{eq:mst-rup}
\end{align}
where $(z)_n=\Gamma(z+n)/\Gamma(z)$ is the Pochhammer symbol.
We write the MST coefficients as $a_n^\nu$ throughout this
appendix to distinguish them from the Jaff\'e coefficients introduced in
Appendix~\ref{app:jaffe}.

\paragraph{Ingoing solution.}
The IN solution is expressed as a bilateral series of Gauss hypergeometric
functions:
\begin{align}
R^{\rm in}_{\ell m\omega}&=e^{\ii\epsilon\kappa x}
(-x)^{-s-\ii\epsilon_+}(1-x)^{\ii\epsilon_-}
\nonumber\\
&\times\sum_{n=-\infty}^{\infty}{a_n^\nu}
{}_2F_1\!\left(\begin{matrix}
n+\nu+1-\ii\tau,\,-n-\nu-\ii\tau\\
1-s-\ii\epsilon-\ii\tau
\end{matrix};x\right).
\label{eq:mst-rin}
\end{align}

\paragraph{Asymptotic amplitudes.}
The amplitudes required by the radial Green's function can be written in
terms of
\begin{align}
A_+^\nu={}&2^{-1+s-\ii\epsilon}e^{-\pi\epsilon/2}
e^{\ii\pi(\nu+1-s)/2}
\nonumber\\
&\times
\frac{\Gamma(\nu+1-s+\ii\epsilon)}
     {\Gamma(\nu+1+s-\ii\epsilon)}
\sum_{n=-\infty}^{\infty}{a_n^\nu},
\label{eq:mst-aplus}\\
A_-^\nu={}&2^{-1-s+\ii\epsilon}e^{-\pi\epsilon/2}
e^{-\ii\pi(\nu+1+s)/2}
\nonumber\\
&\times
\sum_{n=-\infty}^{\infty}(-1)^n
\frac{(\nu+1+s-\ii\epsilon)_n}
     {(\nu+1-s+\ii\epsilon)_n}
{a_n^\nu}.
\label{eq:mst-aminus}
\end{align}
For the asymptotic conventions adopted in the preceding Teukolsky section,
the IN transmission and incidence amplitudes are
\begin{align}
A^{\rm in}_{\rm trans}={}&(2\kappa)^{2s}
\exp\!\left[\ii(\epsilon+\tau)\kappa
\left(\frac12+\frac{\ln\kappa}{1+\kappa}\right)\right]
\sum_{n=-\infty}^{\infty}{a_n^\nu},
\label{eq:mst-ain-trans}\\
A^{\rm in}_{\rm inc}={}&\omega^{-1}
\left[K_\nu-\ii e^{-\ii\pi\nu}
\frac{\sin\pi(\nu-s+\ii\epsilon)}
     {\sin\pi(\nu+s-\ii\epsilon)}K_{-\nu-1}\right]
A_+^\nu
\nonumber\\
&\times e^{-\ii\epsilon[\ln\epsilon-(1-\kappa)/2]}.
\label{eq:mst-ain-inc}
\end{align}

The reflected IN amplitude and the transmitted UP amplitude are
\begin{align}
A^{\rm in}_{\rm ref}={}&\omega^{-1-2s}
\left[K_\nu+\ii e^{\ii\pi\nu}K_{-\nu-1}\right]A_-^\nu
e^{\ii\epsilon[\ln\epsilon-(1-\kappa)/2]},
\label{eq:mst-ain-ref}\\
A^{\rm up}_{\rm trans}={}&\omega^{-1-2s}
e^{\ii\epsilon[\ln\epsilon-(1-\kappa)/2]}A_-^\nu.
\label{eq:mst-aup-trans}
\end{align}
Thus $A^{\rm in}_{\rm inc}$ multiplies the wave incident from infinity,
whereas $A^{\rm in}_{\rm ref}$ multiplies its reflected outgoing component.
The powers of $\omega$ and the use of $A_-^\nu$ in
Eq.~\eqref{eq:mst-ain-ref} follow from the $s$-dependent Teukolsky
normalization at infinity.

The connection coefficient $K_\nu$ relates the near-horizon
hypergeometric series to the asymptotic confluent-hypergeometric series.  A
form useful for numerical evaluation is
\begin{widetext}
\begin{align}
K_\nu={}&
\frac{e^{\ii\epsilon\kappa}(2\epsilon\kappa)^{s-\nu-\xi}
2^{-s}\ii^\xi\Gamma(1-s-2\ii\epsilon_+)\Gamma(\xi+2\nu+2)}
{\Gamma(\xi+\nu+1-s+\ii\epsilon)
 \Gamma(\xi+\nu+1+s+\ii\epsilon)}
\left[
\sum_{n=\xi}^{\infty}
\frac{\Gamma(n+\xi+2\nu+1)}{(-1)^n (n-\xi)!}
\frac{\Gamma(n+\nu+1+s+\ii\epsilon)}
     {\Gamma(n+\nu+1-s-\ii\epsilon)}
\frac{\Gamma(n+\nu+1+\ii\tau)}
     {\Gamma(n+\nu+1-\ii\tau)}
{a_n^\nu}\right]
\nonumber\\
&\times
\left[
\sum_{n=-\infty}^{\xi}
\frac{(-1)^n}{(\xi-n)!(\xi+2\nu+2)_n}
\frac{(\nu+1+s-\ii\epsilon)_n}
     {(\nu+1-s+\ii\epsilon)_n}
{a_n^\nu}\right]^{-1}.
\label{eq:mst-knu}
\end{align}
\end{widetext}
Here $\xi$ is an arbitrary integer chosen for numerical
convenience; after convergence of both sums, $K_\nu$ is independent of $\xi$.

\subsection{Numerical implementation}
\label{app:mst-numerics}

The coefficient sequence and the renormalized angular momentum are obtained
from the MST three-term recurrence.  The recurrence coefficients are
\begingroup\small
\begin{align}
\alpha_n^\nu={}&
\frac{\ii\epsilon\kappa}{2n+2\nu+3}
\frac{n+\nu+1+\ii\tau}{n+\nu+1}
\bigl[(n+\nu+1+s)^2+\epsilon^2\bigr]
\label{eq:mst-alpha}\\
\begin{split}
\beta_n^\nu={}&{-\lambda_{\ell m\omega}}
-s(s+1)+(n+\nu)(n+\nu+1)+\epsilon^2
+\epsilon(\epsilon-mq)
\\
&+\frac{\epsilon(\epsilon-mq)(s^2+\epsilon^2)}
{(n+\nu)(n+\nu+1)},
\end{split}
\label{eq:mst-beta}\\
\gamma_n^\nu={}&-
\frac{\ii\epsilon\kappa}{2n+2\nu-1}
\frac{n+\nu-\ii\tau}{n+\nu}
\bigl[(n+\nu-s)^2+\epsilon^2\bigr].
\label{eq:mst-gamma}
\end{align}
\endgroup
The symbol $q=a/M$ is the dimensionless Kerr spin defined in
Eq.~\eqref{eq:mst-auxiliary-parameters}, and
$\lambda_{\ell m\omega}$ is the separation constant used in the main
text.
The sequence obeys
\begin{equation}
 \alpha_n^\nu a_{n+1}^\nu+\beta_n^\nu a_n^\nu
 +\gamma_n^\nu a_{n-1}^\nu=0.
 \label{eq:mst-three-term}
\end{equation}
The recurrence is invariant under
\begin{equation}
\alpha_{-n}^{-\nu-1}=\gamma_n^\nu,\qquad
\beta_{-n}^{-\nu-1}=\beta_n^\nu,\qquad
\gamma_{-n}^{-\nu-1}=\alpha_n^\nu.
\end{equation}
With $a_0^{-\nu-1}=a_0^\nu=1$, this gives
$a_{-n}^{-\nu-1}=a_n^\nu$ and avoids a second independent coefficient
construction when both $K_\nu$ and $K_{-\nu-1}$ are needed.

Define
\begin{equation}
 R_n=\frac{a_n^\nu}{a_{n-1}^\nu},\qquad
 L_n=\frac{a_n^\nu}{a_{n+1}^\nu}.
\end{equation}
The minimal solutions in the two directions are evaluated from
\begin{align}
R_n&=\cfrac{-\gamma_n^\nu}{\beta_n^\nu-
\cfrac{\alpha_n^\nu\gamma_{n+1}^\nu}{\beta_{n+1}^\nu-
\cfrac{\alpha_{n+1}^\nu\gamma_{n+2}^\nu}{\beta_{n+2}^\nu-\cdots}}},
\label{eq:mst-ratio-r}\\
L_n&=\cfrac{-\alpha_n^\nu}{\beta_n^\nu-
\cfrac{\alpha_{n-1}^\nu\gamma_n^\nu}{\beta_{n-1}^\nu-
\cfrac{\alpha_{n-2}^\nu\gamma_{n-1}^\nu}{\beta_{n-2}^\nu-\cdots}}}.
\label{eq:mst-ratio-l}
\end{align}
With $a_0^\nu=1$, the coefficients follow from
\begin{align}
a_n^\nu=\prod_{j=1}^{n}R_j,n>0; \qquad
{a_n^\nu}=
{\prod_{j=n}^{-1}L_j},n<0.
\label{eq:mst-coefficient-products}
\end{align}
The last product runs upward from $j=n$ to $j=-1$; this
ordering is required by $L_j=a_j^\nu/a_{j+1}^\nu$.

\subsubsection{Renormalized Angular Momentum}

In the MST formalism, the renormalized angular momentum $\nu$ arises as a deformation of the angular index $\ell$, and is determined numerically through a continuation strategy. It is defined implicitly by the requirement that the series coefficients $a_n^\nu$ satisfy the three-term recurrence relation Eq.~(\ref{eq:mst-three-term})
admitting a minimal solution  in both $n \to +\infty$ and $n \to -\infty$. This condition leads to a characteristic equation for $\nu$ in the form of an infinite continued fraction:
\begin{equation}
\beta_n^\nu + \alpha_n^\nu R_{n+1} + \gamma_n^\nu L_{n-1} = 0,
\label{eq:nu_cf}
\end{equation}
which implicitly defines $\nu$ as a function of the complex frequency $\omega$. The recurrence coefficients $\alpha_n^\nu$, $\beta_n^\nu$, and $\gamma_n^\nu$ are given by Eqs.~\eqref{eq:mst-alpha}--\eqref{eq:mst-gamma}.

Numerically, the renormalized angular momenta in our implementation are obtained primarily via the Monodromy method~\cite{Nasipak_2025}, which proves robust across a wide range of input parameters. In addition, we retain the continuation strategy developed in our previous work~\cite{Su:2026fvj}, whereby Eq.~\eqref{eq:nu_cf} is solved along a continuous trajectory in the complex $\omega$ plane. At each frequency step, the value of $\nu$ from the preceding step serves as the initial guess, allowing us to track the same branch of $\nu$ consistently. These two methods typically yield different roots of Eq.~\eqref{eq:nu_cf}. Nevertheless, even when the roots differ, the corresponding asymptotic amplitudes and homogeneous solutions---with the transmitted amplitude normalized to unity---are identically the same. However, we have found empirically that this continuation strategy becomes numerically unstable for frequencies with large absolute values. This explains why the Monodromy method is adopted as the primary approach in our numerical implementation.

\section{Jaff\'e series for the IN solution}
\label{app:jaffe}

Near the event horizon, direct evaluation of the MST UP series can converge
slowly at the complex QNM frequencies.  The Jaff\'e series
\cite{Leaver:1986a} provides a rapidly convergent representation of the IN
solution in this region.  At a QNM frequency ($\omega=\omega_n$), it is $A^{\rm in}_{\rm inc}(\omega_n)=0$, so both physical solutions satisfy the
same outgoing condition at infinity.  With the raw UP
normalization used in the preceding Teukolsky section, comparison of their
outgoing amplitudes gives
\begin{equation}
 R^{\rm in}_{\ell m\omega_n}(r)=
 \frac{A^{\rm in}_{\rm ref}(\omega_n)}
      {A^{\rm up}_{\rm trans}(\omega_n)}
 R^{\rm up}_{\ell m\omega_n}(r).
 \label{eq:qnm-in-up-relation}
\end{equation}
Equivalently, in the unit-transmission normalization used throughout this paper, we have
$R^{\rm up}_{\rm unit}=R^{\rm up}_{\rm raw}/A^{\rm up}_{\rm trans}$ used in the QNM Green's
function.
Thus the radius-independent MST amplitudes supply the conversion factor,
while the finite-radius IN solution is evaluated with the Jaff\'e series.

Introduce
\begin{align}
R_{\ell m\omega}(r)={}&
(\bar r-\bar r_-)^{-s+\frac{\ii}{\bar b}
 (\bar\omega\bar r_- -\bar a m)}
(\bar r-\bar r_+)^{-s-\frac{\ii}{\bar b}
 (\bar\omega\bar r_+ -\bar a m)} \nonumber 
\\
&\times y(\bar r-\bar r_-),
\label{eq:jaffe-radial-redefinition}
\end{align}
where
\begin{align}
 x_0&=\bar b,
 \label{eq:jaffe-x0}\\
 B_1&=(s-1+2\ii\bar\omega)\bar b
 -2\ii(\bar\omega\bar r_+-\bar a m),\\
 B_2&=2(1-s-\ii\bar\omega),\\
 B_3&=(1+\bar b-\bar a^2)\bar\omega^2
 +(2s-1)\ii\bar\omega+\ii s\bar\omega\bar b
 -2s-{}_sA_{\ell m}(a\omega),\\
 \eta&=-(\bar\omega+\ii s).\label{eq:eta}
\end{align}
All quantities in these expressions are dimensionless:
\begin{align}
 \bar r&=\frac{r}{2M},&
 \bar r_\pm&=\frac{r_\pm}{2M},&
 \bar a&=\frac{a}{2M},
 \nonumber\\
 \bar\omega&=2M\omega,&
 \bar b&=\bar r_+-\bar r_-.
 \label{eq:jaffe-dimensionless-definitions}
\end{align}
The Jaff\'e expansion is
\begin{equation}
 y(x)=e^{\ii\bar\omega x}x^{-B_2/2-\ii\eta}
 \sum_{n=0}^{\infty}a_n\left(1-\frac{x_0}{x}\right)^n,
 \label{eq:jaffe-series}
\end{equation}
where $x=\bar r-\bar r_-$, so that $x=x_0$ at the event
horizon and $0\leq1-x_0/x<1$ throughout the exterior.  This $x$ is the
argument of $y$ in Eq.~\eqref{eq:jaffe-radial-redefinition}; using
$\bar r-\bar r_+$ here would make the expansion parameter singular at the
horizon.
The coefficients $a_n$ in Eq.~\eqref{eq:jaffe-series} are distinct from
the MST coefficients $a_n^\nu$ and obey
\begin{align}
 \alpha_0a_1+\beta_0a_0&=0,\\
 \alpha_na_{n+1}+\beta_na_n+\gamma_na_{n-1}&=0,
 \qquad n=1,2,\ldots,
\end{align}
with
\begin{align}
 \alpha_n={}&(n+1)\left(n+B_2+\frac{B_1}{x_0}\right),\\
 \beta_n={}&-2n^2-2\left[B_2+\frac{B_1}{x_0}
 +\ii(\eta-\bar\omega x_0)\right]n
 \nonumber\\
 &-\left(\frac{B_2}{2}+\ii\eta\right)
 \left(B_2+\frac{B_1}{x_0}\right)
 +\ii\bar\omega(B_1+B_2x_0)+B_3,\\
 \gamma_n={}&\left(n-1+\frac{B_2}{2}+\ii\eta\right)
 \left(n+\frac{B_2}{2}+\ii\eta+\frac{B_1}{x_0}\right).
\end{align}

Finally, the overall scale is chosen so that the transmitted amplitude of
the IN solution is unity.  In the conventions above this requires
\begin{align}
a_0^{-1}={}&
\left(\frac{\bar r_-^{\bar r_-}}{\bar r_+^{\bar r_+}}\right)^{
\frac{\ii}{\bar b}(\bar\omega-m\bar a/\bar r_+)}
(\bar r_+-\bar r_-)^{
-\frac{\ii m\bar a\bar r_-}{\bar b}
(1/\bar r_- -1/\bar r_+)+2\ii\bar\omega-1}
\nonumber\\
&\times\exp\Biggl\{-\ii[m\bar a+(\bar r_--2\bar r_+)\bar\omega]
\nonumber\\
&\qquad+\ii\left(\bar\omega-\frac{m\bar a}{\bar r_+}\right)
\left[\frac{\bar r_+}{\bar r_+-\bar r_-}\ln\bar r_+
-\frac{\bar r_-}{\bar r_+-\bar r_-}\ln\bar r_-\right]\Biggr\}.
\label{eq:jaffe-normalization}
\end{align}

\section{Angular eigenfunctions, analytic continuation, and angular cuts}
\label{app:angular}

\subsection{Spectral angular solution}
\label{app:angular-spectral}

The angular Teukolsky equation, the spin-weighted spheroidal harmonics
${}_sS_{\ell m\omega}$, and the separation constant
$\lambda_{\ell m\omega}$ have already been defined in the preceding
Teukolsky section.  
Set $c=a\omega$.  We expand the angular eigenfunction in the
spin-weighted spherical basis,
\begin{equation}
 {}_sS_{\ell m\omega}(\theta)
 =
 \sum_{L=L_{\min}}^{L_{\max}}
 v_{L;\ell m\omega}\,{}_sY_{Lm}(\theta),
 \qquad
 L_{\min}=\max(|m|,|s|).
\end{equation}
This basis already has the correct regular behavior at the polar endpoints
$\theta=0,\pi$.  Note that the spheroidal mode label $\ell$ and the spherical-basis index $L$ are distinct, as in the main text.

Projecting the angular equation onto the same finite spherical basis gives a
standard spectral Galerkin discretization~\cite{Canuto2006Spectral}.  The
result is the frequency-dependent matrix eigenvalue problem
\begin{equation}
 \mathsf M(\omega)v_{\ell m\omega}
 =\eta_{\ell m\omega}v_{\ell m\omega},
 \qquad
 \mathsf M(\omega)=M_1-\ii\omega M_2-\omega^2M_3 .
 \label{eq:angular-pencil}
\end{equation}
whose three matrices are specified below.  \blue{Here
$v_{\ell m\omega}$ is the coefficient vector of the spheroidal eigenfunction
in the spherical basis $\{|L\rangle\}$; its $L$th component is denoted by
$v_{L;\ell m\omega}$.}  Define the coefficients
\begin{equation}
 B_L=-\frac{ms}{L(L+1)},\qquad
 C_L=\frac{\sqrt{(L^2-m^2)(L^2-s^2)}}
 {L\sqrt{(2L-1)(2L+1)}}
 \label{eq:angular-BC}
\end{equation}
The action of multiplying a basis function by $\cos\theta$ is represented by
the real symmetric matrix $X$ defined by
\begin{equation}
 \cos\theta\,|L'\rangle
 =
 \sum_L X_{LL'}|L\rangle ,
\end{equation}
with the matrix $X_{LL'}$ being
\begin{equation}
 X_{LL'}=C_{L'}\delta_{L,L'-1}+B_{L'}\delta_{L,L'}
          +C_{L'+1}\delta_{L,L'+1}.
 \label{eq:angular-X}
\end{equation}
After some derivation, we find that the matrix elements of $M_1$, $M_2$, and $M_3$ are
\begin{align}
 (M_1)_{LL'}&=\frac{L(L+1)-s^2}{2}\,\delta_{LL'},
 \label{eq:angular-M1}\\
 (M_2)_{LL'}&=\ii a s\,X_{LL'},
 \label{eq:angular-M2}\\
 (M_3)_{LL'}&=\frac{a^2}{2}X^{(2)}_{LL'},
 \qquad
 X^{(2)}_{LL'}=\langle L|\cos^2\theta|L'\rangle.
 \label{eq:angular-M3}
\end{align}
\green{Explicitly}, the pentadiagonal matrix in the last line is
\begin{align}
 X^{(2)}_{LL'}={}&C_{L'}C_{L'-1}\delta_{L,L'-2}
 \nonumber\\
 &+C_{L'}(B_{L'-1}+B_{L'})\delta_{L,L'-1}
 \nonumber\\
 &+(C_{L'}^2+B_{L'}^2+C_{L'+1}^2)\delta_{L,L'}
 \nonumber\\
 &+C_{L'+1}(B_{L'}+B_{L'+1})\delta_{L,L'+1}
 \nonumber\\
 &+C_{L'+1}C_{L'+2}\delta_{L,L'+2}.
 \label{eq:angular-X2}
\end{align}
Coefficients with indices below $L_{\min}$ are absent.  The analytic matrix
elements in Eqs.~\eqref{eq:angular-M1}--\eqref{eq:angular-X2} are formed
before restricting the row and column indices to $L\leq L_{\max}$. In particular, the last diagonal element of $X^{(2)}$ contains the
contribution $C_{L_{\max}+1}^2$, which represents coupling through the first
mode outside the retained basis.  This contribution would be lost if one
first truncated the $\cos\theta$ matrix $X$ to $L\leq L_{\max}$ and then
formed the finite-dimensional product $X^2$.

The auxiliary matrix eigenvalue appearing in
Eq.~\eqref{eq:angular-pencil} is denoted by \(\eta_{\ell m\omega}\):
\begin{equation}
 \eta_{\ell m\omega}
 =\frac{\lambda_{\ell m\omega}+s-a^2\omega^2+2am\omega}{2}
 =\frac{{}_sA_{\ell m}(a\omega)+s}{2}.
 \label{eq:angular-mu-convention}
\end{equation}
At $c=0$, the spin-weighted spherical harmonics give separated seeds labelled
by $\ell$.  The matrix is diagonalized at successively larger basis cutoff
until both $\lambda_{\ell m\omega}$ and the invariant angular contractions
are stable.

\begin{bluerevision}
The spheroidal harmonic and its derivative are reconstructed spectrally from
this coefficient vector:
\begin{align}
 {}_sS_{\ell m\omega}(\theta)
 &=\sum_Lv_{L;\ell m\omega}\,{}_sY_{Lm}(\theta),
 \label{eq:Spheroidal_Spherical_Expansion}
 \\
 \partial_\theta{}_sS_{\ell m\omega}(\theta)
 &=\sum_Lv_{L;\ell m\omega}\,
 \partial_\theta{}_sY_{Lm}(\theta).
 \label{eq:harmonic-reconstruction}
\end{align}
\end{bluerevision}
For the equatorial source we need only
$S_0={}_{-2}S_{\ell m\omega}(\pi/2)$ and
$S_1=\partial_\theta{}_{-2}S_{\ell m\omega}|_{\pi/2}$, together with the
observer value.  This is sufficient because the angular source operator
reduces to a linear combination of $S_0$, $S_1$, and
$\lambda_{\ell m\omega}$.

At complex $c$, $\mathsf M$ is symmetric under transpose but not Hermitian.
The left eigenvector associated with a simple eigenvalue is therefore
$v_{\ell m}(c)^{\mathsf T}$, not $v_{\ell m}(c)^\dagger$.  Hence the analytic
continuation of the angular normalization uses the bilinear product
$v_{\ell m}(c)^{\mathsf T}v_{\ell m}(c)$, rather than the Hermitian product
$v_{\ell m}(c)^\dagger v_{\ell m}(c)$.  Equivalently, the continued angular
factor contains two copies of the same analytic eigenfunction, not a
pointwise complex conjugate.  A literal complex conjugate would depend on
$c^*$ and would not define an analytic contour integrand.

\subsection{Continuation algorithm in \texorpdfstring{$c=a\omega$}{c=a omega}}
\label{app:angular-continuation}

Correct continuation transports an eigenstate from the spherical seed rather
than sorting eigenvalues independently at every target point.  The procedure
is summarized in Algorithm~\ref{alg:angular-continuation}.

\blue{At step $k$, $v_i^{(k)}$ denotes the normalized coefficient vector of
the $i$th tracked eigenstate, whereas $w_j^{(k+1)}$ denotes the $j$th
candidate eigenvector returned by the unordered eigensolve at $c_{k+1}$.
The norm in the overlap is the Euclidean norm of a
complex coefficient vector,
$\|z\|_2=(z^\dagger z)^{1/2}$.}

\begin{prdalgorithm}{Overlap continuation of an angular eigenstate}
\label{alg:angular-continuation}
\begin{algorithmic}[1]
\Require Path $c_0,c_1,\ldots,c_N$ beginning at $c=0$ or at a
well-separated real-frequency anchor; tracked eigendata at $c_0$;
overlap and eigenvalue-gap tolerances.
\Ensure Analytically tracked $\lambda_{\ell m\omega}$, coefficient vector,
bilinear projector, path identifier, and diagnostics at $c_N$.
\For{$k=0,1,\ldots,N-1$}
  \State Solve the complete unordered angular spectrum at $c_{k+1}$.
  \State Form the direction-overlap matrix
  \Statex \hspace{1.4em}$\displaystyle
  O_{ij}=\frac{|(v_i^{(k)})^\dagger w_j^{(k+1)}|}
  {\|v_i^{(k)}\|_2\|w_j^{(k+1)}\|_2}$.
  \State Obtain a one-to-one mode assignment by maximizing the total
  overlap over all tracked states.
  \If{the minimum assigned overlap or a neighboring eigenvalue gap fails
  its tolerance}
    \State Bisect the step $[c_k,c_{k+1}]$ and repeat it.
  \Else
    \State Align each accepted vector's phase with its predecessor.
    \State Store the assigned eigenvalue, coefficient vector, bilinear projector,
    path identifier, and continuation diagnostics.
  \EndIf
\EndFor
\State \Return the tracked endpoint eigendata.
\end{algorithmic}
\end{prdalgorithm}

The Hermitian overlaps in Algorithm~\ref{alg:angular-continuation} are used as stable numerical direction matchers.  The returned angular state is used
consistently by the homogeneous radial solution, asymptotic amplitude, source term, and QNM
calculations at that frequency.

For a spectrally closed cluster $\mathcal{C}$, the Riesz projector is defined by:
\begin{equation}
\Pi_{\mathcal C}(c)=\frac{1}{2\pi\ii}
\oint_\Gamma (z-\mathsf M(c))^{-1}\dd z,
\label{eq}
\end{equation}
where $\Gamma$ encloses the eigenvalues in $\mathcal{C}$ and remains separated from the rest of the spectrum as $c$ varies. Under this condition, $\Pi_{\mathcal{C}}(c)$ is analytic even when the individual eigenprojectors are not. For the exchanging pair $\mathcal{C}=\{2,3\}$, one has $\Pi_{\mathcal{C}}=\Pi_2+\Pi_3$ away from the exceptional point. Analytic continuation around the exceptional point interchanges $\Pi_2$ and $\Pi_3$, so their square-root discontinuities cancel in the sum. Consequently, the closed-cluster contribution to the angularly reconstructed (3+1)-dimensional Green's function is single-valued across this internal angular cut \cite{PhysRevD.94.124053}. By contrast, the isolated $(\ell,m)=(2,2)$ contribution can retain a nonzero discontinuity. In the present work, we show only the $(2,2)$ contribution; the neighboring $\ell$ sectors are retained solely to close and diagnose the angular cluster. Since the waveform is evaluated along the real-frequency axis, its construction requires no deformation across the off-axis angular cuts. Any future complex-frequency contour reconstruction of the isolated $(2,2)$ contribution must either include the discontinuities across every angular cut crossed by the contour or be formulated in terms of the corresponding closed-cluster projector.

\section{Time-domain simulations}
\label{SimulatorAppendix}

This appendix describes the time-domain solver used as an independent check
of the time-domain Green's function construction.  We evolve the spin-$s$
Teukolsky equation in the horizon-penetrating, hyperboloidally compactified
(HPHC) coordinates, and drive the system with a
narrow Gaussian approximation to an impulsive source, and extract the
retarded field directly at future null infinity $\mathcal I^+$. The HPHC coordinates used below are related to the Boyer-Lindquist variables used in the main text by the coordinate transformations of ~\cite{Ripley:2022ypi}:

\begin{align}\label{eq:HPHC}
dT &= dt + \frac{2Mr}{\Delta} dr + \left(-1 - \frac{4M}{r}\right) dr, \\
R &= \frac{\mathcal{L}_{\rm H}^2}{r}, \\
%\theta &= \Theta, \\
d\varphi &= d\phi + \frac{a}{\Delta} dr,
\end{align}  

where $\mathcal{L}_{\rm H}$ is a constant introduced for dimensional consistency, $(T,R,\theta,\varphi)$ denote the HPHC coordinates, and $(t,r,\theta,\phi)$ denote the BL coordinates. Within this framework, the gravitational waveform can be extracted naturally at null infinity, enabling direct comparison with waveforms computed via contour integral methods.

Also, consistent with the notation introduced in the main text, we use $L$ for the spin-weighted spherical-harmonic index throughout this appendix.

\subsection{Continuum equation and evolved field}
\label{app:tdgf-continuum}

We use the $\mathcal{L}_{\rm H}=1$ specialization of that transformation, so that $R=1/r$.
The compact radial domain is
\begin{equation}
 R\in[0,R_+],\qquad R_+=\frac{1}{r_+},\qquad
 r_+=M+\sqrt{M^2-a^2}.
 \label{eq:tdgf-compact-domain}
\end{equation}
The endpoints $R=0$ and $R=R_+$ represent $\mathcal I^+$ and the future
event horizon, respectively; no finite-radius extrapolation is required.

Let $\psi$ be the Boyer--Lindquist Teukolsky master scalar of spin weight
$s$.  Following Ref.~\cite{Ripley:2022ypi}, the evolved field is the
radially rescaled variable $\Psi$ defined by
\begin{equation}
 \psi=r^{-1}\Delta^{-s}\Psi,
 \qquad\text{or equivalently}\qquad
 \Psi=r\Delta^s\psi.
 \label{eq:tdgf-field-rescaling}
\end{equation}
This rescaling regularizes the equation at the future event horizon and
removes the leading radiative falloff at future null infinity.  For
gravitational perturbations we set $s=-2$ and denote the evolved field by
$\Psi_4$.  If $\psi_4^{\rm K}$ is the Weyl scalar in the Kinnersley tetrad
and $\varrho_{\rm NP}=-(r-\ii a\cos\theta)^{-1}$, then
\begin{equation}
 \Psi_4=r\Delta^{-2}\varrho_{\rm NP}^{-4}\psi_4^{\rm K}.
 \label{eq:tdgf-psi4-relation}
\end{equation}
Thus $\Psi_4$ is a rescaled curvature variable rather than the strain.  Its
limit at $\mathcal I^+$ is proportional to $r\psi_4^{\rm K}$; recovering
$h_+-\ii h_\times$ requires the two retarded time integrations specified by
the curvature convention used in the main text.

In HPHC coordinates, the homogeneous equation implemented by the solver is
\begin{align}
0={}&
 \bigl[8M(2M-a^2R)(1+2MR)
       -a^2\sin^2\theta\bigr]\partial_T^2\Psi
 \nonumber\\
&-2\bigl[1-(8M^2-a^2)R^2
          +4a^2MR^3\bigr]\partial_T\partial_R\Psi
 \nonumber\\
&-(1-2MR+a^2R^2)R^2\partial_R^2\Psi
 +2aR^2\partial_R\partial_\varphi\Psi
 \nonumber\\
&+2a(1+4MR)\partial_T\partial_\varphi\Psi
 \nonumber\\
&+2\Bigl\{2M\bigl[-s+2M(2+s)R-3a^2R^2\bigr]
 -a^2R+\ii sa\cos\theta\Bigr\}\partial_T\Psi
 \nonumber\\
&+2R\bigl[-(1+s)+(s+3)MR-2a^2R^2\bigr]\partial_R\Psi
 \nonumber\\
&+2aR\partial_\varphi\Psi
 +2\bigl[(1+s)MR-a^2R^2\bigr]\Psi
 -{}_s\!\mathcal A\Psi,
 \label{eq:tdgf-hphc-pde}
\end{align}
where
\begin{equation}
{}_s\!\mathcal A\Psi=
\frac{1}{\sin\theta}\partial_\theta
 (\sin\theta\,\partial_\theta\Psi)
+\left[s-\frac{(s\cos\theta-\ii\partial_\varphi)^2}
{\sin^2\theta}\right]\Psi.
\label{eq:tdgf-angular-full}
\end{equation}

For a fixed azimuthal number $m$, write
$\Psi=e^{\ii m\varphi}\Psi_m$, so that
$\partial_\varphi\to\ii m$.  With a source $S_m$, the equation takes the
operator form
\begin{equation}
A\,\partial_T^2\Psi_m+\mathcal D_T\partial_T\Psi_m
+\mathcal D_0\Psi_m=S_m,
\label{eq:tdgf-operator-split}
\end{equation}
where
\begin{align}
A(R,\theta)&=8M(2M-a^2R)(1+2MR)-a^2\sin^2\theta,\\
\mathcal D_T&=B(R)\partial_R+D_m(R,\theta),\\
B(R)&=-2\left[1-(8M^2-a^2)R^2+4a^2MR^3\right],\\
D_m(R,\theta)&=2\ii am(1+4MR)
 +2\Bigl\{2M[-s+2M(2+s)R
 \nonumber\\
&\hspace{7.5em}{}-3a^2R^2]-a^2R+\ii sa\cos\theta\Bigr\},\\
\mathcal D_0&=C(R)\partial_R^2+E_m(R)\partial_R+F_m(R)
-{}_s\!\mathcal A_m,\\
C(R)&=-(1-2MR+a^2R^2)R^2,\\
E_m(R)&=2R[-(1+s)+(s+3)MR-2a^2R^2]+2\ii amR^2,\\
F_m(R)&=2[(1+s)MR-a^2R^2]+2\ii amR,
\label{eq:tdgf-split-coefficients}
\end{align}
and
\begin{equation}
{}_s\!\mathcal A_m f=
\frac{1}{\sin\theta}\partial_\theta(\sin\theta\,\partial_\theta f)
+\left[s-\frac{(s\cos\theta+m)^2}{\sin^2\theta}\right]f.
\label{eq:tdgf-spin-angular-operator}
\end{equation}

\subsection{Spectral discretization and evolution}
\label{app:tdgf-spectral}

The radial dependence is expanded in Chebyshev polynomials and the angular
dependence in spin-weighted spherical harmonics:
\begin{equation}
\Psi_m(T,R,\theta)=
\sum_{L=L_{\min}}^{L_{\max}}
\sum_{n=0}^{n_{\max}}c_{n L}(T)T_n[x(R)]
\,{}_sY_{L m}(\theta,0),
\label{eq:tdgf-spectral-expansion}
\end{equation}
where
\begin{equation}
 x(R)=\frac{2R-R_+}{R_+},\qquad
 L_{\min}=\max(|s|,|m|).
\end{equation}
The representative Kerr calculation uses $s=-2$, $m=2$,
$2\leq L\leq15$, and $0\leq n\leq1700$.  The first-order state
$\mathbf y=(\mathbf c,\mathbf v)^{\mathsf T}$ therefore has dimension
$2\times14\times1701=47628$.  Vanishing initial data are imposed, so the
evolved field is generated entirely by the source.

Angular projection uses Gauss--Legendre quadrature in
$\zeta=\cos\theta$:
\begin{align}
(S_{\rm grid})_{j L}&={}_sY_{L m}(\arccos\zeta_j,0),\\
(S_{\rm mode})_{L j}&=2\pi w_j
\overline{{}_sY_{L m}(\arccos\zeta_j,0)}.
\end{align}
In this normalization, $S_{\rm mode}S_{\rm grid}$ approaches the identity
as the quadrature order is increased.  Multiplication by $\cos\theta$
couples adjacent $L$ modes, whereas multiplication by $\cos^2\theta$
has bandwidth $|\Delta L|\leq2$.

Radial grid values and Chebyshev coefficients are related by a DCT-I.
Radial derivatives and multiplication operators are constructed using the
ultraspherical conversion chain
$C^{(0)}\to C^{(1)}\to C^{(2)}$, which gives sparse banded radial
operators. Let $\mathbf S_m(T)$ denote the spectral coefficient vector obtained by
projecting the continuum source $S_m(T,R,\theta)$ onto the retained
Chebyshev-spherical basis and $\mathbf{M}_0$ and $\mathbf{M}_1$ represent $\mathcal D_0$ and $\mathcal D_T$,
respectively, the semidiscrete equations are: 
\begin{align}
 \dot{\mathbf c}=\mathbf v,
\qquad
\dot{\mathbf v}
=A^{-1}
\left(
\ \mathbf{S}_m-\mathbf M_1\mathbf v-\mathbf M_0\mathbf c
\right).
 \label{eq:tdgf-semidiscrete}
\end{align}
We do not form a dense spectral representation of $A^{-1}$.  Instead, the
quantity in parentheses is transformed to the $(R,\theta)$ collocation
grid, divided pointwise by $A(R,\theta)$, and projected back to spectral
space.

\subsection{Gaussian Source and waveform extraction}
\label{app:tdgf-source-extraction}

The radial and temporal delta distributions are approximated by normalized
Gaussians.  For a source at $r_0$, let $R_0=1/r_0$ and
\begin{align}
g_R(R)&=\frac{1}{\sqrt{2\pi}\sigma_R}
\exp\left[-\frac{(R-R_0)^2}{2\sigma_R^2}\right],\\
g_T(T)&=\frac{1}{\sqrt{2\pi}\sigma_T}
\exp\left[-\frac{(T-T_0)^2}{2\sigma_T^2}\right].
\end{align}
The source is injected into one retained angular mode,
\begin{equation}
S_m(T,R,\theta)=C_{\rm src}g_T(T)g_R(R)
{}_sY_{L_{\rm inj}m}(\theta,0),
\label{eq:tdgf-source-form}
\end{equation}
and is set to zero for $T\geq0.5M$.  In the representative runs,
\begin{equation}
C_{\rm src}=\frac{8\pi/R_0^3}{16M^2(1+2MR_0)}.
\label{eq:tdgf-source-normalization}
\end{equation}
This normalization is to be interpreted together with the definition of
$S_m$ in Eq.~\eqref{eq:tdgf-operator-split}.  Convergence toward the
distributional-source limit is tested by varying $\sigma_R$, $\sigma_T$,
and the radial and angular truncations.

The first-order system is evolved using an adaptive fifth-order explicit
Runge--Kutta method.  Sparse operator applications and
spectral--collocation transforms are evaluated with \texttt{CUDA.jl} on an
NVIDIA GeForce RTX 4090 GPU.  The solution is sampled at
$\Delta T=0.1M$.  Since $x(0)=-1$, the waveform at $\mathcal I^+$ is read
directly from the Chebyshev coefficients:
\begin{equation}
\begin{aligned}
\Psi_{4,L m}^{\mathcal I^+}(T_k)
&=\sum_{n=0}^{n_{\max}}(-1)^n c_{n L}(T_k),\\
\end{aligned}
\label{eq:tdgf-waveform-extraction}
\end{equation}
with $T_k =0,0.1M,\ldots,200M$.

\section{Large-frequency angular reconstruction and the split time in Kerr}
\label{app:kerr_split_time}

In this appendix we discuss how the ``split time'' between the
 direct and scattered parts of the Kerr Green function should be
understood.  One subtlety, compared with the Schwarzschild case,
is that the angular basis functions in Kerr are frequency-dependent
spin-weighted spheroidal harmonics.  Therefore, the spheroidal harmonics contribute frequency dependent factors. Additionally, the separation constant ${}_sA_{\ell m \omega}$ that appears in the radial equation is also frequency dependent and has additional frequency-dependent contributions to the radial solutions. Here we discuss its consequences for the large arc contour integral in the radial Green's function as well as the 3+1 Green's function.

Recall that in the Schwarzschild radial problem one may split the
radial Green function into two pieces whose large-frequency phases are
controlled by
\begin{equation}
\begin{split}
    \tilde{G}^- &\sim  e^{-i\omega[t-t'-(r_*-r_*')]},\\
    \tilde{G}^+ &\sim  e^{-i\omega[t-t'-(r_*+r_*')]} .
\end{split}
\label{eq:schw_radial_phases}
\end{equation}
The corresponding radial split times are
\begin{equation}
    t-t'=r_*-r_*',
    \qquad
    t-t'=|r_*|+|r_*'|.
\label{eq:schw_radial_split_times}
\end{equation}
In the full 3+1 dimensional Schwarzschild Green's function, the order of performing the $(\ell,m)$ sum and the contour integral is consequential. Doing the $(\ell,m)$ mode sum first would alter the large frequency asymptotics. Therefore, we instead choose to perform the contour integration first, fixing the $(\ell, m)$ mode. Each mode in the sum will have the split times in~\eqref{eq:schw_radial_split_times}. This may seem at odds with the causality of the full 3+1 Green's function, but the interference from the mode sum  would restore causality.

To study the transition times in Kerr, we study the large frequency asymptotics of the retarded Green's funtion using WKB analysis. Similar to the Schwarzschild case, we fix the spheroidal harmonic $(\ell,m)$ mode while doing the contour integral. The separated retarded Green function has the schematic form (see Eq.~\eqref{TotalKerrGF})
\begin{equation}
\begin{split}
    G_{\rm ret}(x,x')
    =&\sum_{\ell m}
    \int_{\mathcal C}d\omega\,
    e^{-i\omega(t-t')}e^{im(\phi-\phi')}  \\
    &\times
    \frac{1}{\alpha_\ell (a\omega)}
    {}_sS_{\ell m\omega}(\theta)
    {}_sS_{\ell m\omega}(\theta')
    \widetilde G_{\ell m\omega}(r,r') .
\end{split}
\label{eq:sep_green_large_omega}
\end{equation}
Here $\mathcal C$ is the original Fourier contour.  Overall
normalization factors and spin-dependent prefactors have been
suppressed, since they do not affect the leading large-frequency
exponential behavior.  In what follows, $\omega$ denotes a large
complex frequency on the large contour arc. Thus the large frequency dependent exponential contributions to the large arc contour integral come from the spheroidal harmonics ${}_s S_{\ell m \omega}$ and the radial green's function. Notably, the spheroidal harmonics can generate a $\theta$-dependent contribution to the asymptotics.

Furthermore, we note that the WKB analysis of the \emph{geometrical optics} limit where $m \sim \omega$, is very different from the fixed $(\ell,m)$ analysis. In the main text, we consider the spheroidal harmonics for fixed $m$ only, and do not sum over $m$ before doing the contour integral. Consequently, we may take the $m\ll\omega$ limit for the analysis.
%It should not be identified with a quasinormal-mode frequency.

\subsection{Angular Teukolsky equation at large frequency}
\label{app:angular_teukolsky_large_omega}

We now study the large-frequency asymptotics of the angular Teukolsky equation
given in~\eqref{eq:angular_teukolsky_full}.  On the large-frequency contour the
spheroidicity parameter \(c=a\omega\) is complex, and it is useful to regard the
angular problem as depending directly on \(c\).  We denote the phase of \(c\) by
\(\alpha_c=\arg c\), to avoid confusion with the Boyer--Lindquist azimuthal
coordinate.  With the angular-eigenvalue convention used here,
\[
  {}_sA_{\ell m}(c)\equiv 2\eta_{\ell m\omega}-s,
  \qquad c=a\omega ,
\]
the large-\(|c|\) spectrum is expected to exhibit the two standard asymptotic
scalings known from the oblate and prolate limits~\cite{Berti:2005gp,Cook:2026gpm}:
\[
  {}_sA_{\ell m}\sim -c^2 ,
  \qquad
  {}_sA_{\ell m}\sim \pm i(2\bar L+1)c ,
\]
where \(\bar L\) labels the prolate asymptotic branch.  These asymptotic
scalings are properties of eigenvalue branches in the complex \(c\)-plane, and
should not be identified simply with real or pure-imaginary values of \(c\).

\begin{figure*}[t]
  \centering
  \includegraphics[width=\linewidth]{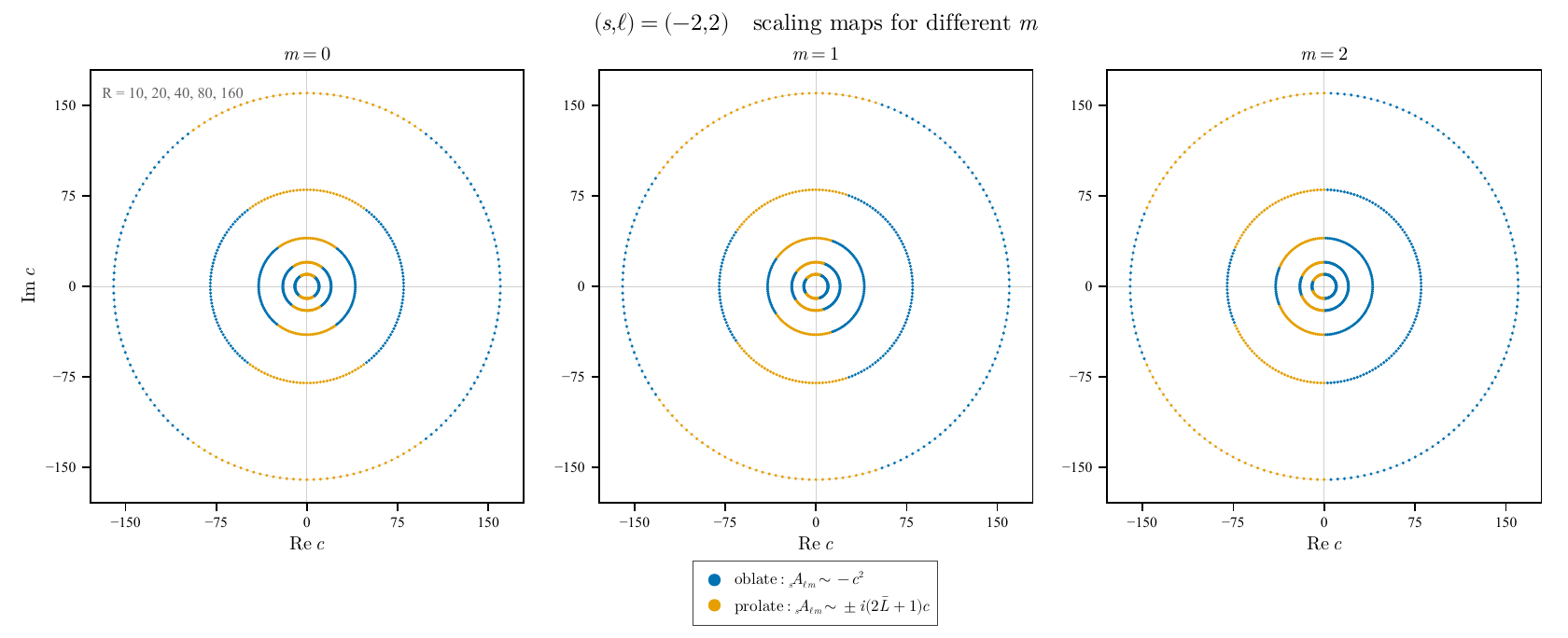}
  \caption{
   Scaling maps of the spin-weighted angular eigenvalue
  \({}_sA_{\ell m}(c)\), evaluated with
  \(c=a\omega\), for fixed \((s,\ell)=(-2,2)\) and \(m=0,1,2\).
  The axes show \(\mathrm{Re}\,c\) and \(\mathrm{Im}\,c\).  The five
  concentric circles correspond to \(|c|=10,20,40,80,160\) in
  \(c=|c| e^{i\alpha_c}\).  Blue points denote the oblate scaling
  \({}_sA_{\ell m}\sim -c^2\), while orange points denote the prolate scaling
  \({}_sA_{\ell m}\sim \pm i(2\bar L+1)c\).
%\MC{Sounds good (in fact, instead of introducing $R$, $\alpha_c$ and \(c=R e^{i\alpha_c}\) maybe we could just use $|c|$ instead of $R$?)}  
}
  
  \label{fig:sector-polar-atlas}
\end{figure*}

As a numerical check of this picture, we computed scaling maps in the complex
\(c\)-plane for fixed \((s,\ell)=(-2,2)\) and \(m=0,1,2\);
see Fig.~\ref{fig:sector-polar-atlas}.  The maps use the radii
\(R=10,20,40,80,160\), with \(c=R e^{i\alpha_c}\), to compare the apparent
large-\(|c|\) behavior along different complex directions.  In the tested
modes and radii, the eigenvalue branches are well described by the known
oblate and prolate scalings, and we did not find a stable open region requiring
a distinct leading power.  The result also shows a visible \(m\)-dependence:
for \(m\neq0\), the prolate region is shifted in the complex \(c\)-plane.
Thus a single \((s,m,\ell)\) mode can encounter both oblate-type and
prolate-type asymptotic behavior as the large semicircle in the complex
\(\omega\)-plane passes through different directions of \(c=a\omega\).

We emphasize that this is a numerical experiment, not a proof of a complete
sector classification.  In particular, the absence of an additional sector in
Fig.~\ref{fig:sector-polar-atlas} should be understood as negative numerical
evidence within the tested range, rather than a rigorous mathematical exclusion.
A deeper theoretical derivation of the complex-\(c\) sector structure is still
lacking, and the subsequent analysis therefore assumes the two observed
asymptotic scalings while keeping this limitation in mind.

In what follows we use this two-balance description of the angular asymptotics.
The numerical maps do not constitute a mathematical proof that no further
asymptotic sectors exist on all sheets, but they provide no evidence for an
additional large-arc sector in the modes and radii tested here.

We now discuss the two fixed-mode asymptotic sectors in turn.  In each
case the axis named below is a representative ray on which the WKB
geometry is real; the defining distinction is the eigenvalue scaling,
which extends to complex $c$~\cite{Berti:2005gp,Cook:2026gpm}.

\paragraph*{Oblate-type sector: ${}_sA_{\ell m}\sim-c^2$.}
Branches in this sector satisfy ${}_sA_{\ell m}(c)=-c^2+O(c)$ as
$|c|\to\infty$, including along generic complex paths.  On the positive
real-axis representative, the more detailed expansion is
\begin{equation}
    {}_sA_{\ell m}(c)
    =
    -c^2+2\,{}_sq_{\ell m}\,c+O(1),
\label{eq:A_oblate_large_c}
\end{equation}
where the coefficient ${}_sq_{\ell m}$ is fixed by the zero-counting
condition of the real-axis oblate asymptotic solution~\cite{Casals:2004zq,Casals:2018cgx}.  The next-to-leading term is not known in this form for general complex $c$; the important point for the present discussion is the sector-defining leading term $-c^2$.

On this representative ray, the scaling can be understood from the fixed-mode WKB momentum
\begin{equation}
    P_\theta^2(\theta)
    =
    c^2\cos^2\theta
    -\frac{m^2}{\sin^2\theta}
    +{}_sA_{\ell m}(c)+O(c),
\label{eq:fixed_lm_ptheta}
\end{equation}
where the $O(c)$ terms include the spin-dependent pieces.  If one tried
${}_sA_{\ell m}=-\alpha c^2+o(c^2)$ with $\alpha<1$, an oscillatory
region of order-unity angular width would remain and the number of nodes
would grow with $|c|$.  If $\alpha>1$, no such finite oscillatory well is
available.  The fixed-$(\ell,m)$ spectrum therefore lies at the boundary
between these cases, giving Eq.~\eqref{eq:A_oblate_large_c}.

Near the North pole, set
\begin{equation}
    y=\sqrt{c}\,\theta,
    \qquad
    u=y^2\simeq c\sin^2\theta .
\label{eq:north_boundary_layer_y}
\end{equation}
Keeping the scalar leading terms, Eq.~\eqref{eq:fixed_lm_ptheta} becomes
\begin{equation}
    P_\theta^2
    \simeq
    c\left(
    \beta-y^2-\frac{m^2}{y^2}
    \right),
    \qquad
    \beta=2\,{}_sq_{\ell m}
\label{eq:north_boundary_layer_p}
\end{equation}
up to spin-dependent shifts of the $O(c)$ coefficient.  The turning
points are determined by
\begin{equation}
    \beta-y^2-\frac{m^2}{y^2}=0,
\end{equation}
or
\begin{equation}
    u_\pm
    =
    \frac{\beta\pm\sqrt{\beta^2-4m^2}}{2} .
\label{eq:u_turning_points}
\end{equation}
Thus the North-pole turning points scale as
\begin{equation}
    \theta_\pm^{(N)}
    \simeq
    \sqrt{\frac{u_\pm}{c}} .
\label{eq:north_turning_points}
\end{equation}
The allowed region is the thin island
\begin{equation}
    \theta_-^{(N)}<\theta<\theta_+^{(N)},
\label{eq:north_allowed_region}
\end{equation}
not the interval from the pole to the first turning point.  This is
because $P_\theta^2\to -\infty$ as $\theta\to0$ due to the centrifugal
singularity.  The corresponding allowed-region action is finite,
\begin{equation}
    \int_{\theta_-^{(N)}}^{\theta_+^{(N)}}
    P_\theta\,d\theta
    \simeq
    \int_{\sqrt{u_-}}^{\sqrt{u_+}}
    dy\,
    \left(
    \beta-y^2-\frac{m^2}{y^2}
    \right)^{1/2},
\label{eq:north_action_finite}
\end{equation}
and this finite action quantizes the linear coefficient.  The South-pole
boundary layer is obtained by replacing $\theta$ with $\chi=\pi-\theta$.
For spin-weighted harmonics the singular indices are shifted near the
two poles, schematically $m^2\to(m+s)^2$ near $\theta=0$ and
$m^2\to(m-s)^2$ near $\theta=\pi$, together with additional spin-dependent
$O(c)$ shifts.  These changes modify ${}_sq_{\ell m}$ but not the leading
$-c^2$ behavior.

%\begin{figure*}[t]
%   \centering
%   \includegraphics[width=0.92\textwidth]{turning.png}
%    \caption{Schematic angular structure for fixed $(\ell,m)$ and large
%    real $|c|=|a\omega|$.  The oblate eigenvalue
%    ${}_sA_{\ell m}(c)\simeq -c^2+\beta c$ produces two thin allowed
%    islands near the poles, separated both from the poles and from each
%    other by forbidden regions.  The allowed islands have width
%    $O(c^{-1/2})$ and provide the finite action that quantizes the
%    coefficient $\beta$.  Away from these islands, the angular WKB
%    solution is exponentially growing or decaying.}
%    \label{fig:fixed_lm_turning_points}
%\end{figure*}

For angles away from the thin allowed islands, the oblate fixed-mode
momentum satisfies
\begin{equation}
    P_\theta^2=-c^2\sin^2\theta+O(c),
\label{eq:forbidden_middle_p}
\end{equation}
so that
\begin{equation}
    P_\theta\simeq \pm i c\sin\theta .
\label{eq:forbidden_middle_momentum}
\end{equation}
Consequently the local branches contain the exponential factors
\begin{equation}
    {}_sS_{\ell m}(\theta;c)
    \propto
    \exp\left[\pm c\cos\theta\right]
\label{eq:oblate_fixed_lm_angular_exponential}
\end{equation}
up to algebraic and subleading exponential corrections.  Equivalently,
a branch decaying away from the North-pole island behaves as
\begin{equation}
    {}_sS_{\ell m}^{(N)}(\theta;c)
    \sim
    \mathcal A_N(\theta;c)
    \exp[-c(1-\cos\theta)],
\label{eq:oblate_north_decay}
\end{equation}
while a branch decaying away from the South-pole island behaves as
\begin{equation}
    {}_sS_{\ell m}^{(S)}(\theta;c)
    \sim
    \mathcal A_S(\theta;c)
    \exp[-c(1+\cos\theta)] .
\label{eq:oblate_south_decay}
\end{equation}
For non-real $c$ in the oblate-type sector, the WKB actions are complex
and the dominant/subdominant assignment is determined by their real
parts and by Stokes switching; the literal pole-localization picture
above is the positive-real-axis representative.

\paragraph*{Prolate-type sector: ${}_sA_{\ell m}=O(c)$.}
For branches in this sector with $\operatorname{Im}c\ne0$, the leading
behavior is
\[
    {}_sA_{\ell m}(c)
    \sim
    \begin{cases}
       -ic(2\bar L+1), & \operatorname{Im}c>0,\\
       +ic(2\bar L+1), & \operatorname{Im}c<0,
    \end{cases}
\]
where $\bar L$ is a zero-based asymptotic index for the prolate-type
family~\cite{Cook:2026gpm}. It need not agree globally with the
spherical-limit ordering after eigenvalue crossings.  On the
pure-imaginary representative, write $c=i c_I$ and $C\equiv|c_I|$; then
\begin{equation}
    {}_sA_{\ell m}(ic_I)
    =
    (2\bar L+1)C+O(1).
\label{eq:A_prolate_large_c}
\end{equation}  The
linear scaling on this representative ray follows from an inner solution near the equator.  With
$x=\cos\theta$ and
\begin{equation}
    u=\sqrt{2C}\,x,
\label{eq:prolate_inner_u}
\end{equation}
the leading inner equation reduces to a parabolic-cylinder problem.  Its
regular solutions have a fixed number of zeros in the equatorial region,
and this quantizes the coefficient of $C$ in
Eq.~\eqref{eq:A_prolate_large_c}.

For $x=\cos\theta$ away from both the equatorial inner region and the
endpoints, the dominant term in the angular equation is
$(cx)^2=-C^2x^2$.  Thus
\begin{equation}
    P_\theta^2\simeq -C^2\cos^2\theta,
\label{eq:prolate_outer_momentum}
\end{equation}
and the outer WKB branches contain
\begin{equation}
    \exp\left[\pm C\sqrt{1-x^2}\right]
    =
    \exp\left[\pm C\sin\theta\right] .
\label{eq:prolate_outer_exp}
\end{equation}
After matching to the equatorial inner solution and choosing the branch
that is exponentially suppressed away from the equatorial localization
region, the relative behavior can be written schematically as
\begin{equation}
    {}_sS_{\ell m}(\theta;ic_I)
    \sim
    \mathcal A_P(\theta;C)
    \exp[-C(1-\sin\theta)] .
\label{eq:prolate_relative_decay}
\end{equation}
Thus, on the pure-imaginary representative, prolate-type spheroidal
harmonics are localized near the equator, in contrast to the
positive-real oblate representative where the thin allowed islands lie
near the poles.  For non-imaginary $c$ in the prolate-type sector, the
corresponding actions are complex and this localization statement is
understood through their Stokes sectors rather than as a literal real
turning-point geometry.

\paragraph*{General complex $c$.}
For generic complex $c$, the oblate- and prolate-type limits are not tied
to the real and imaginary axes.  Rather, at fixed $(s,m)$ the
large-$|c|$ spectrum separates into branches with either
${}_sA_{\ell m}\sim-c^2$ or
${}_sA_{\ell m}\sim\pm ic(2\bar L+1)$.  Along a single generic ray,
different branches may lie in different sectors; as $|c|\to\infty$,
finite subsets of the two sectors separate, while at finite $|c|$
eigenvalue crossings and branch points can occur.
The corresponding WKB saddles and Stokes lines change with $\arg c$.
Therefore a large semicircle in the complex $\omega$ plane cannot be characterized by one branch-independent angular exponential of the form $\exp[i\omega f(\theta)]$ with a unique function $f(\theta)$. The additional angular factor is therefore $\theta-$ and sector-dependent.

\subsection{Radial Teukolsky equation at large frequency}
\label{app:radial_teukolsky_large_omega}

The radial Teukolsky equation, given in Eq.~\eqref{eq:radial_teukolsky_equation}, can equivalently be transformed to a Schr\"odinger-like
radial equation. The leading high-frequency equation is
\begin{align}
    \frac{d^2u_r}{dr_*^2}
    +Q_{\ell m\omega}(r)u_r=0,
\end{align}
where
\begin{align}
    Q_{\ell m\omega}(r)
    &=
   \omega^2 - \frac{\Delta\lambda_{\ell m\omega}}{(r^2+a^2)^2}.
\label{eq:Qrad_large_omega}
\end{align}

Now, in the oblate sectors, $\lambda_{\ell m\omega}$ is only $O(\omega)$.  Thus, one obtains
$Q_{\ell m\omega}=\omega^2+O(\omega)$ in the large $\omega$ limit, and the radial homogeneous
solutions behave as
\begin{equation}
    R_{\ell m\omega}^{\rm up}\sim e^{+i\omega r_*},
    \qquad
    R_{\ell m\omega}^{\rm down}\sim e^{-i\omega r_*},
\label{eq:radial_fixed_lm_up_down}
\end{equation}
up to subexponential factors.  This gives the familiar radial phases
$r_*-r_*'$ and $r_*+r_*'$ for the fixed-$(\ell,m)$ radial Green function.

In the prolate sector we expect that $\lambda_{\ell m\omega} \sim a^2 \omega^2$ and 
\begin{align}
Q_{\ell m\omega} =\omega^2 -\frac{a^2\omega^2 \Delta}{(r^2+a^2)^2} +\mathcal{O}(\omega)\,.
\end{align}

As a result, we can define
\begin{align}
    f(r) = \int dr_* \sqrt{1-\frac{a^2\Delta}{(r^2+a^2)^2}}
\end{align}
and the radial homogeneous
solutions behave as
\begin{equation}
    R_{\ell m\omega}^{\rm up}\sim e^{+i\omega f(r)},
    \qquad
    R_{\ell m\omega}^{\rm down}\sim e^{-i\omega f(r)},
\label{eq:radial_fixed_lm_up_down_prolate}
\end{equation}
up to subexponential factors. Note that in the Schwarzschild limit Eq.~\eqref{eq:radial_fixed_lm_up_down_prolate} reduces to  Eq.~\eqref{eq:radial_fixed_lm_up_down}.

Consequently, the radial solutions to Kerr black holes also have a nontrivial, sector-dependent phase factor in the large $|\omega|$ limit.

\subsection{Large-arc convergence}
Suppressing algebraic prefactors and branch labels, the large-frequency integrands for $\widetilde G^+$ and $\widetilde G^-$ can be schematically  written as
\begin{equation}
\mathcal I(\omega)
\sim
\exp\left(
-i\omega\left[
t-t'
-\tau_{\rm Schw}(r',r)
-\mathcal{T}\left(
r',\theta',r,\theta, \vartheta_\omega
\right)
\right]
\right),
\end{equation}
where $\tau_{\rm Schw}$ denotes the corresponding Schwarzschild phase, including the appropriate direct or scattered contribution, and
\begin{equation}
\mathcal T\left(
r',\theta',r,\theta,\vartheta_\omega
\right)=O(a),
\end{equation}
where $\vartheta_\omega=\mathrm{Arg}(\omega)$ encodes the fact that different portions of the large complex-frequency arc lie in different prolate- or oblate-type asymptotic sectors. In particular, there is no reason for the Kerr correction to have the same sign, or even the same asymptotic form, in every sector.

Provided that $\mathcal T$ remains uniformly $O(a)$ on the relevant sectors, this sector dependence only shifts the Schwarzschild convergence boundary by an $O(a)$ amount. Once $t-t'$ lies more than $O(a)$ away from the corresponding Schwarzschild transition time, the Schwarzschild part of the phase dominates the bounded Kerr correction in every sector, and the large arc can again be closed in the appropriate half-plane. There is therefore only an $O(a)$ time window around the Schwarzschild transition in which the sector dependence can be important. Inside this window, some sectors of the large arc may be exponentially suppressed while others are exponentially enhanced. Consequently, neither an upper- nor a lower-half-plane closure need be uniformly valid there, and the large-arc contribution cannot in general be discarded on the basis of the present asymptotic analysis alone.

\section{Subtleties in QNMs}
\label{sec:qnm-subtleties}

\begin{figure*}[t]
  \centering
  \includegraphics[width=\linewidth]{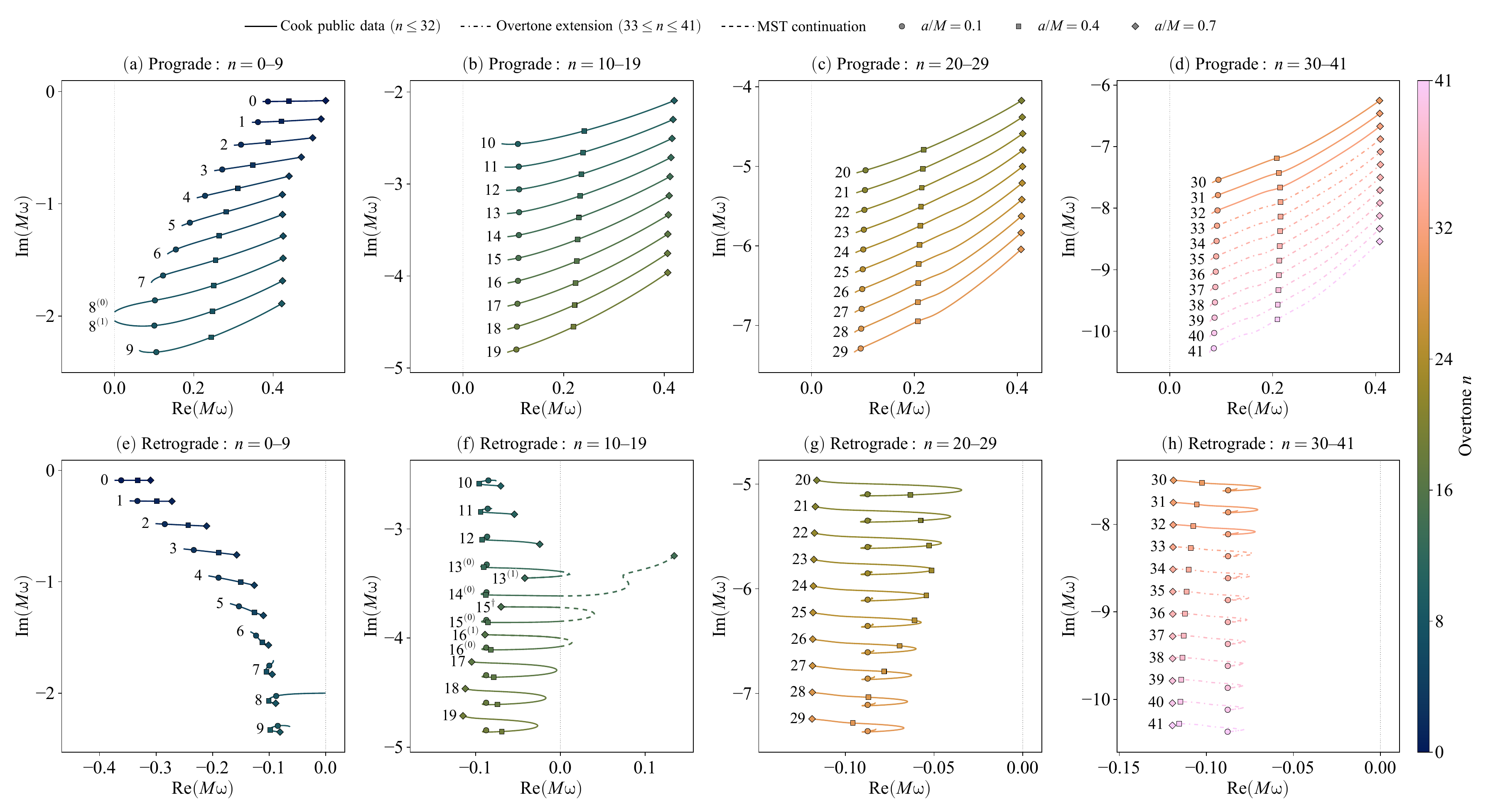}
  \caption{
       Trajectories of the QNM frequencies in the complex \(M\omega\) plane as \(a/M\) varies from \(0\) to \(0.7\), the largest dimensionless spin considered in this paper. The upper and lower rows show prograde and retrograde modes, respectively. Circles, squares, and diamonds mark the QNM frequencies at \(a/M=0.1,\ 0.4,\) and \(0.7\), respectively, the three spins used in our calculations. Solid curves show Cook’s public data through \(n=32\). Dash-dotted curves show the overtone extension to \(33\leq n\leq41\), and dashed curves show segments reconstructed using the MST method. Superscripts \((0)\) and \((1)\) identify the branches stored in Cook’s HDF5 files. The label \(15^\dagger\) marks the retrograde \(n=15\) continuation stored in Cook’s files as \(n=14\), branch 1.
  }
  \label{fig:QNM_spectrum}
\end{figure*}
%%%
%%%   
%%%
We obtain the QNM frequency initial guess from Cook’s tabulated Kerr QNM data \cite{Cook_2025_KerrModes} and use  our MST code to check and refine the frequencies. For the \(s=-2\), \(\ell=m=2\) mode studied in this paper, the prograde frequencies are taken from  Cook's \(m=2\) sequences, whereas the retrograde frequencies are computed by the  relation:
%%%
%%%   
%%%
\begin{equation}
     \omega_{\ell m n} =-\left[\omega_{\ell (-m) n}\right]^*,
\end{equation}
where $n$  is the overtone number. The QNM condition is
\begin{equation}
     \mathcal A(s,\ell,m,a,\omega) \equiv \frac{A_{\mathrm{inc}}^{\mathrm in}(s,\ell,m,a,\omega)}{A_{\mathrm{trans}}^{\mathrm in}(s,\ell,m,a,\omega)} =0.
\label{eq:qnm_condition}
\end{equation}
In our implementation, $A_{\mathrm{inc}}^{\mathrm in}$ and $A_{\mathrm{trans}}^{\mathrm in}$ are obtained from Eq.~\eqref{eq:mst-ain-inc} and Eq.~\eqref{eq:mst-ain-trans},
with the angular separation constant evaluated on the continuously followed angular branch.   
In practice, we start from a reliable point in Cook's dataset. At each subsequent step, we use the QNM frequency at the previous value of \(a\) to predict the frequency at a neighboring value of \(a\), and then correct this prediction using the MST code. The initial root calculations use 320-bit arithmetic and 320 terms in each direction of the recurrence defined in Eq.~\eqref{eq:nu_cf}. The angular equation is solved using the expansion in Eq.~\eqref{eq:Spheroidal_Spherical_Expansion}, with its upper angular-basis cutoff set to \(L_{\max}=50\). Numerical convergence is checked by recomputing the roots at 384-bit precision with 400 recurrence terms in each direction and \(L_{\max}=60\).

\subsection{Radial sheet consistency}

A subtlety arises when a trajectory reaches the NIA branch cut of the radial solution. A root can continue across the branch cut onto an adjacent radial sheet, even though it remains a smooth trajectory in the complex-frequency plane.

For the QNM frequencies considered here, we evaluate the $\mathcal{A}(s,\ell,m,a,\omega)$ using its standard physical-sheet convention in the right half of the complex-frequency plane. To obtain the physical-sheet amplitude at a frequency with \(\operatorname{Re}\omega<0\), we use the mentioned symmetry. For convenience, we define the following reflected parameters
$ (\widetilde\omega,\widetilde m)
= (-\omega^*,-m)$.  
We first calculate the standard amplitudes \(\widetilde A_{\mathrm{inc}}^{\mathrm{in}}\) and \(\widetilde A_{\mathrm{trans}}^{\mathrm{in}}\) at these reflected parameters, and then obtain
\begin{equation}
\mathcal A_{\mathrm{phys}}(s,\ell,m,a,\omega)
= \left[ \frac{\widetilde A_{\mathrm{inc}}^{\mathrm{in}}}
              {\widetilde A_{\mathrm{trans}}^{\mathrm{in}}} \right]^*.
\end{equation}
Tracking a QNM continuously across the radial branch cut requires a different prescription. Instead of switching to the physical-sheet amplitude on the other side of the cut, we continue the original $\mathcal A(s,\ell,m,a,\omega)$ (i.e. the left hand side of Eq.~\eqref{eq:qnm_condition}) from MST expression Eq.~\eqref{eq:mst-ain-inc} and Eq.~\eqref{eq:mst-ain-trans} along the QNM trajectory. We denote the resulting amplitude by \(\mathcal A_{\mathrm{cont}}\). After crossing the cut, this continued amplitude generally differs from \(\mathcal A_{\mathrm{phys}}\).
We distinguish the two radial branches by evaluating both amplitudes at the tracked root. If \(\mathcal A_{\mathrm{cont}}=0\) but \(\mathcal A_{\mathrm{phys}}\ne0\), the root lies on a different radial Riemann sheet.

Radial-sheet transitions have long been discussed in connection with the \(n=8\) overtone family near the Schwarzschild algebraically special frequency \cite{Leung_2003,Cook_2014}. Recent studies have explicitly traced the relevant pole trajectories across Riemann sheets \cite{kubota2026poleskippingavoidedcrossing,chen2026completequasinormalmodestyped}. Our MST continuation also resolves such transitions for the retrograde overtones \(13\leq n\leq16\) as shown in Fig.~\ref{fig:QNM_spectrum}. The latter are not shown here and are excluded from the physical-sheet QNM sum used in our \(a/M=0.4\) kernel calculation.

Our calculations also resolve an overtone-label mistake in Cook's retrograde dataset, which was already noted by Cook and Zalutskiy \cite{Cook_2025_KerrModes}. As shown in panel (f) of Fig.~\ref{fig:QNM_spectrum}, the continuously tracked retrograde \(n=15\) trajectory enters an adjacent radial sheet and subsequently returns to the physical sheet. The returning segment is catalogued as ``branch 1'' of \(n=14\). By following the trajectory continuously from its Schwarzschild-anchored \(n=15\) starting branch, our MST calculation identifies this segment as the continuation of \(n=15\), rather than of \(n=14\). In contrast, the continuously tracked \(n=14\) branch remains on the adjacent radial sheet at \(a/M=0.7\), as shown in the same panel.

\subsection{Angular sheet consistency}

\begin{figure*}[t]
  \centering
  \includegraphics[width=0.95\linewidth]{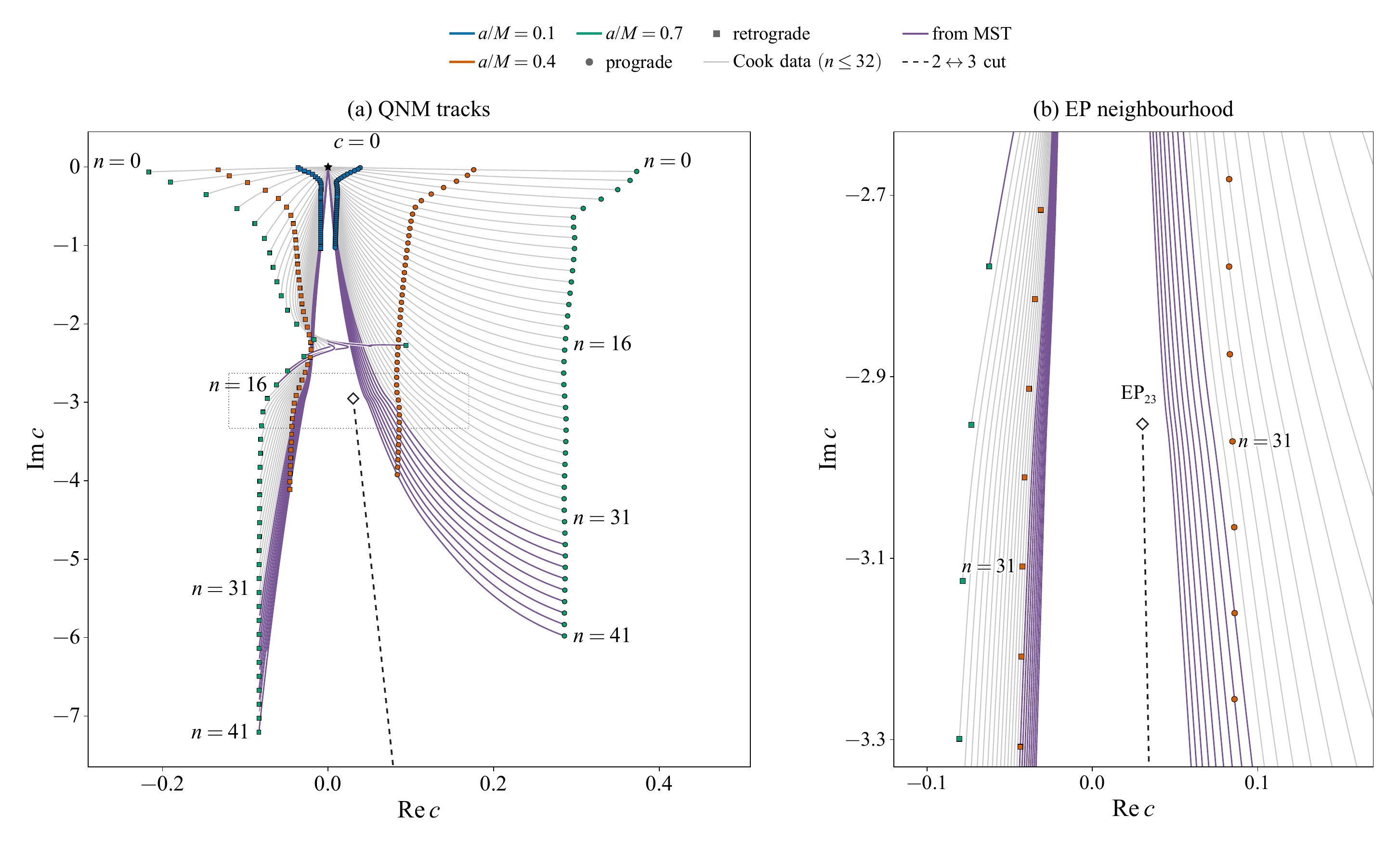}
\caption{
QNM trajectories in the \(c=a\omega\) plane for the
\((s,\ell,m)=(-2,2,2)\) channel over \(0\le a/M\le0.7\).
(a) Trajectories for \(n=0,\ldots,41\). Segments obtained from
Cook's data~\cite{Cook_2025_KerrModes} are shown in gray, while
MST-computed completion segments ($n=13,14,15,16$) and independently computed full
trajectories ($n>32$) are shown in purple.
(b) Enlarged view near the \(2\leftrightarrow3\) angular exceptional
point (diamond) and the adopted cut (dashed line).
No trajectory crosses the cut.
}
  \label{fig:angular-sheet-consistency}
\end{figure*}

In this subsection, we provide numerical evidence that the QNMs used
in this paper do not cross the angular branch cut and therefore remain on the same angular Riemann sheet.

For complex frequencies, the angular functions depend on the
spheroidicity \(c=a\omega\). For the
\((s,\ell,m)=(-2,2,2)\) QNMs, the physical angular sheet is the
analytic branch connected to the spherical-limit state
\({}_{-2}Y_{22}\) at \(c=0\). Prograde and retrograde overtones share
this angular seed, although their frequencies and angular
eigenfunctions at finite \(c\) differ.

Figure~\ref{fig:angular-sheet-consistency} (a) shows the available QNM trajectories \(c_n(a)=a\omega_n(a)\) for \(n=0,\ldots,41\), as the spin varies over \(0\le a/M\le0.7\), the range considered in this paper. The figure combines data from Ref.~\cite{Cook_2025_KerrModes} with our MST-computed completion segments and independently computed full trajectories. It also marks the frequencies used at \(a/M=0.1,0.4,0.7\).

The \(2\leftrightarrow3\) angular exceptional point is located at
\(c_{\rm EP}\simeq0.0303725-2.9520981\,i\), and the displayed cut
follows our convention
\begin{equation}
 \mathcal C_{23}
 =\left\{
 c_{\rm EP}+t\frac{c_{\rm EP}}{|c_{\rm EP}|}:t\ge0
 \right\}.
\end{equation}
The enlarged view in Fig.~\ref{fig:angular-sheet-consistency}(b)
shows that the high-overtone trajectories remain separated from the
selected angular cut. None of the displayed trajectories crosses the
cut over the spin range considered. This provides numerical evidence
that the higher overtones used in the numerical calculations in this
paper remain on the same physical angular sheet connected to
\({}_{-2}Y_{22}\) at \(c=0\).

This angular-sheet identification is distinct from the radial-sheet
classification discussed in the previous subsection.

\begin{figure*}[h]
  \centering
  \includegraphics[width=1.05\linewidth]{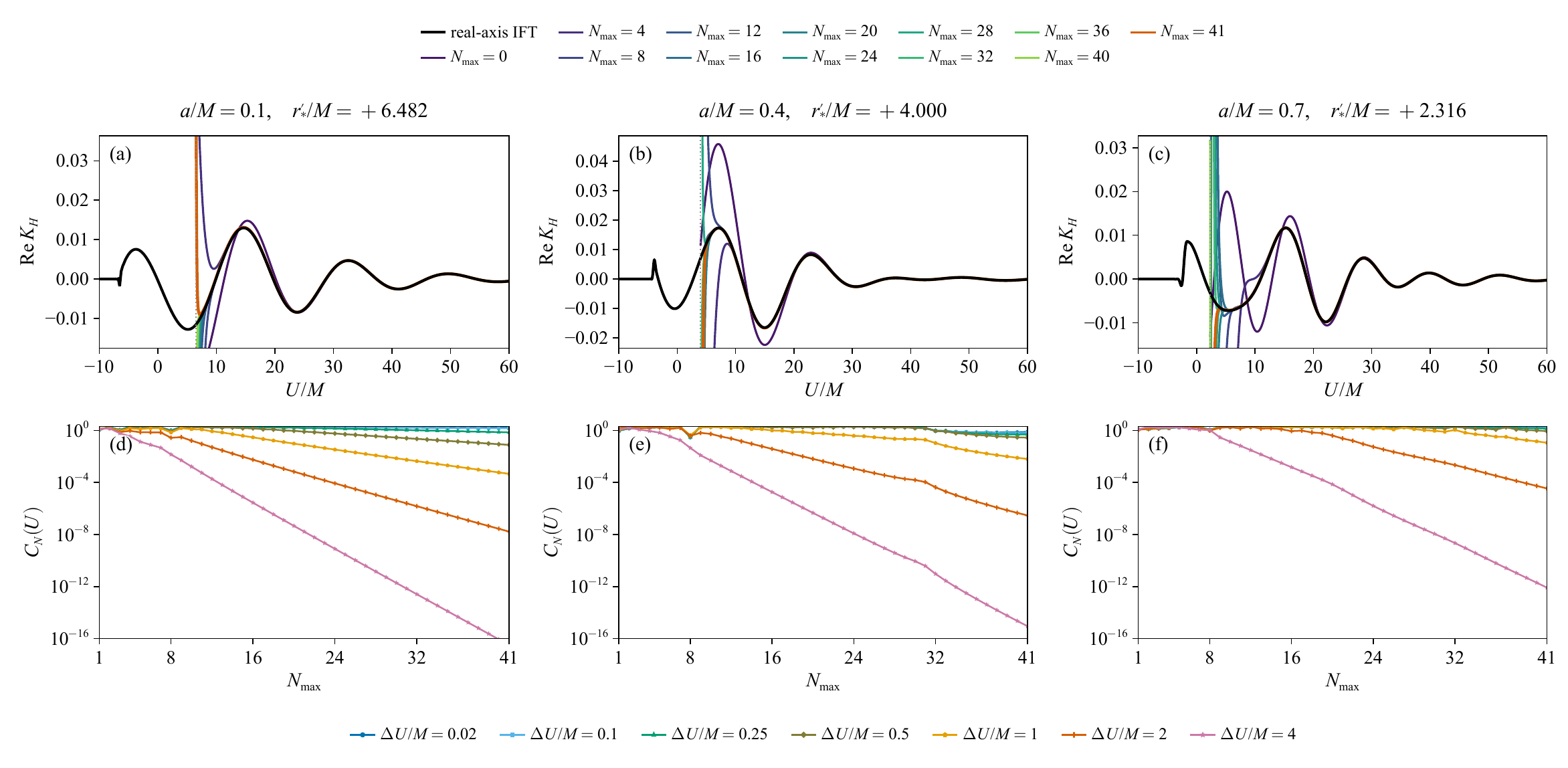}
\caption{ 
  Convergence of the QNM sum for the metric radiative kernels shown in Fig.~\ref{fig:kernel-direct}. \textit{Top panel}: Comparison between the QNM waveform reconstructed with the QNM sum truncated at different overtone numbers and the waveform obtained from the real-frequency inverse Fourier transform (IFT). The vertical dashed line marks the time when $U = |r_*'|$ (named as \emph{reference front}). \textit{Bottom panel}: Normalized successive increment as defined in Eq.~\eqref{eq:normalized_successive_increment}.
  The data point corresponding to $N_{\rm max}=15$ at $a=0.7$ is excluded because the $n=14$ QNM moves onto the adjacent radial Riemann sheet. 
}
  \label{fig:PositiveSourceQNMConvergence}
\end{figure*}
\begin{figure*}[h]
  \centering
  \includegraphics[width=1.05\linewidth]{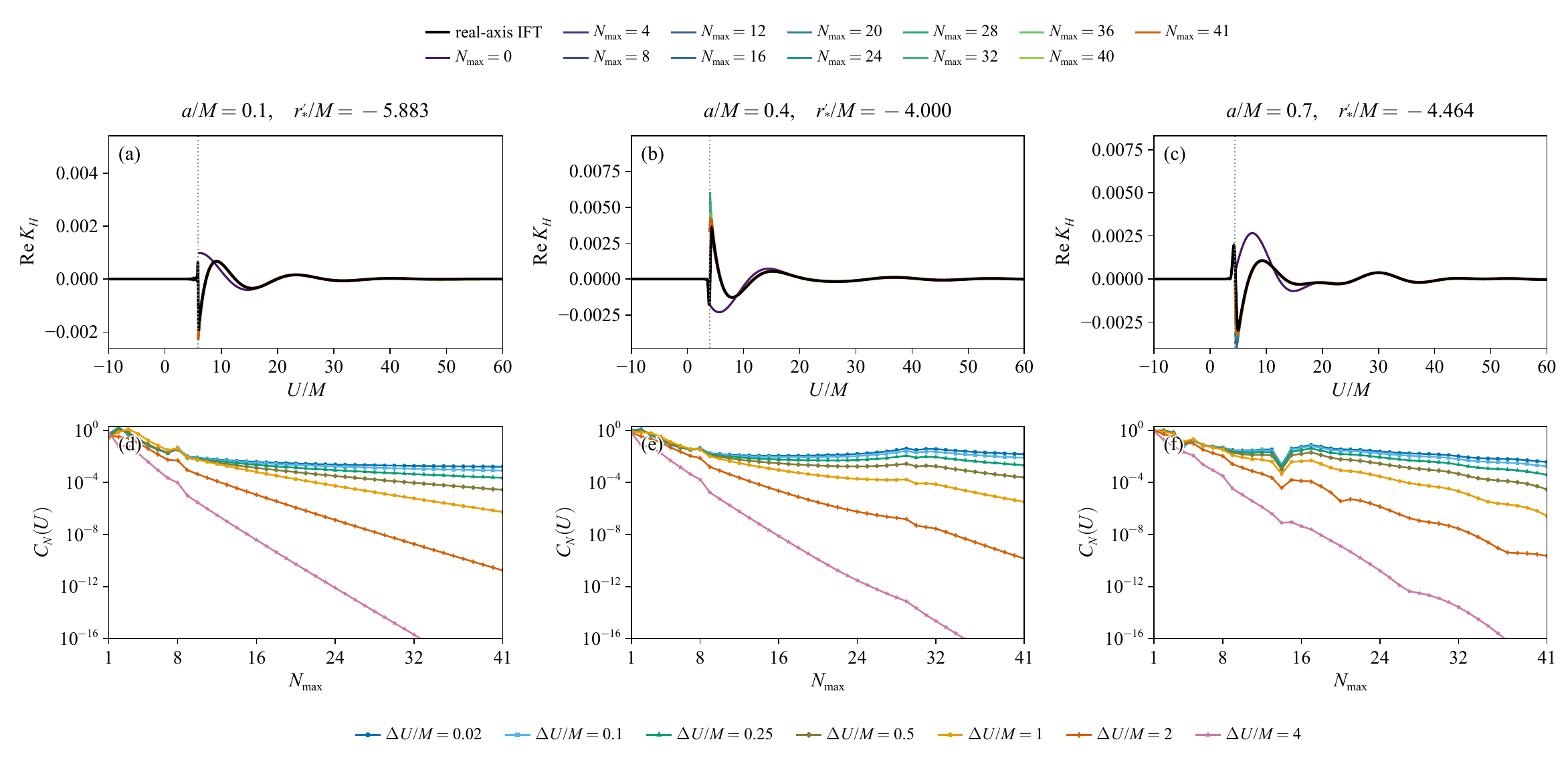}
\caption{ 
Same as Fig.~\ref{fig:PositiveSourceQNMConvergence}, but for the metric radiative kernels with negative source positions shown in Fig.~\ref{fig:kernel-no-direct}, at different black hole spins.
}
  \label{fig:NegativeSourceQNMConvergence}
\end{figure*}

\subsection{QNM sum convergence}\label{sec:QNMconv}

To examine the convergence of the QNM sum, we define $K_N(U)$
as the QNM contribution truncated at overtone label $N$:
\begin{equation}
 K_N(U)=\sum_{n=0}^{N}\sum_b A_{n,b}e^{-i\omega_{n,b}U},
 \label{eq:naive-qnm-contribution}
\end{equation}
where, for each $n$, $b$ runs over the prograde and
retrograde poles with that overtone label. We count each distinct
pole once.

Cook and Zalutskiy~\cite{Cook_2014} discuss the prograde $n=8$
doublet near the Schwarzschild algebraically special frequency
$M\omega=-2i$; see their Fig.~23. At the spins considered here,
the two prograde modes with $n=8$ have distinct frequencies.
We therefore compute their frequencies and residues separately
and retain both contributions. Overtone multiplets can also occur
in other $(\ell,m)$ sectors, so the sum allows more than one pole
for a given overtone label.

To make the prograde and retrograde contributions explicit, we
write Eq.~\eqref{eq:naive-qnm-contribution} as
\begin{equation}
 K_N(U)=
 \sum_{\sigma\in\{\mathrm{pro},\mathrm{retro}\}}
 \sum_{n=0}^{N}
 \sum_{b\in\mathcal B_{n,\sigma}}
 A_{n,b}^{\sigma}\,e^{-i\omega_{n,b}^{\sigma}U}.
\end{equation}
For a fixed overtone label $n$ and sector $\sigma$,
$\mathcal B_{n,\sigma}$ is the set of indices $b$ assigned to the
distinct physical-sheet poles retained in that group. Each index
$b$ identifies one pole with frequency $\omega_{n,b}^{\sigma}$.
If no pole is retained for a given $(n,\sigma)$, the set is empty
and the inner sum contributes zero. The coefficients
$A_{n,b}^{\sigma}$ include the source dependence and the time-domain
normalization of the metric kernel.

For the prograde $n=8$ doublet, $\mathcal B_{8,\mathrm{pro}}=\{0,1\}$,
so the two poles contribute
\begin{equation}
 A_{8,0}^{\mathrm{pro}}e^{-i\omega_{8,0}^{\mathrm{pro}}U}
 +
 A_{8,1}^{\mathrm{pro}}e^{-i\omega_{8,1}^{\mathrm{pro}}U}.
\end{equation}
Both terms enter $K_N(U)$ for $N\geq8$, together with the
corresponding retrograde contribution at $n=8$.

At fixed $U$, we measure the change between successive QNM partial
sums using the normalized increment
\begin{equation}
 C_N(U)=
 \frac{|K_N(U)-K_{N-1}(U)|}
 {\max\!\left\{|K_N(U)|,\,|K_{N-1}(U)|,\,
               10^{-8}K_{\mathrm{scale}}\right\}},
\label{eq:normalized_successive_increment}
\end{equation}
where $N=1,\ldots,41$.
For each spin and source position, we use a single reference scale
\begin{equation}
 K_{\mathrm{scale}}
 = \max_{\substack{j\\0\leq k\leq41}} |K_k(U_j)|,
\end{equation}
where $U_j$ runs over the fixed sampling times used in the
convergence test. The floor $10^{-8}K_{\mathrm{scale}}$ keeps the
denominator from becoming too small when both partial sums are
close to zero.

Figures~\ref{fig:PositiveSourceQNMConvergence} and \ref{fig:NegativeSourceQNMConvergence} show how the QNM sums change with the overtone cutoff for fixed source points at positive and negative $r_*'$, respectively. The quantity $N_{\max}$ shown
in the figures is the cutoff $N$ in $K_N(U)$ and $C_N(U)$:
$K_N(U)$ includes the retained physical-sheet poles with overtone
labels $0\leq n\leq N$. The largest cutoff considered is $N=41$.

At $N=41$, the prograde sum contains 43 poles for each spin
because it includes both roots labeled $n=8$. The retrograde
sum contains 42 poles at $a/M=0.1$ and $0.4$. At $a/M=0.7$,
it contains 41 poles because we exclude the retrograde $n=14$
mode, which is on the adjacent radial Riemann sheet.
We still plot $C_{14}(U)$ for this spin because
$K_{14}(U)-K_{13}(U)$ includes the prograde $n=14$ contribution.

In the lower-row panels, we evaluate $C_N(U)$ at fixed offsets
\begin{equation}
 \Delta U=U-U_0,\qquad U_0=|r_*'|,
\end{equation}
with $\Delta U/M=0.02,0.1,0.25,0.5,1,2,4$. We use $U_0$ as a
radial reference time for comparing the different source positions,
without assuming that it coincides with the onset of convergence
of the angularly resolved QNM expansion.

At the smallest offset, $\Delta U/M=0.02$, the negative-source
value of $C_{41}$ is smaller than the corresponding positive-source
value at each spin. This also holds for $a/M=0.4$: although the
increments in panel~(e) of Fig.~\ref{fig:NegativeSourceQNMConvergence}
do not decrease monotonically with $N$, the final increment is
$C_{41}\simeq0.015$, compared with $0.78$ for the positive source.
At $\Delta U/M=4$, the final increment is much smaller for both
signs of $r_*'$.

The diagnostic $C_N$ measures the change between successive partial
sums and shows their sensitivity to the overtone cutoff near the
front. Small changes at the largest values of $N$ considered here,
however, do not establish convergence of the infinite series.
For related work on QNM excitation and ringdown reconstruction,
see Refs.~\cite{oshita2024reconstructionringdownexcitationfactors,
Oshita:2026vxh}. Determining when the QNM sum converges for the
Kerr kernel considered here would require large-overtone estimates
of both the QNM frequencies and the full residue weights, including
the radial Green's functions and angular factors.

\bibliography{References}

\end{document}